\documentclass[3p, nopreprintline]{elsarticle}

\usepackage{framed,multirow}

\usepackage{amssymb} % useful mathematical symbols
\usepackage{amsmath} % \text in mathmode
\usepackage{float}
\usepackage{latexsym}
\usepackage{MnSymbol} % some fancy symbols
\usepackage{wrapfig} % embed figures in text
\usepackage{subfigure} % multiple images in figure
\usepackage[table]{xcolor}		% color in tables
\usepackage{booktabs}			% fancy tables
\usepackage{acronym}
\usepackage{cancel} 	% durchstreichen in gleichungen

\newcommand{\avg}[1]{\ensuremath{
		\{\!\!\{ #1 \}\!\!\}
	}
}

\newcommand{\norm}[1]{\ensuremath{
		\| #1 \|
	}
}

\newcommand{\diff}[1]{\ensuremath{
		\lsem #1 \rsem
	}
}
\newcommand\FV{\mathrm{FV}}
\newcommand\DG{\mathrm{DG}}
\newcommand\EC{\mathrm{EC}}
\newcommand\ES{\mathrm{ES}}
\newcommand\spvec[1]{{\boldsymbol{ #1}}} %spatial vector
\newcommand\vpsi{{\spvec{\psi}}}
\newcommand\vn{{\spvec{n}}}
\newcommand\va{{\spvec{a}}}
\newcommand{\eux}{u_x} %1d velocity for 1d euler eq
\newcommand{\ux}{u_x} %first component in 2D/3D euler examples
\newcommand{\uy}{u_y} %first component in 2D/3D euler examples
\newcommand{\uz}{u_z} %first component in 2D/3D euler examples

\newcommand\changed[1]{#1}
\newcommand{\changeremoved}{}
\newcommand{\changehiddentext}[1]{}
\begin{document}

\newacro{dg}[DG]{Discontinuous Galerkin}
\newacro{dgsem}[DGSEM]{Discontinuous Galerkin Spectral Element Method}
\newacro{lgl}[LGL]{Legendre-Gauss-Lobatto}
\newacro{lg}[LG]{Legendre-Gauss}
\newacro{fv}[FV]{finite volumes}
\newacro{sbp}[SBP]{summation-by-parts}
\newacro{ibp}[IBP]{Integration by Parts}
\newacro{nfvse}[NFVSE]{Native Finite Volumes Sub Elements}
\newacro{rhs}[RHS]{Right-Hand Side}

\newacro{ec}[EC]{Entropy Conservative}
\newacro{es}[ES]{Entropy Stable}
\newacro{les}[LES]{Large Eddy Simulation}
\newacro{fsp}[FSP]{free stream preservation}

\newacro{eoc}[EOC]{Experimental Order of Convergence}

\newacro{hpc}[HPC]{high performance computing}

\newacro{bc}[BC]{Boundary Condition}

\newacro{dof}[DOF]{degrees of freedom}

\newacro{tvd}[TVD]{total variation diminishing}

\newacro{ppm}[PPM]{piecewise parabolic method}
%\verso{Sebastian Hennemann \textit{et al.}}

\begin{frontmatter}

\title{A provably entropy stable subcell shock capturing approach for high order split form DG for the compressible Euler Equations}

\author[1]{Sebastian Hennemann}\corref{cor1}
\ead{sebastian.hennemann@dlr.de}
\cortext[cor1]{Corresponding author: 
  Tel.: +49 2203 601 3306;}

\author[2]{Andrés M. Rueda-Ramírez}
\author[3]{Florian J. Hindenlang}
\author[2]{Gregor J. Gassner}

\address[1]{German Aerospace Center (DLR), Linder Höhe, 51147 Cologne, Germany}
\address[2]{Department of Mathematics and Computer Science/Center for Data and Simulation Science, Universität zu Köln, Weyertal 86-90, 50937 Cologne, Germany}
\address[3]{Max Planck Institute for Plasma Physics, Boltzmannstraße 2, 85748 Garching, Germany}

\begin{abstract}
The main result in this paper is a provably entropy stable shock capturing approach for the high order entropy stable \ac{dgsem} based on a hybrid blending with a subcell low order variant. \changed{Since it is possible to rewrite} a high order \ac{sbp} operator into an equivalent conservative finite \changed{volume} form, \changed{we were able} to design a low order scheme directly with the \ac{lgl} nodes that is compatible to the discrete entropy analysis used for the proof of the entropy stable \ac{dgsem}. \changed{Furthermore, we present} a hybrid low order/high order discretisation where it is possible to seamlessly blend between the two approaches, while still being provably entropy stable. With tensor products and careful design of the low order scheme on curved elements, we are able to extend the approach to three spatial dimensions on unstructured curvilinear hexahedral meshes. We validate our theoretical findings and demonstrate convergence order for smooth problems, conservation of the primary quantities and discrete entropy stability for an arbitrary blending on curvilinear grids. In practical simulations, we connect the blending factor to a local troubled element indicator that provides the control of the amount of low order dissipation injected into the high order scheme. We modified \changed{an existing shock indicator, which is based on the modal polynomial representation}, to our provably stable hybrid scheme. The aim is to reduce the impact of the parameters as good as possible. We describe our indicator in detail and demonstrate its robustness in combination with the hybrid scheme, as it is possible to compute all the different test cases without changing the indicator. The test cases include e.g. the double Mach reflection setup, forward and backward facing steps with shock Mach numbers up to 100. The proposed approach is relatively straight forward to implement in an existing entropy stable \ac{dgsem} code \changeremoved as only modifications local to an element are necessary.  
\acresetall

\end{abstract}

\end{frontmatter}

\newcommand\blfootnote[1]{%
	\begingroup
	\renewcommand\thefootnote{}\footnote{#1}%
	\addtocounter{footnote}{-1}%
	\endgroup
}

\blfootnote{Formal publication: https://doi.org/10.1016/j.jcp.2020.109935}

\blfootnote{©2020. Licensed under Creative Commons CC-BY-NC-ND 4.0, creativecommons.org/licenses/by-nc-nd/4.0/}

\section{Introduction}

The \ac{dg} methodology provides a straight forward strategy to construct high order methods on unstructured grids for advection dominated problems, such as the compressible Euler and the compressible Navier-Stokes equations. Based on a Galerkin type local finite element ansatz and a weak formulation of the underlying non-linear problem, ideas from the finite volume methodology are applied via the so-called numerical fluxes to add stabilization for advection dominated flow problems. However, due to e.g. variational crimes caused by the insufficient integration of the highly non-linear volume and surface terms in the weak formulation and the finite dimensional approximation space, the \ac{dg} scheme may still suffer critical stability issues. We roughly distinguish two stability issues: (i) the variational crimes or  \changed{aliasing-driven} instabilities that may cause crashes of the simulation especially in underresolved vortical driven flow fields such as e.g. turbulence; (ii) the oscillations caused by Gibbs phenomena when approximating very steep gradients or even discontinuities with high order polynomials. 

Aliasing instabilities can be reasonably controlled by so-called polynomial de-aliasing (or `over-integration' or `consistent integration'), e.g., \cite{Kirby2003,Mengaldo201556,Gassner:2013qf,FLD:FLD3943,Kopriva2017}, or by discrete entropy stable split formulations, e.g., \cite{carpenter_esdg,Parsani2016,chan2018,Parsani:2015:ESD:2784985.2785151,gassner2016,gassner2018}, or by a space-time ansatz in entropy variables with consistent integration, e.g., \cite{Hiltebrand2013,murman2016}. Due to the positive properties of the split form based approach, e.g., \cite{winters2018compareSplit}, and the reduced computational complexity in comparison to a full space-time ansatz, the work presented in this paper is based on discretely entropy stable collocation \ac{dgsem} with \ac{lgl} nodes. \changed{The method can be found in, e.g. \cite{carpenter_esdg,gassner2016, gassner2018}.}

Oscillations are typically much harder to control as they are not caused by `variational crimes', but are inherent in the ansatz and construction of the \ac{dg} scheme with local polynomial approximations. Oscillations are as critical as the aliasing issues because they may cause crashing of the simulation due to undershoots in density and temperature, i.e. unphysical negative values may arise. It seems that such artificial oscillations at steep gradients can only be addressed by direct interaction with the high order operator or the general idea of the \ac{dg} methodology. The task gets even more complicated, as we aim to retain the nice theoretical properties of the discretely entropy stable \ac{dgsem}. In \cite{Hiltebrand2013}, artificial dissipation powered shock capturing was designed for the space-time approach such that it is compatible with the entropy analysis. Shock capturing based on artificial dissipation that is compatible with the discretely entropy stable split forms was constructed in, e.g.,  \cite{wintermeyer2018entropy,bohmthesis}. In the latter works, a troubled element indicator based on Persson and Peraire \cite{persson2006} was adopted. While artificial dissipation works reasonably well in practice, some additional issues are related to this approach: (i) by changing the character of the underlying partial differential equation, new boundary conditions have to be specified and adjusted \changed{\cite{lodato2019}}; (ii) at strong shocks, the amount of artificial dissipation may be high enough to negatively impact the explicit time step typically used for advection dominated problems \changed{\cite{kloeckner2011}}. 

As a side node, a very interesting approach based on tailored subcell \ac{fv} in combination with \ac{lg} powered collocation \ac{dgsem} was presented by Sonntag and Munz \cite{sonntag2017}. The idea is to split the \ac{dg} element in a subcell for each \ac{lg} node. The authors choose to span the subcell around the \ac{lg} node and apply a standard second order \ac{tvd}-type reconstruction on this subcell grid. As the local number of \ac{dof} match between the two schemes, it is straight forward to construct projection and reconstruction operators that connect the two approximation spaces. Again, using a troubled element indicator, a hard switch was introduced: in case the \ac{dg} element is troubled, project the data and use subcell \ac{tvd} \ac{fv} in this element\changed{, otherwise} keep using \ac{dgsem}. An even more general idea based on subcell type shock capturing was recently introduced by Vilar \cite{Vilar2019}. 

As we aim to keep discrete entropy stability in our approach, we adopt, extend and modify the subcell idea accordingly. First, we focus on the LGL nodes and the discretely entropy stable \ac{dgsem} as our high order baseline scheme. Next, we show how to construct a robust low order subcell scheme, that is fully compatible with the discrete entropy analysis, i.e. that is entropy stable for collocated entropy variables. We make heavy use of the very important findings presented first in \cite{fisher2013} for high order diagonal norm \ac{sbp} operators, where it is proven that these operators can be equivalently represented as specific subcell conservative finite \changed{volume} approximations. Lastly, instead of implementing a hard switch, we will show that it is possible to seamlessly blend the high order and low order variant\changed{s} local to each element with a blending factor $\alpha$ in a way such that the hybrid scheme is still entropy stable for arbitrary values of $\alpha$.

Thus, the key result of this paper is the first subcell based shock capturing technique for high order \ac{dg} that is discretely entropy stable. For clarity, we present the main ideas and derivations of the one dimensional case in the next two sections of the paper. However, extension to the fully three dimensional case on curvilinear hexahedral meshes is available and all associated proofs are collected in \ref{sec:3d_es}. We invest substantial work in modifying and assessing a troubled element indicator inspired by Persson and Peraire \cite{persson2006} and present detailed description in section \ref{sec:BlendingAndShockIndicator}. Here, our goal is to determine robust values of the parameters as good as possible and demonstrate that the proposed choice of the described indicator is very robust in combination with our shock capturing, as it is possible to compute all the different numerical experiments without change. Besides presenting the numerical examples in the last part of the paper, we also present numerical validations of our theoretical findings to demonstrate that the method is fully conservative and discretely entropy stable for all blending factors. Conclusions are drawn in the final section.

\section{The Hybrid Scheme}
\changed{Although the paper focuses on the application to the compressible Euler Equations, the theory presented in the following chapters is applicable to any hyperbolic conservation law for which a two point numerical flux exists.}

\subsection{The High Order Scheme}

The starting point is a high order \ac{lgl} \ac{dgsem} in split form, see e.g. \changed{\cite{carpenter_esdg, gassner2016}}. We split the domain $\Omega$ into elements with width $\Delta x_i$ and use a local polynomial Lagrange ansatz with degree $N$. For this nodal approach, the unknowns are the nodal values collected in a vector $\underline{u}$. With collocation, we can directly compute the nodal values of non-linear functions, such as e.g. the the nodal values of the flux $\underline{f}$. The standard strong form \ac{dgsem}, e.g., \cite{Gassner:2010pd} of the generic conservation law
\begin{equation}
u_t + f(u)_x = 0,
\end{equation} 
for an element $i$ reads as 
\begin{equation}
J_i \underline{\dot{u}}_i + \underline{R}^{\DG}_i := 
J_i \underline{\dot{u}}_i + D\underline{f}_i + M^{-1} B (\underline{f}^*_i - \underline{f}_i) = 0. \label{DGSEMconsLaw}
\end{equation}
We call $\underline{R}^{\DG}$ the high order \ac{dgsem} \changed{residual} operator,
$ \dot{u} =\partial u / \partial t$ is the temporal derivative, $D$ is the polynomial derivative matrix, $M = \text{diag}(w_0,\ldots,w_N)$ is the mass matrix with the \ac{lgl} quadrature weights as entries on the diagonal. The surface evaluation matrix $B = \text{diag}(-1,0,\ldots,0,1)$ and the Jacobian of the local element mapping $J_i=\Delta x_i / 2$. The numerical flux vector is
\begin{equation}
\underline{f}^*_i = ( f^*(u_{i-1,N}, u_{i,0}), 0, \ldots, 0, f^*(u_{i,N}, u_{i+1,0}) )^T,
\end{equation}
with the two point numerical interface flux
\begin{equation}
	f^*=f^*(u_\mathrm{Left}, u_\mathrm{Right}).
\end{equation}
To obtain an entropy stable \ac{dgsem}, one of the necessary ingredients is to use an entropy stable two point flux\changed{, e.g., as found the context on approximative Riemann solvers \cite{toro1999yq}.} We focus on a single element, drop the element index and introduce the element centric notation
\begin{equation}
u_L := u_{-1} = u_{i-1,N} \text{ and } u_R := u_{N+1} = u_{i+1,0}  \label{index_pair}
\end{equation}
to describe the face states from the neighbour elements of the element $i$.

It was shown in \cite{gassner_skew_burgers} that the \ac{lgl} \ac{dgsem} operators $M$ and $D$ form a \ac{sbp} operator introduced in the finite difference community, e.g., \cite{strand1994}. This is an important property, as in combination with the theoretical findings of Fisher and Carpenter \cite[page 10]{fisher2013}, it was possible to show \changed{\cite{fisher2013,carpenter_esdg,gassner2016}} that many split form \ac{dgsem}\changed{, including kinetic energy preserving and entropy conserving forms,} can be rewritten in an element local conservative flux differencing formula on a complementary grid (see Figure~\ref{fig:complementarygrid1}) with specific choices of symmetric two point fluxes  $f^{*S}$
\begin{equation}
J \dot{u}_j + 2\sum_{l=0}^{N}D_{jl}f^{*S}_{(j,l)} + \frac{\delta_{jN}}{w_\changed{N}}  (f^*_{(N,R)} - f_N) - \frac{\delta_{j0}}{w_0}  (f^*_{(L,0)} - f_0) = 0 \label{eq:DGSEMconsLaw_nodal_symflux}, \quad j \in \{0,\ldots,N\}.
\end{equation}
It is possible to write the volume operator in a conservative flux difference form
\begin{equation}
2\sum_{l=0}^{N}D_{jl}f^{*S}_{(j,l)} = \frac{1}{w_j} (\bar{f}_{j+1} - \bar{f}_j), \label{eq:split_form_volume_as_fv}
\end{equation}
with

\begin{align*}
\bar{f}_0 &= f_0, \\
\bar{f}_j &= \sum_{k=j}^{N} \sum_{l=0}^{j-1} 2 Q_{lk} f^{*S}_{(l,k)} \quad j\in \{1,\ldots, N\}, \\
\bar{f}_{N+1} &= f_N,
\end{align*}
where $Q := M D$ is the high order differencing operator with the \ac{sbp} property 
\begin{equation}
	Q + Q^T = B,
\end{equation}
which leads to the additional properties
\begin{align}
	\sum_{k=0}^{N} Q_{jk} &= 0,\\
	\sum_{j=0}^{N} Q_{jk} &= \tau_k := \delta_{kN} - \delta_\changed{{k0}} = B_{kk},
\end{align}
that we will use later in the proofs. 

We can rewrite the \ac{fv} like formulation \eqref{eq:split_form_volume_as_fv} in matrix vector notation
\begin{equation}
J \underline{\dot{u}} 
+ M^{-1}\left[\Delta \underline{\bar{f}} + B(\underline{f}^* - \underline{f}) \right]  = 0, \label{DG_as_ho_fv_volume_integral}
\end{equation}
with the standard \ac{fv} differencing matrix
\begin{equation}
\Delta :=
\left(\begin{matrix}
-1 & 1 & 0 & \ldots & \ldots & 0 \\
0 & -1 & 1 & 0 		& \ldots & 0 \\
\vdots & \ddots & \ddots & \ddots & \ddots & \vdots \\
0 & \ldots & 0 & -1 & 1 & 0 \\
0 & \ldots &\ldots & 0 & -1 & 1 
\end{matrix}\right) \in \mathbb{R}^{(N+1)\times (N+2)}. \label{differenceMatrix}
\end{equation}

\subsection{The Low Order Scheme}

\begin{figure}[!htbp]
	\centering
	\includegraphics[width=0.7\linewidth]{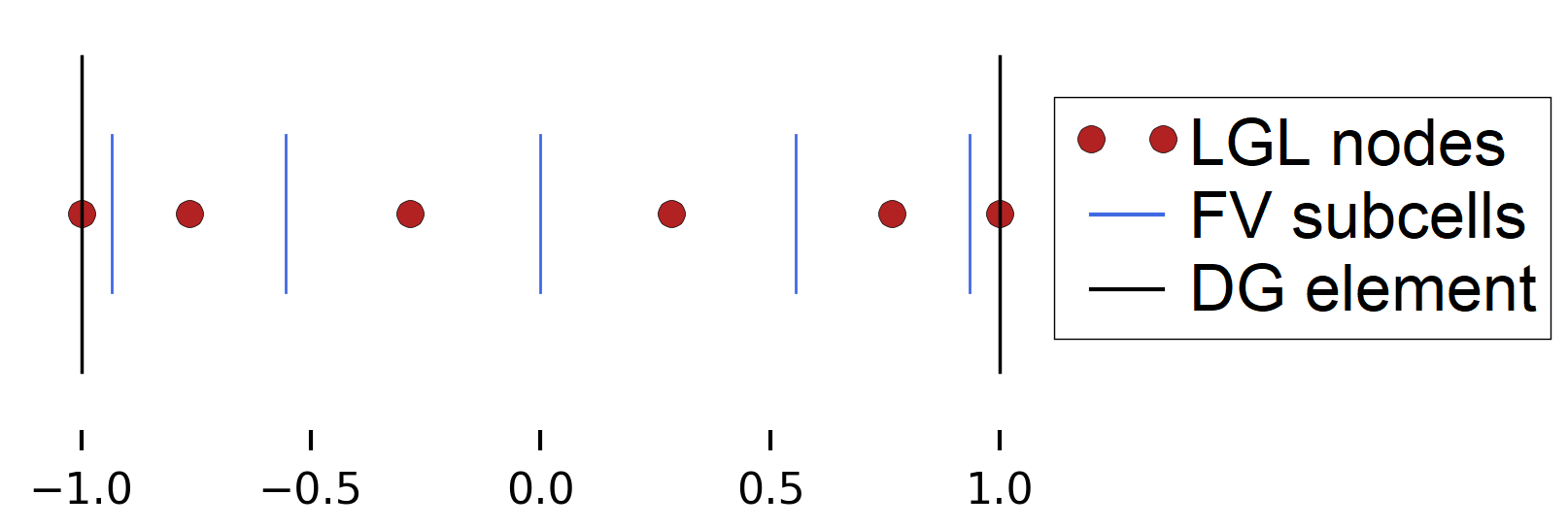}
	\caption{Example of a 6th order \ac{dgsem} \ac{lgl} element with \ac{fv} subcells}
	\label{fig:complementarygrid1}
\end{figure}

The goal is to find a low order scheme that is compatible to the discrete entropy analysis performed in the next section for the split form \ac{dgsem} to show its entropy stability. The key is that for the discrete analysis the discretisation is multiplied with a collocated nodal value of the entropy variables. Thus, we decided to construct a low order scheme, to be precise, a \ac{fv} type discretisation on a subcell grid that directly uses the nodal \ac{lgl} values of the high order \ac{dgsem} ansatz. We observe from representation \eqref{eq:split_form_volume_as_fv} that a natural associated subcell grid is spanned by the respective \ac{lgl} weights $w_j$ as depicted in Figure~\ref{fig:complementarygrid1}.
As we need a direct correlation of the high order and the low order \ac{dof} for the discrete entropy analysis, we directly use the \ac{lgl} nodal values and interpret these as the respective subcell 'average' values of the low order \ac{fv} method. Besides allowing to analyse both schemes' discrete entropy evolution simultaneously, the choice gives further advantages: (i) using compatible \ac{dof} and a compatible subcell grid, we are able to show conservation of the hybrid scheme (note that conservation is not guaranteed, when reconstructing the \ac{fv} representation and blending both \changed{residual} operators, see \ref{sec:otherFVrepresentations}); (ii) no additional memory is needed to store the \ac{dof} of the low order scheme; (iii) we are able to construct the low order \ac{fv} such that the contribution at the element interfaces (not the subcell interfaces) \changed{is} identical to the surface contribution of the high order \ac{dgsem}. \changed{Therefore,} only the volume term contributions are blended, compare formulation \eqref{eq:blended_fv_form}. The low order \ac{fv} discretisation on the compatible subcell grid with the nodal \ac{dof} choice reads as
\begin{equation}
J \dot{u}_{j} + \frac{1}{w_j} (f^*_{(j,j+1)} - f^*_{(j-1,j)}) = 0, \label{eq:FVstandard}
\end{equation}
where the \ac{lgl} \ac{dgsem} nodal value $u_j$ is interpreted as a mean value in the subcells $j \in \{0,\ldots,N\}$ of the element. At the internal interface between the subcells $j$ and $j+1$, the numerical flux approximation is $f^*_{(j,j+1)} = f^*(u_j, u_{j+1})$. Again, it is possible to rewrite the scheme in matrix vector notation
\begin{equation}
J \underline{\dot{u}} + \underline{R}^{\FV} 
:= J \underline{\dot{u}} +  M^{-1} \Delta \underline{f}^{\FV} = 0, \label{RHS_FV_vector}
\end{equation}
where $\Delta$ is the \ac{fv} differencing operator defined in \eqref{differenceMatrix} and $\underline{f}^{\FV} = (f^*_{(L,0)}, f^*_{(0,1)}, \ldots, f^*_{(N-1,N)}, f^*_{(N,R)})^{\mathrm{T}} \in \mathbb{R}^{N+2}$. We can mimic a structure of the low order discretisation analogous to the \ac{dgsem} structure by introducing the fluxes

\begin{align*}
\bar{f}_0^{\FV} &= f_0, \\
\bar{f}_j^{\FV} &= f_j^{\FV} = f^*_{(j-1,j)} \quad j\in \{1,\ldots,N\}, \\
\bar{f}_{N+1}^{\FV} &= f_N, \label{FV_as_volume_surface_split_flux}
\end{align*} 
to rewrite \eqref{RHS_FV_vector} as 
\begin{equation}
J \underline{\dot{u}} 
+  M^{-1} \left[ \Delta \underline{\bar{f}}^{\FV} + B (\underline{f}^* - \underline{f}) \right] = 0, \label{FV_as_volume_surface_split}
\end{equation}
which looks very similar to \eqref{DG_as_ho_fv_volume_integral} and shows how to split the subcell \ac{fv} operator into an element surface contribution and a local volume contribution. 

\subsection{The Hybrid Scheme}

As the low order discretisation directly uses and updates the DOF of the \ac{lgl} nodal ansatz, it is possible to directly blend the low order operator with the high order operator
\begin{equation}
\underline{R} := \alpha \underline{R}^{\FV} + (1-\alpha) \underline{R}^{\DG}, \label{eq:RHS_blended}
\end{equation}
where $\alpha\in\changed{[0,1]}$ is a blending function that depends on the shock indicator discussed in detail in section \ref{sec:BlendingAndShockIndicator}. Using the natural subcell forms \eqref{DG_as_ho_fv_volume_integral} and \eqref{FV_as_volume_surface_split} we get 
\begin{equation}
J \underline{\dot{u}} 
+  M^{-1} \left[ \Delta (\alpha \underline{\bar{f}}^{\FV} + (1-\alpha) \underline{\bar{f}}) + B (\underline{f}^* - \underline{f}) \right] = 0.\label{blended_as_volume_surface_split}
\end{equation}
By construction of the low order discretisation, the blending of the surface term recovers the original \ac{dgsem} strong surface contribution. Thus, actual blending is only effective in the volume terms. Consequently, no change of data dependency and connectivity is introduced into the existing code framework and e.g. no additional boundary conditions, or a different MPI communication has to be implemented. 

Introducing the compact notation of the blended flux as
\begin{align}
\begin{split}
\bar{f}^\alpha_0 =& f_0, \\
\bar{f}^\alpha_j :=& \alpha f^*_{(j-1,j)} +  (1-\alpha) \bar{f}_j \quad j\in \{1,\ldots,N\}, \\
\bar{f}^\alpha_{N+1} =& f_N, \label{blended_flux}
\end{split}
\end{align}
gives us the equivalent form of the hybrid scheme
\begin{equation}
J \underline{\dot{u}} 
+  M^{-1} \left[ \Delta \underline{\bar{f}}^\alpha + B (\underline{f}^* - \underline{f}) \right] = 0.\label{blended_compact}
\end{equation} 
Alternatively, we can choose a pure flux differencing formulation to get
\begin{align}
\begin{split}
J \underline{\dot{u}} 
&= -M^{-1}\left[\Delta \underline{\bar{f}}^\alpha + B(\underline{f}^* - \underline{f})\right]  
\\ &= M^{-1}\left(
\begin{matrix}
f^*_{(L,0)} - f_0 - \bar{f}^\alpha_1 + \bar{f}^\alpha_0 
\\ - \bar{f}^\alpha_2 + \bar{f}^\alpha_1 
\\ \vdots 
\\ - \bar{f}^\alpha_N + \bar{f}^\alpha_{N-1} 
\\ f_N - f^*_{(N,R)} - \bar{f}^\alpha_{N+1} + \bar{f}^\alpha_N
\end{matrix}\right) 
= M^{-1}\left(
\begin{matrix}
f^*_{(L,0)} - \bar{f}^\alpha_1 
\\ \bar{f}^\alpha_1 - \bar{f}^\alpha_2 
\\ \vdots 
\\ \bar{f}^\alpha_{N-1} - \bar{f}^\alpha_N 
\\ \bar{f}^\alpha_N - f^*_{(N, R)}
\end{matrix}\right).
\end{split} \label{eq:blended_fv_form}
\end{align}
%This form is especially intriguing, as it shows a form and setup very similar to the flux-corrected transport (FCT) approach of Boris \cite{Boris1997}. 

\section{Discrete Entropy Stability of the Hybrid Scheme}
\subsection{The Entropy Inequality}
\label{sec:EntropyCondition}
For a generic conservation law 
\begin{equation}
u_t + f(u)_x = 0,
\end{equation}
we introduce a convex function, the entropy, $\eta=\eta(u)$ and define the set of entropy variables $v=v(u)=\frac{\partial\eta}{\partial u}$. The entropy and a corresponding flux function $q=q(u)$ define an entropy pair $(\eta(u), q(u))$ \changed{if} the chain rule $\eta'(u)f'(u) = q'(u)$ holds. It follows \changed{\cite{Harten1982}} that for smooth solutions the entropy variables $v$ contract the conservation law to 
\begin{equation}
	\frac{\partial \eta}{\partial t} + \frac{\partial q}{\partial x} = 0.
	\label{eq: entropy_conservation_law}
\end{equation}
But for discontinuous solutions we can only satisfy an entropy inequality in the distributional sense\footnote{\cite[page 338]{ray2013}, notations of variables adapted to match this paper notation} 
\begin{equation}
	\frac{\partial \eta}{\partial t} + \frac{\partial q}{\partial x} \le 0. \label{eq: entropy_inequality}
\end{equation}
\changeremoved We define a numerical discretisation as entropy stable or entropy dissipative, if it is possible to \changed{prove} a discrete entropy inequality for the scheme.

\subsection{Derivation of the Discrete Entropy Evolution}

For the discrete entropy evolution analysis, we introduce the typical \ac{dg} notation for an arithmetic average and the jump difference of two values
\begin{align}
	 \avg{\cdot}_{(j,m)}  &:= \frac{1}{2} \left[(\cdot)_j + (\cdot)_m\right], \\
	 \diff{\cdot}_{(j,m)} &:= (\cdot)_{m} - (\cdot)_j.
\end{align}
We first look at the low order discretisation  \eqref{eq:FVstandard} and mimic the continuous entropy analysis and contract the low order discretisation of the conservation laws with the discrete entropy variable, $v_j$, which in this particular case are the collocated nodal values at the corresponding \ac{lgl} node $j$
\begin{align}
	v_j J \dot{u}_j   &= - \frac{1}{w_j} v_j  (f^*_{(j,j+1)} - f^*_{(j-1,j)}) 
	\\ = J \dot{\eta}_j &= - \frac{1}{w_j} \left[(q^*_{(j,j+1)} - q^*_{(j-1,j)}) - \frac{1}{2} (r_{(j,j+1)} + r_{(j-1,j)})\right], \label{FVentropyBalance}
\end{align}
where we \changed{used continuity in time and the fact that the mass matrix is diagonal} to contract the time derivative of the entropy. We skipped the intermediate algebra and added the proof of the above entropy balance equation in \ref{subsec: ProofFVentropyBalance}. We define the discrete numerical entropy flux as
\begin{equation}
	q^*_{(j,j+1)} := \avg{v}_{(j,j+1)} f^*_{(j,j+1)} - \avg{\psi}_{(j,j+1)},
\end{equation}
the numerical entropy production term as
\begin{equation}
	r_{(j,j+1)}:=  \diff{v}_{(j,j+1)} f^*_{(j,j+1)} - \diff{\psi}_{(j,j+1)},
\end{equation} 
with the entropy flux potential 
\begin{equation}
\psi(v) :=  v f(u(v)) - q(u(v)).
\end{equation} 
We get the discrete total entropy evolution by applying the \ac{lgl} quadrature rule for the element
\begin{align}
\begin{split}
\sum_{j=0}^{N} J w_j \dot{\eta}_{j} 
&=-\sum_{j=0}^{N} (q^*_{(j,j+1)} - q^*_{(j-1,j)}) - \frac{1}{2} (r_{(j,j+1)} + r_{(j-1,j)})
\\	&  = -q^*_{(N,R)} + q^*_{(L,0)} + \frac{1}{2} \sum_{j=0}^N  r_{(j,j+1)} + r_{(j-1,j)}
\\	&  = -q^*_{(N,R)} + q^*_{(L,0)} + \frac{1}{2} r_{(N,R)} + \frac{1}{2} r_{(L,0)} + \sum_{j=0}^{N-1} r_{(j,j+1)}.
\label{eq:FV_ES}
\end{split}
\end{align}
We rewrote the discrete entropy evolution in this specific form to make it easier \changed{to compare} with the high order entropy evolution derived next. 

We note that a proof for \ac{dgsem} on general curvilinear hexahedral element in three space dimensions is available in, e.g., \changed{\cite{fisher_phdthesis2012,carpenter2016,gassner2018}}. In this paper, we present an alternative proof that is more similar to the approach used for the low order method. We multiply the split form \ac{dgsem} with collocated nodal values of the entropy variables and apply the \ac{lgl} quadrature to obtain the discrete entropy evolution of the high order method as 
\begin{equation}
	\sum_{j=0}^N J w_j \dot{\eta}_j
	= -q^*_{(N,R)} + \frac{1}{2} r_{(N,R)} + q^*_{(L,0)} + \frac{1}{2} r_{(L,0)}  + \sum_{j,k=0}^N Q_{jk} r_{(j,k)}.
	\label{eq:DG_entropyChange}
\end{equation}
We skip the details of this proof and collect all the steps in \ref{subsec: ProofDGSEMentropyStab}.

Finally, we consider the blended hybrid scheme and derive its discrete entropy evolution. Now, \changed{using the} careful construction presented above, everything fits together such that a common discrete entropy analysis of the hybrid scheme \eqref{eq:RHS_blended} is feasible. Technically, again, we multiply the hybrid discretisation with the collocated nodal values of the entropy variables and apply the \ac{lgl} quadrature to derive the discrete total entropy balance
\begin{align}
\begin{split}
\sum_{j=0}^N J w_j \dot{\eta}_j &= \sum_{j=0}^N J w_j v_j \dot{u}_j  
\\&=- \sum_{j=0}^N w_j v_j (\alpha R^{\FV}_j + (1- \alpha) R^{\DG}_j)
\\&= -\left( \alpha \sum_{j=0}^N w_j v_j R^{\FV}_j + (1- \alpha) \sum_{j=0}^N w_j v_j R^{\DG}_j \right)
\\&=  \alpha \left[ -q^*_{(N,R)} + q^*_{(L,0)} + \frac{1}{2} r_{(N,R)} + \frac{1}{2} r_{(L,0)} + \sum_{j=0}^{N-1} r_{(j,j+1)}^{\FV} \right]
\\&+ (1 - \alpha) \left[ -q^*_{(N,R)} + \frac{1}{2} r_{(N,R)} + q^*_{(L,0)} + \frac{1}{2} r_{(L,0)} + \sum_{j,k=0}^N Q_{jk} r_{(j,k)}^{\DG} \right]
\\&= -q^*_{(N,R)} + \frac{1}{2} r_{(N,R)} + q^*_{(L,0)} + \frac{1}{2} r_{(L,0)} + \alpha \sum_{j=0}^{N-1} r_{(j,j+1)}^{\FV} + (1-\alpha) \sum_{j,k=0}^N Q_{jk} r_{(j,k)}^{\DG},  \label{eq: entropy_balance_law_blended}
\end{split}
\end{align}
where we used \eqref{eq:FV_ES} and \eqref{eq:DG_entropyChange} and assumed that both methods used the same numerical flux at the element interface. 
This derivation shows that both entropy production terms\changed{, $r^{\FV}$ and $r^{\DG}$, determine if the scheme is entropy conservative or entropy stable.}

\subsection{Discrete Entropy Conservation and Stability of the Hybrid Scheme} \label{sec:EntropyConsStabHybrid}

The discrete entropy evolution of the low order \ac{fv} type discretisation leads us to the definition of an \ac{ec} numerical flux function $f^* = f^{*\EC}$. If the numerical flux function brings entropy production at each interface to zero, i.e.
\begin{equation}
r_{(j,j+1)} =  \diff{v}_{(j,j+1)} f^{*\EC}_{(j,j+1)}  -  \diff{\psi}_{(j,j+1)} = 0,
\end{equation}
the resulting low order \ac{fv} scheme is virtually (entropy) dissipation free. Analogously, an \ac{es} numerical flux function $f^*= f^{*\ES}$ that \changed{guarantees entropy dissipation} at each interface satisfies
\begin{equation}
r_{(j,j+1)} =  \diff{v}_{(j,j+1)} f^{*\ES}_{(j,j+1)}  -  \diff{\psi}_{(j,j+1)} \le 0.
\end{equation}

By carefully choosing the numerical fluxes in our hybrid scheme, it is possible to discretely satisfy either entropy conservation or entropy stability. As an intermediate hybrid scheme, we choose all occurring numerical fluxes as an \ac{ec} flux and directly get that the hybrid scheme is by construction virtually dissipation free and conserves the total discrete entropy. We will apply this intermediate scheme later to demonstrate our theoretical findings in numerical experiments. Our goal however is to construct a provably entropy stable discretisation. First, we choose an \ac{ec} numerical flux in the \ac{dg} volume terms
\begin{equation}
\sum_{j=0}^N J w_j \dot{\eta}_j 
= -q^*_{(N,R)} + \frac{1}{2} r_{(N,R)} + q^*_{(L,0)} + \frac{1}{2} r_{(L,0)} + \alpha \sum_{j=0}^{N-1} r_{(j,j+1)}^{\FV},
\end{equation}
which means that in the pure high order configuration, the scheme only dissipates through the \ac{dg} element interfaces. Next, we choose an \ac{es} numerical flux in the volume parts of the low order \ac{fv} scheme, as we on purpose want to add guaranteed entropy dissipation 
\begin{equation}
\sum_{j=0}^N J w_j \dot{\eta}_j 
\le -q^*_{(N,R)} + \frac{1}{2} r_{(N,R)} + q^*_{(L,0)} + \frac{1}{2} r_{(L,0)},
\end{equation} 
when activating and blending the low order scheme, i.e. for values $\alpha>0$. This additional volume type dissipation should act as a mechanism to control oscillations e.g. at shock fronts. Lastly, we choose an \ac{es} numerical flux at the element interfaces, to add a small amount of dissipation 
\begin{equation}
\sum_{j=0}^N J w_j \dot{\eta}_j 
\le -q^*_{(N,R)} + q^*_{(L,0)},
\end{equation}
even in a pure high order configuration, i.e. $\alpha=0$, \changed{which} is desirable for automatic de-aliasing \changed{\cite{winters2018compareSplit}}.

In summary, for all non-negative blending factors $\alpha\geq 0$, the novel hybrid discretisation is provably entropy stable. While these derivations are for the one dimensional case, it is possible to extend this analysis to the three dimensional case on curvilinear hexahedral elements. As the proof is very technical and algebraically involved, details are collected in \ref{sec:3d_es}. It is furthermore interesting to note that all the derivations and constructions directly carry over to general diagonal norm \ac{sbp} operators, such as e.g. \ac{sbp} finite difference schemes. Instead of using the \ac{lgl} weights to define the subcell grid, the analogue is to use the diagonal entries of the corresponding \ac{sbp} norm matrix to define the low order subcell grid.

\section{The Blending Function for Shock Capturing}
\label{sec:BlendingAndShockIndicator}
\subsection{Finding the Troubled Element}
%\subsubsection{Modal Polynomial Base}

In order to estimate the amount of underresolution, we follow ideas presented in Persson and Peraire \cite{persson2006} and compare the 'modal energy' of the highest polynomial modes to the overall modal energy of the quantity. To clearly separate the contributions, we transform our troubled element indicator quantity from a (collocated) nodal representation to a hierarchical modal representation with Legendre polynomials. We define the modal energy of a 1D polynomial as
\begin{equation}
\langle \changed{\epsilon, \epsilon} \rangle_{L^2} = \langle \sum_{j=0}^{N} m_j \tilde{L}_j, \sum_{j=0}^{N} m_j \tilde{L}_j \rangle_{L^2} =
\sum_{i,j=0}^{N} m_i m_j \langle \tilde{L}_i, \tilde{L}_j \rangle_{L^2} = \sum_{j=0}^{N} m_j^2,
\end{equation}
where $\{m_j\}_{j=0}^{N}$ are the modal coefficients. We compute for each \ac{dg} element how much energy is contained in the highest modes relative to the total energy of the polynomial as following
\begin{equation}
\mathbb{E} = \max\left(\frac{m_N^2}{\sum_{j=0}^{N} m_j^2}, \frac{m_{N-1}^2}{\sum_{j=0}^{N-1} m_j^2}\right),
\end{equation}
where we used the highest and second highest mode to avoid odd/even effects when approximating element local functions. In the numerical results section, we will focus on the compressible Euler equations as an example. The indicator variable used to estimate the missing resolution for the compressible Euler equation is the product of the collocated density and collocated pressure, \changed{$\epsilon = \rho p$}. This product reacts to jumps in both density and pressure and is thus a very robust choice, as both quantities, density and pressure need to retain positivity throughout the simulation. 

The next step is to decide if the value of the highest mode energy is critical or not, i.e. if there is a shock in the element or not. We define a threshold value  $\mathbb{T}=\mathbb{T}(N)$, that is used to decide if a shock is detected. Motivated by the discussion in Persson and Peraire \cite{persson2006} on the expected decay proportional to $\propto 1/N^4$ we propose the ansatz
\begin{equation}
\mathbb{T}(N) = a \cdot 10^{-c (N+1)^\frac{1}{4}}.
\end{equation}
With the insight from many different numerical experiments, the parameters $a$ and $c$ are predetermined as
\begin{equation}
\mathbb{T}(N) = 0.5\cdot 10^{-1.8 (N+1)^\frac{1}{4}}
\end{equation}
and since then never adjusted or changed. All below presented numerical results use this decision threshold. It is however clear, that there is no proof and claim for optimality and that there might be alternative forms of the threshold function that could give an improvement.

\subsection{The Blending Function $\alpha$}
\label{sec:shock_indicator_tunings}
For each element, the highest mode energy indicator $\mathbb{E} \in [0,1]$ and the threshold  $\mathbb{T}$ \changed{are} computed. 
\changed{These values need} to be translated into a blending value $\alpha\in[0,\alpha_{\max}]$, where $0\leq\alpha_{\max}\leq1$ defines the maximum amount of low order discretisation that can be blended to the high order \ac{dgsem}. \changeremoved

In a first step, we define the blending function $\alpha \in [0,1]$, with
\begin{equation}
\alpha = \frac{1}{1+\exp(\frac{-s}{\mathbb{T}}(\mathbb{E}-\mathbb{T}))},
\end{equation}
as a smooth mapping. The sharpness factor $s$ was chosen so that the low order \ac{fv} scheme gets the blending weight with $\alpha= 0.0001\approx0$ for a highest energy indicator $\mathbb{E}=0$, i.e. 
\begin{equation}
	0.0001 = \alpha(\mathbb{E}=0) =  \frac{1}{1+\exp(\frac{-s}{\mathbb{T}}(0-\mathbb{T}))} =
	\frac{1}{1+\exp(s)},
\end{equation}
and hence
\begin{equation}
	\Rightarrow s = \ln \left( \frac{1 - 0.0001}{0.0001} \right) \approx \changed{9.21024}.
\end{equation}

For computational efficiency, we can adjust the cases close to pure low order or high order configurations, by clipping of the values of $\alpha$ as
\begin{equation}
\tilde{\alpha} := \begin{cases}
0, 		& \text{if } \alpha < \alpha_{\min}	\\
\alpha, 	& \text{if } \alpha_{\min} \le \alpha \le 1-\alpha_{\min} \\
1, 		& \text{if } 1-\alpha_{\min} < \alpha
\end{cases}, \label{blending_function}
\end{equation}
with \changed{$\alpha_{\min} = 0.001$}. \changed{The choice $\alpha_{\max}=1$ seems natural, as it corresponds to pure first order \ac{fv}.} Although this \changeremoved offers a very robust scheme, we also assessed the accuracy of this approach in many investigations and observed that we can substantially improve the numerical results by first clipping of the value of alpha 
\begin{equation}
\alpha = \min(\changed{\tilde \alpha}, \alpha_{\max}),
\end{equation}
with the value $\alpha_{\max}=0.5$ for polynomial order $N=4$  chosen in this work and used for all test cases presented below. Furthermore, accuracy could be improved by diffusing the element wise values of $\alpha$ by a single sweep of 
\begin{equation}
	\alpha^\mathrm{final} = \max\limits_{E}\{\alpha, 0.5 \alpha_E\},
\end{equation}
where $E$ denotes all elements sharing a face.

\section{The Numerical Setup for the Compressible Euler Equations}
\label{numerical_results}

For numerical validation and assessment, we consider the compressible Euler equations. For brevity, we discuss the equations in one spatial dimension as
\begin{equation}
\frac{\partial}{\partial t} \left(\begin{array}{c}
\rho \\
\rho \eux \\
\rho E
\end{array}\right) 
+ \frac{\partial}{\partial x} \left(\begin{array}{c}
\rho \eux \\
\rho \eux^2 + p \\
\eux (\rho E + p) 
\end{array}\right)
=0,
\end{equation}
where $\rho$ is the density of the fluid, $\eux$ its velocity, $E$ its specific total energy and $p$ the pressure. The system is closed with the perfect gas assumption
\begin{equation}
p = \left(\rho \left(E - \frac{\eux^2}{2}\right)\right)(\gamma-1),
\end{equation}
where $\gamma$ is the specific heat capacity ratio.

We consider the entropy pair 
\begin{equation}
	\eta = - \frac{\rho s}{\gamma - 1}, \qquad q = - \frac{\rho \eux s}{\gamma - 1},
\end{equation}
where s is the thermodynamic entropy given by
\begin{equation}
	s = \ln(p) - \ln(\rho \gamma) + \mathrm{const.} = -(\gamma - 1) \ln(\rho) - \ln(\beta) + \mathrm{const.}
\end{equation}
and the inverse temperature $\beta = 1 / 2RT$. The corresponding entropy variables are 
\begin{equation}
v = \left(\frac{\gamma -s}{\gamma - 1} - \beta \eux^2, 2\beta \eux, -2\beta \right)^T.
\label{eq:entropy_variables}
\end{equation}

There are several \ac{ec} numerical flux functions available. In this work, we use the one presented by Chandrashekar \cite[page 11]{chandrashekar2012}, which reads in one dimension as
\begin{equation}
f^{*\EC}_{(j,k)} = \left(\begin{matrix}
\rho^{\ln} \avg{\eux} \\ 
\rho^{\ln} \avg{\eux}^2 + \hat{p} \\
\rho^{\ln} \avg{\eux} \hat{h}
\end{matrix}\right),
\label{flux_ec_chandra}
\end{equation}
where the logarithmic average of a positive quantity is defined as
\begin{equation}
	(\cdot)^{\ln} = (\cdot)^{\ln}_{(j,k)} := \frac{(\cdot)_k - (\cdot)_j}{(\ln \cdot)_k - (\ln \cdot)_j},
\end{equation}
and
\begin{equation}
	\hat{p} := \frac{\avg{\rho}}{2\avg{\beta}}, \qquad
	\hat{h} := \frac{1}{2 \beta^{\ln} (\gamma - 1)} - \frac{1}{2} \avg{\eux^2} + \frac{\hat{p}}{\rho^{\ln}} + \avg{\eux}^2.
\end{equation}
A three dimensional version can be found, e.g., in \cite[page 12]{gassner2016}. Adding an explicit dissipation term that is guaranteed entropy dissipative \changed{\cite{chandrashekar2012}} gives an \ac{es} numerical flux
\begin{equation}
f^{*\ES}_{(j,k)} = \left(\begin{matrix}
\rho^{\ln} \avg{\eux} \\ 
\rho^{\ln} \avg{\eux}^2 + \hat{p} \\
\rho^{\ln} \avg{\eux} \hat{h}
\end{matrix}\right) - \frac{\lambda_{\max}}{2} \left(\begin{matrix}
\diff{\rho} \\ 
\diff{\rho \eux} \\
(\frac{1}{2 \beta^{\ln}(\gamma - 1)} + \frac{1}{2} (\eux)_j (\eux)_k) \diff{\rho} + \avg{\rho} \avg{\eux} \diff{\eux} + \frac{\avg{\rho}}{2(\gamma - 1)\diff{\frac{1}{\beta}}}
\end{matrix}\right),
\label{flux_es_chandra}
\end{equation}
with $\lambda_{\max} := \max\{ |(\eux)_j| + c_j, |(\eux)_k| + c_k\} $ is the maximum eigenvalue of the two states and $c$ is the speed of sound with $\rho c^2 = \gamma p$.

So far, only the semi discrete spatial discretisation is described. In the numerical experiments, we integrate the resulting system of coupled ordinary differential equations in time with a \changed{fourth-order} accurate explicit five stage Runge-Kutta method by Carpenter and Kennedy \cite[page 13]{carpenter1994}. The explicit time step restriction is estimated as 
\begin{equation}
	\Delta t = CFL \frac{\Delta x_{\min}}{\lambda_{\max}} \frac{1}{(N + 1)^2} , \label{CFLcondition}
\end{equation}
where $\Delta x$ is the typical element size, $\lambda_{\max}$ is the maximal wave speed and $CFL=1$.

The scheme is implemented in the three dimensional curvilinear DLR inhouse turbomachinery code TRACE 9.3.106 \changed{\cite{traceUserGuide} and in FLUXO \cite{fluxo}, which implements the \ac{dgsem} for general advection-diffusion equations}. TRACE is a three dimensional code framework, hence all the two dimensional computations presented below are done in three dimensions, with only one element layer in the third dimension. The global explicit time step is \changed{pre-computed and fixed for the whole simulation.} All the computations use a polynomial degree of $N=4$, as this is the polynomial degree we envision for our future applications with the turbulent compressible Navier-Stokes equations. \changed{All boundary conditions are imposed weakly using a Riemann solver, see e.g. \cite{hindenlang2019}.}

\section{Numerical Validation}
\subsection{Experimental Order of Convergence}
\label{test:convected_vortex}

\begin{figure}[!htbp]
	\centering
	\includegraphics[width=0.48\linewidth]{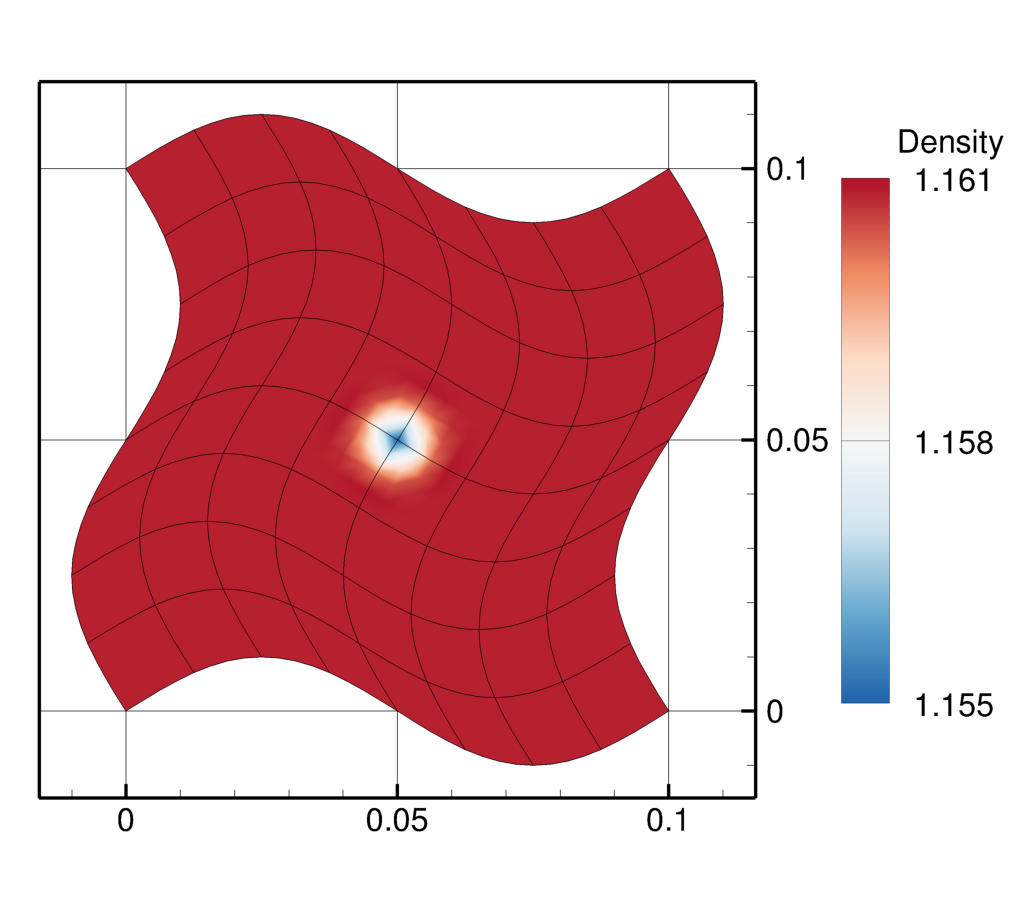}
	\caption{\label{fig:isentoicvortex} Initial condition of the vortex transport. All boundaries are fully periodic. Resolution of $8$ elements per space direction with degree $N=4$ gives a total of $1600$ DOF. The vortex advects through the domain at constant speed and after one period reaches this initial  position again.}
\end{figure}

The goal of this numerical test is to ensure that the full spatial convergence order is recovered for smooth problems, as the shock capturing mechanism should not trigger in this case. We consider the smooth isentropic vortex convection problem with the basic setup from \cite{hiocfd5}. However, instead of a Cartesian straight sided mesh, we consider a fully periodic curved mesh as depicted in Figure~\ref{fig:isentoicvortex}.
To generate the curved domain, we map a unit square $\Omega = [0, 1]^2$ with the transformation
\begin{align}
	%X:\Omega \rightarrow f(\Omega) \\ 
	X(\xi,\eta) = \left(\begin{array}{c}
	\xi  L_x - A_x L_y \sin(2 \pi \eta) \\
	\eta L_y + A_y L_x \sin(2 \pi \xi)  
	\end{array}\right)
\end{align}
to distort a square of $[0, L_x] \times [0, L_y]$ with sines of amplitudes $A_x, A_y$. We choose $L_x = L_y = 0.1$ and $A_x = A_y = 0.1$. The parameters are chosen, so that we get a smooth mesh deformation and the initial vortex gets well resolved in the domain. All boundaries are set to periodic. A vortex with radius $R_v = 0.005$ is initialized in the curved domain in the location $(x_v, y_v) = (L_x / 2, L_y / 2)$. The specific heat ratio $\gamma = 1.4$ and the gas constant $R_\mathrm{gas} = 287.15$. The free stream state is defined by the Mach number $M_0 = 0.5$,  temperature $T_0 = 300$, pressure $p_0 = 10^5$, velocity $u_0 = M_0 \sqrt{\gamma R_\mathrm{gas} T_0}$, and density $\rho_0 = \frac{p_0}{R_\mathrm{gas} T_0}$. The initial vortex is given by
\begin{align}
\begin{split}
u_x(x,y) =& u_0\left(1-\beta \frac{y-y_v}{R_v} e ^{\frac{-r^2}{2}}\right), \\
u_y(x,y) =& u_0 \beta \frac{x-x_v}{R_v} e ^{\frac{-r^2}{2}},\\
T(x,y) =& T_0 - \frac{(u_0 \beta)^2}{2 C_p} e^{-r^2},\\
\rho(x,y) =& \rho_0 \left(\frac{T}{T_0}\right)^\frac{1}{\gamma - 1},\\
p(x,y) =& \rho\changed{(x,y)} R_\mathrm{gas} T\changed{(x,y)},
\end{split}
\end{align}
where $C_p = R_\mathrm{gas} \gamma / (\gamma - 1)$ is the heat capacity at constant pressure and $r = \sqrt{ (x-x_v)^2 + (y-y_v)^2} / R_v$ is the relative vortex radius and $\beta = 0.2$ is the vortex strength. We simulate the propagation for one time period $t_p= L_x / u_0$ and compare the result with the initial condition for the mesh resolutions $8,16,32,64,128$ in each direction and the polynomial degree $N=4$. The discrete norms are defined as
\begin{equation}
	\|f\|_{\mathbb{L}^p(\Omega)} := \left(\frac{\int_\Omega |f(x,y)|^p dx dy}{\int_\Omega 1\ dx dy}\right)^\frac{1}{p}
\end{equation}
for $p\in \{1,2\}$ and
\begin{equation}
	\|f\|_{\mathbb{L}^\infty(\Omega)} := \sup_{x\in\Omega} |f(x)|.
\end{equation}
We collect the results in Table~\ref{tab: eoc_vortex_transport} and get the expected convergence behaviour of approximately $N+1 = 5$. It is worth pointing out that the shock capturing is not triggered in any of the configurations.

\begin{table}[H]
	\subfigure[$\rho$]{
		\resizebox{0.5\textwidth}{!}{
    \begin{tabular}{clrlrlr} \toprule
        & \multicolumn{2}{c}{$\mathbb{L}^{\infty}$} & \multicolumn{2}{c}{$\mathbb{L}^1$} & \multicolumn{2}{c}{$\mathbb{L}^2$} \\
        \cmidrule(lr){2-3} \cmidrule(lr){4-5} \cmidrule(lr){6-7}
        \# Elements & Error & EOC & Error & EOC & Error & EOC \\ \midrule
        $8$ & $ 1.51 \cdot 10^{-3}$ & $$ & $ 8.35 \cdot 10^{-5}$ & $$ & $ 1.80 \cdot 10^{-4}$ & $$ \\
        $16$ & $ 2.05 \cdot 10^{-4}$ & $ 2.89$ & $ 5.07 \cdot 10^{-6}$ & $ 4.04$ & $ 1.80 \cdot 10^{-5}$ & $ 3.32$ \\
        $32$ & $ 8.86 \cdot 10^{-6}$ & $ 4.53$ & $ 1.31 \cdot 10^{-7}$ & $ 5.28$ & $ 5.35 \cdot 10^{-7}$ & $ 5.07$ \\
        $64$ & $ 3.16 \cdot 10^{-7}$ & $ 4.81$ & $ 3.90 \cdot 10^{-9}$ & $ 5.07$ & $ 2.05 \cdot 10^{-8}$ & $ 4.70$ \\
        $128$ & $ 1.16 \cdot 10^{-8}$ & $ 4.77$ & $ 1.33 \cdot 10^{-10}$ & $ 4.87$ & $ 7.08 \cdot 10^{-10}$ & $ 4.86$ \\
        \bottomrule
    \end{tabular}}}
	\subfigure[$\rho \ux$]{
	\resizebox{0.5\textwidth}{!}{
    \begin{tabular}{clrlrlr} \toprule
        & \multicolumn{2}{c}{$\mathbb{L}^{\infty}$} & \multicolumn{2}{c}{$\mathbb{L}^1$} & \multicolumn{2}{c}{$\mathbb{L}^2$} \\
        \cmidrule(lr){2-3} \cmidrule(lr){4-5} \cmidrule(lr){6-7}
        \# Elements & Error & EOC & Error & EOC & Error & EOC \\ \midrule
        $8$ & $ 6.51 \cdot 10^{00}$ & $$ & $ 1.61 \cdot 10^{-1}$ & $$ & $ 4.60 \cdot 10^{-1}$ & $$ \\
        $16$ & $ 5.66 \cdot 10^{-1}$ & $ 3.52$ & $ 9.06 \cdot 10^{-3}$ & $ 4.15$ & $ 3.43 \cdot 10^{-2}$ & $ 3.75$ \\
        $32$ & $ 1.41 \cdot 10^{-2}$ & $ 5.33$ & $ 1.28 \cdot 10^{-4}$ & $ 6.15$ & $ 7.52 \cdot 10^{-4}$ & $ 5.51$ \\
        $64$ & $ 8.33 \cdot 10^{-4}$ & $ 4.08$ & $ 5.61 \cdot 10^{-6}$ & $ 4.51$ & $ 3.47 \cdot 10^{-5}$ & $ 4.44$ \\
        $128$ & $ 3.61 \cdot 10^{-5}$ & $ 4.53$ & $ 2.09 \cdot 10^{-7}$ & $ 4.74$ & $ 1.31 \cdot 10^{-6}$ & $ 4.73$ \\
        \bottomrule
    \end{tabular}}}
	\subfigure[$\rho \uy$]{
	\resizebox{0.5\textwidth}{!}{
    \begin{tabular}{clrlrlr} \toprule
        & \multicolumn{2}{c}{$\mathbb{L}^{\infty}$} & \multicolumn{2}{c}{$\mathbb{L}^1$} & \multicolumn{2}{c}{$\mathbb{L}^2$} \\
        \cmidrule(lr){2-3} \cmidrule(lr){4-5} \cmidrule(lr){6-7}
        \# Elements & Error & EOC & Error & EOC & Error & EOC \\ \midrule
        $8$ & $ 5.38 \cdot 10^{00}$ & $$ & $ 2.01 \cdot 10^{-1}$ & $$ & $ 5.43 \cdot 10^{-1}$ & $$ \\
        $16$ & $ 3.97 \cdot 10^{-1}$ & $ 3.76$ & $ 7.53 \cdot 10^{-3}$ & $ 4.74$ & $ 2.80 \cdot 10^{-2}$ & $ 4.28$ \\
        $32$ & $ 1.43 \cdot 10^{-2}$ & $ 4.80$ & $ 1.22 \cdot 10^{-4}$ & $ 5.95$ & $ 7.29 \cdot 10^{-4}$ & $ 5.26$ \\
        $64$ & $ 7.81 \cdot 10^{-4}$ & $ 4.19$ & $ 5.48 \cdot 10^{-6}$ & $ 4.47$ & $ 3.41 \cdot 10^{-5}$ & $ 4.42$ \\
        $128$ & $ 3.96 \cdot 10^{-5}$ & $ 4.30$ & $ 2.05 \cdot 10^{-7}$ & $ 4.74$ & $ 1.28 \cdot 10^{-6}$ & $ 4.73$ \\
        
    \bottomrule
    \end{tabular}}}
	\subfigure[$\rho E$]{
	\resizebox{0.5\textwidth}{!}{
    \begin{tabular}{clrlrlr} \toprule
        & \multicolumn{2}{c}{$\mathbb{L}^{\infty}$} & \multicolumn{2}{c}{$\mathbb{L}^1$} & \multicolumn{2}{c}{$\mathbb{L}^2$} \\
        \cmidrule(lr){2-3} \cmidrule(lr){4-5} \cmidrule(lr){6-7}
        \# Elements & Error & EOC & Error & EOC & Error & EOC \\ \midrule
        $8$ & $ 8.12 \cdot 10^{02}$ & $$ & $ 3.89 \cdot 10^{01}$ & $$ & $ 9.63 \cdot 10^{01}$ & $$ \\
        $16$ & $ 1.34 \cdot 10^{02}$ & $ 2.60$ & $ 2.37 \cdot 10^{00}$ & $ 4.04$ & $ 8.83 \cdot 10^{00}$ & $ 3.45$ \\
        $32$ & $ 4.02 \cdot 10^{00}$ & $ 5.06$ & $ 4.66 \cdot 10^{-2}$ & $ 5.66$ & $ 2.16 \cdot 10^{-1}$ & $ 5.35$ \\
        $64$ & $ 1.85 \cdot 10^{-1}$ & $ 4.44$ & $ 1.50 \cdot 10^{-3}$ & $ 4.96$ & $ 8.42 \cdot 10^{-3}$ & $ 4.68$ \\
        $128$ & $ 7.30 \cdot 10^{-3}$ & $ 4.66$ & $ 5.00 \cdot 10^{-5}$ & $ 4.90$ & $ 2.74 \cdot 10^{-4}$ & $ 4.94$ \\
        
        \bottomrule
    \end{tabular}}}
    \caption[Convergence results for the isentropic vortex convection test.]{Convergence results for the isentropic vortex convection test. Errors and EOC for all four conservative quantities are shown. Error is calculated after one convection period. With the polynomial degree $N=4$, the expected convergence rate of $N+1=5$ reasonably well obtained. In none of the cases is the shock capturing mechanism activated. }
    \label{tab: eoc_vortex_transport}
\end{table}

\subsection{Conservation of the Primary Quantities Mass, Momentum and Energy}
\label{test:conservation}

\begin{figure}[!htbp]
	\centering
	\subfigure[Density, $t=1.2$] {
		\includegraphics[width=0.45\linewidth]{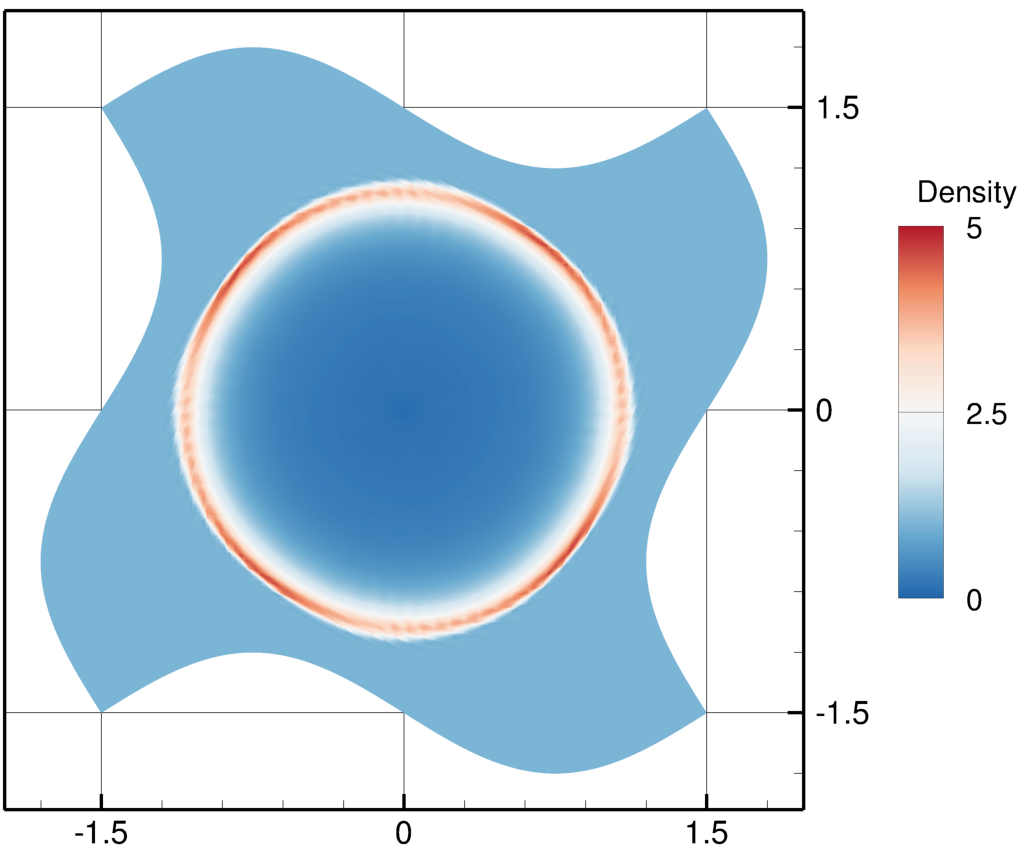}}
	\subfigure[Alpha and mesh lines, $t=1.2$] {
		\includegraphics[width=0.45\linewidth]{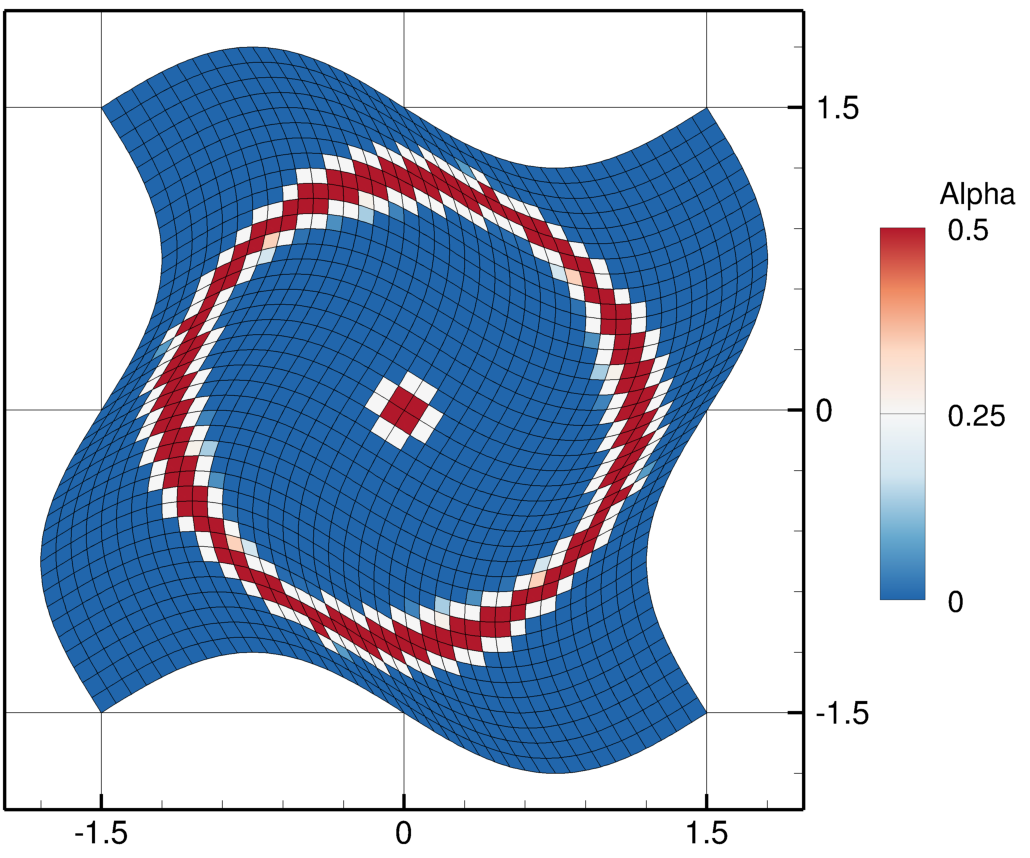}}
	\subfigure[Density, $t=8$] {
		\includegraphics[width=0.45\linewidth]{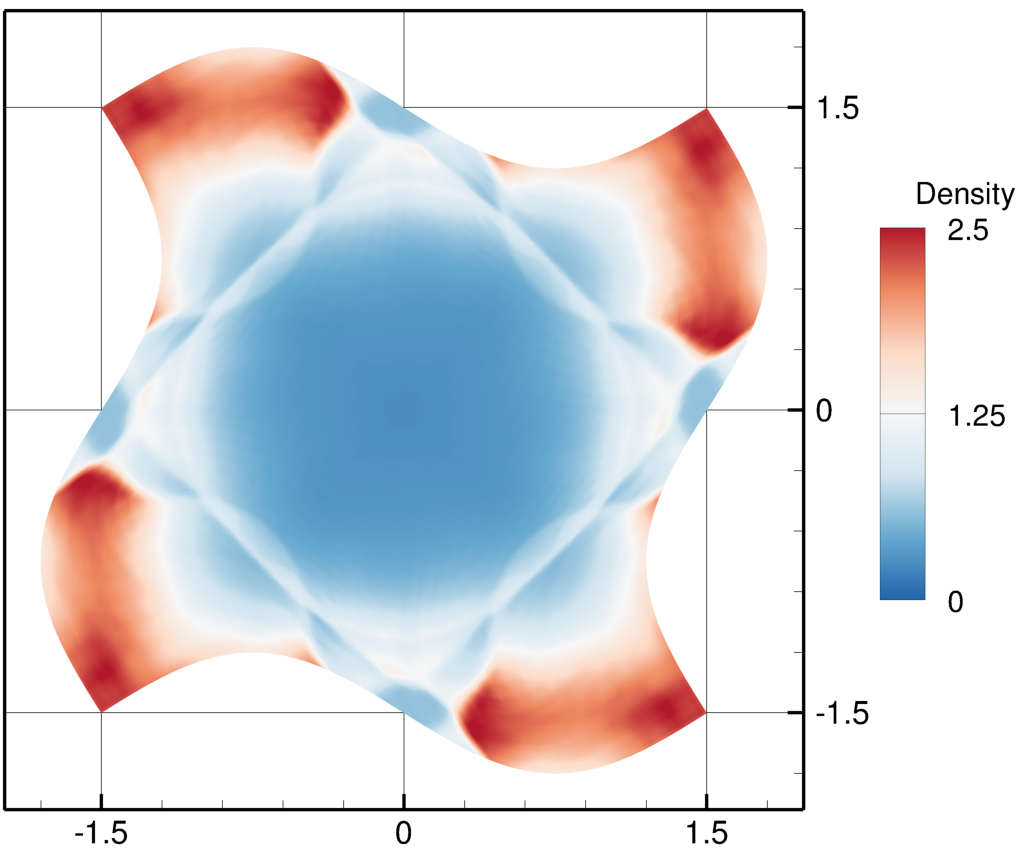}}
	\subfigure[Alpha and mesh lines, $t=8$] {
		\includegraphics[width=0.45\linewidth]{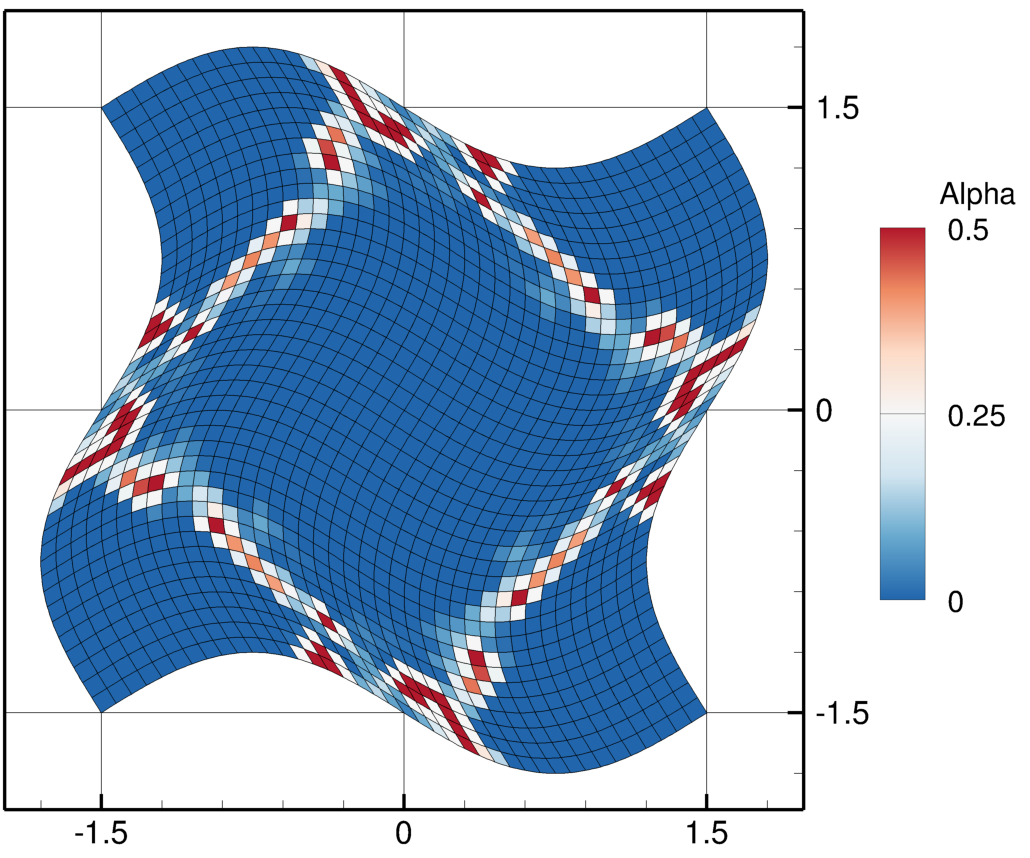}}
	\caption{Blast wave on curvilinear domain with $40\times 40$ elements (40k \ac{dof}). The contour plots illustrate the density distribution and the values of the blending factor $\alpha$ at different times $t$. For all primary quantities, conservation up to machine precision errors are observed throughout the simulation test.}
	\label{fig:conservation}
\end{figure}

The goal of this subsection is to demonstrate that the proposed hybrid scheme is fully conservative as stated in equation \eqref{eq:blended_fv_form}. To assess the conservation, we consider a circular blast wave problem in the periodic and curvilinear domain from the previous subsection \ref{test:convected_vortex} scaled to the domain size $\pm 1.5$. At time equal to zero, a mass and energy peak is initialized in the centre of the domain

\begin{align*}
\rho(x,y) &= \rho_0 + \frac{\changed{m}_\mathrm{ejecta}}{2 \pi \sigma_\mathrm{ejecta}^2} e^{-\frac{1}{2} \frac{x^2+y^2}{\sigma_\mathrm{ejecta}^2} },
& \ux(x,y) =\uy(x,y)&=0,
& p(x,y) &= \frac{p_0}{\gamma - 1} + \frac{E_\mathrm{blast}}{2 \pi \sigma_\mathrm{blast}^2} e^{-\frac{1}{2} \frac{x^2+y^2}{\sigma_\mathrm{blast}^2} },
\end{align*}
with the parameters $\rho_0 = 1$,  $\changed{m}_\mathrm{ejecta} = 0.5$,  $\sigma_\mathrm{ejecta} = 3 \cdot 10^{-2}$, $ p_0 = 10^{-5}$,  $E_\mathrm{blast} = 1$, $\sigma_\mathrm{blast} = 2 \cdot 10^{-2}$, and $\gamma = 1.4$. The simulation end time is $t_{\max} = 8$ to allow for complex interactions of shock waves and multiple triggering of the shock capturing mechanism. The numerical results are collected in Table~\ref{tab:conservation} where the maximum deviation of the total mass, momentum and energy throughout the simulation is listed. Visualisation of the flow fields and the blending factor $\alpha$ is shown in Figure~\ref{fig:conservation}.
\begin{table}[!htbp]
	\centering
	\resizebox{0.5\textwidth}{!}{
		\begin{tabular}{cccc} \toprule
			$\rho$ & $\rho \ux$ & $\rho \uy$ & $\rho E$ \\ \midrule
			$ 9.99 \cdot 10^{-15}$ & $ 7.25 \cdot 10^{-16}$ & $ 7.56 \cdot 10^{-16}$ & $ 1.26 \cdot 10^{-15}$ \\
			\bottomrule
		\end{tabular}
	}
	\caption{For any conservative quantity $u$ we show $\max\limits_{t \in T} \{ |\int_\Omega u(t) dV - \int_\Omega u(0) dV|\}$, where $T=\{8 \cdot \frac{i}{4210} | i \in \{1,\ldots, 4210\}\}$ is the set of time samples.}
	\label{tab:conservation}
\end{table}

\def\d{\mathrm{d}}

\subsection{Entropy Conservation and Stability}

The goal of this subsection is to evaluate the theoretical findings from \changed{section \ref{sec:EntropyConsStabHybrid}} by investigating if the hybrid method with the \ac{ec} numerical fluxes is indeed discretely entropy conservative and if the hybrid scheme with the \ac{es} numerical fluxes is guaranteed entropy stable. We compute the global change of the discrete entropy change in three spatial dimensions as 
\begin{equation}
\dot{\eta}_\Omega := \int_\Omega \dot{\eta} dV \approx \sum_{Elements (\Omega)} \sum_{i,j,k=0}^N J_{ijk} w_{ijk} \dot{\eta}_{ijk}= \sum_{Elements (\Omega)}  - \sum_{i,j,k=0}^N w_{ijk} v_{ijk} R_{ijk},
\end{equation}
where $R$ is the \changed{residual} operator of the blended method (see \eqref{blended_compact} in 1D or \eqref{eq:blendend_rhs_3d} in 3D). The entropy variables $v$ for the Euler equations were defined in \eqref{eq:entropy_variables}.

\begin{figure}[htb]
	
	\subfigure[Density] {
		\includegraphics[width=0.45\linewidth]{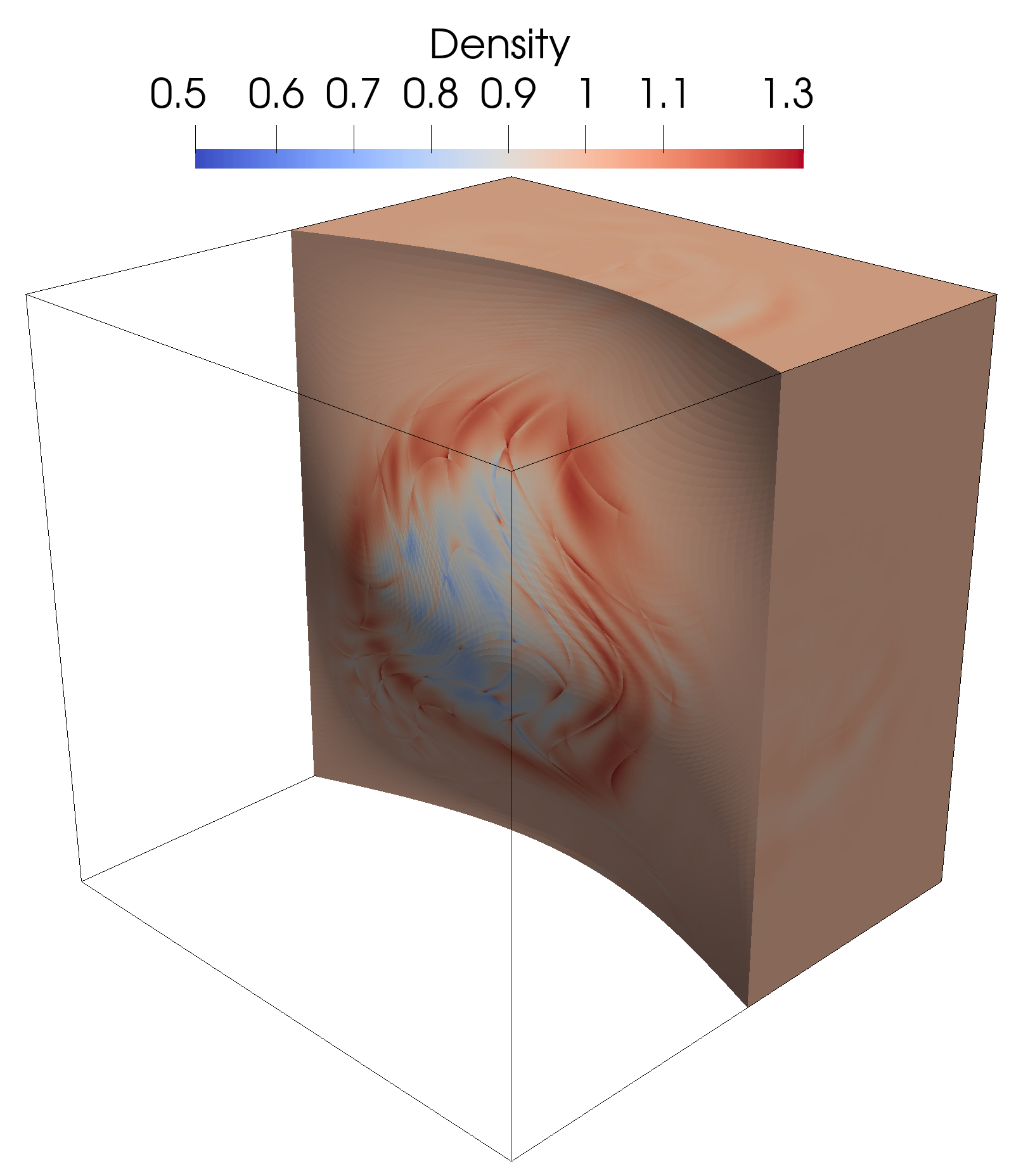}}
	\subfigure[Alpha and mesh lines] {
		\includegraphics[width=0.45\linewidth]{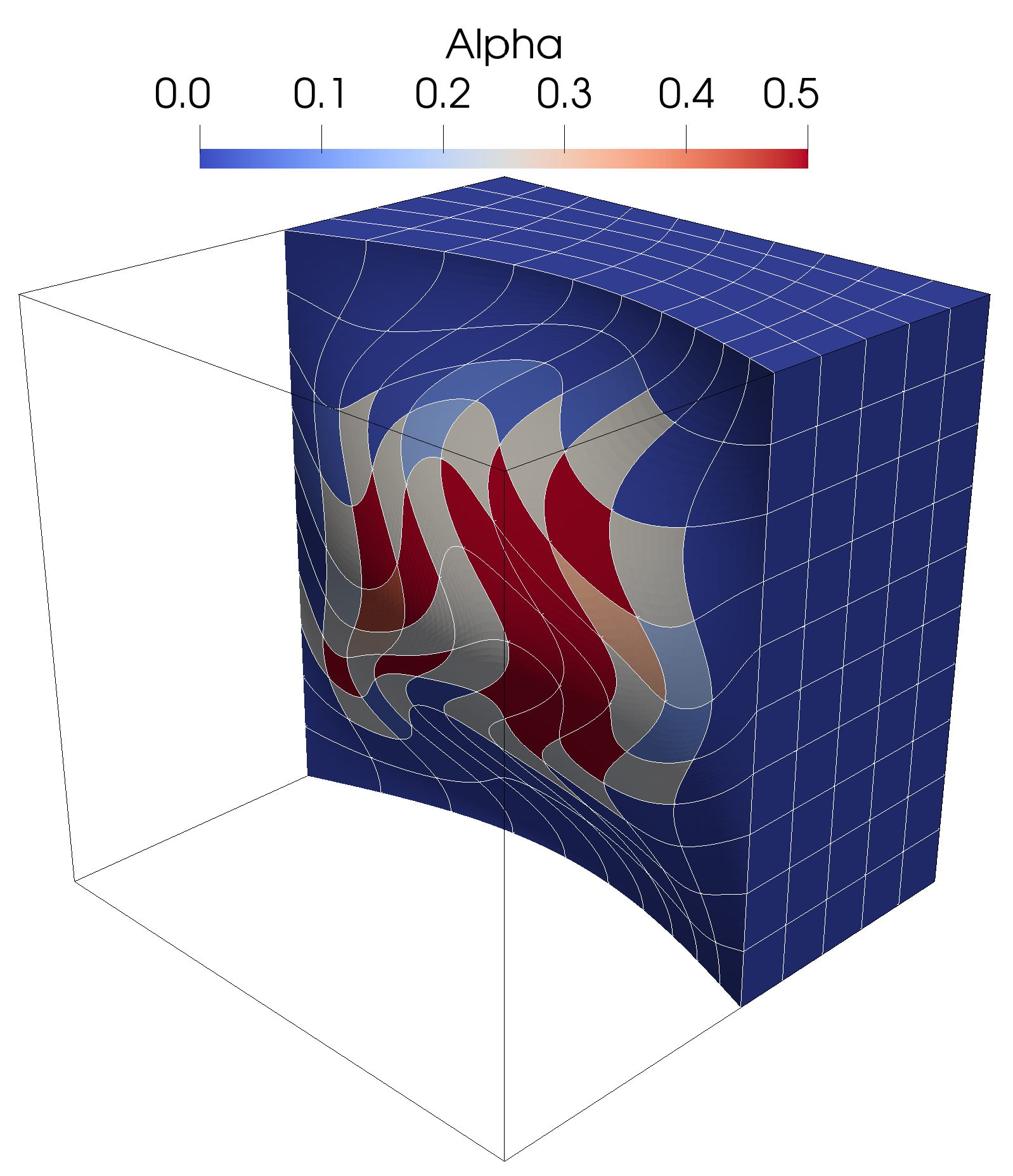}}
\caption{Mach 1.2 shock wave on heavily warped mesh with 10 elements in each direction ($125$k DOF) at $t = 0.4$. The images illustrate the entropy conservation test. Without dissipation terms and for such a warped mesh the simulation is visibly numerically distorted.}
\label{fig:warped3Dmesh}
\end{figure}

We use a 3D heavily warped mesh adapted from \cite{chan2019efficient}.
We start with the cube $\Omega=[0,3]^3$ and apply the transformation
\begin{equation}
X(\xi,\eta,\zeta) = (x,y,z): \Omega \rightarrow f(\Omega)
\end{equation}
such that
\begin{align}
y &= \eta + 
\frac{1}{8} L_y
\cos \left( \frac{3}{2} \pi \frac{2\xi - L_x}{L_x} \right) 
\cos \left( \frac{\pi}{2} \frac{2\eta- L_y }{L_y} \right)
\cos \left( \frac{\pi}{2} \frac{2\zeta- L_z }{L_z} \right), \\
x &= \xi + 
\frac{1}{8} L_x 
\cos \left( \frac{\pi}{2} \frac{2\xi- L_x}{L_x} \right)
\cos \left( 2 \pi \frac{2y- L_y }{L_y} \right)
\cos \left( \frac{\pi}{2} \frac{2\zeta- L_z}{L_z} \right), \\
z &= \zeta + 
\frac{1}{8} L_z 
\cos \left( \frac{\pi}{2} \frac{2x- L_x}{L_x} \right)
\cos \left( \pi \frac{2y- L_y }{L_y}  \right)
\cos \left( \frac{\pi}{2} \frac{2\zeta- L_z}{L_z} \right),
\end{align}
where $L_x = L_y = L_z = 3$. 
The mesh, which can be seen in Figure \ref{fig:warped3Dmesh}, was generated with the HOPR package \cite{hindenlang2015mesh}.
All boundaries are set to periodic.

We first test that the scheme fulfils \ac{fsp}, because, as is shown in \ref{sec:3d_fv_entropy_balance}, this property is necessary to ensure entropy conservation and stability.
The blending function is selected randomly in each element of the domain and the initial condition is set to a uniform flow,

\begin{equation}
    \rho = p = \ux = 1, \ \ \uy = \uz = 0.
\end{equation}

Table \ref{tab:3DcurvedFSP} shows the mean rate of change, in the $\mathbb{L}_2$ norm, of all conservative quantities for the uniform flow (or free stream) condition when using the entropy conservative (EC) and entropy stable (ES) surface numerical fluxes. 
As expected, the rate of change of the conservative variables is near machine precision.
The same results are obtained for the EC and ES fluxes since, in the absence of jumps in the solution, the dissipation term is equal to zero.

\begin{table}[htb]
\centering
\begin{tabular}{c|ccccc} \toprule
 & $\norm{\rho_t}_{\mathbb{L}^2}$ & $\norm{(\rho \ux)_t}_{\mathbb{L}^2}$ & $\norm{(\rho \uy)_t}_{\mathbb{L}^2}$ & $\norm{(\rho \uz)_t}_{\mathbb{L}^2}$ & $\norm{(\rho E)_t}_{\mathbb{L}^2}$ \\
\midrule
$f^{*\EC}$ &
$4.38\cdot 10^{-13}$ & $8.75\cdot 10^{-13}$ & $3.71\cdot 10^{-13}$ & $4.06\cdot 10^{-13}$ & $1.75\cdot 10^{-12}$ \\
$f^{*\ES}$ & 
$4.38\cdot 10^{-13}$ & $8.75\cdot 10^{-13}$ & $3.71\cdot 10^{-13}$ & $4.06\cdot 10^{-13}$ & $1.75\cdot 10^{-12}$ \\
\bottomrule
\end{tabular}
\caption{Mean rate of change of the conservative variables for the uniform flow}
\label{tab:3DcurvedFSP}
\end{table}

Now, to test entropy conservation/stability, instead of the strong blast wave from section \ref{test:conservation}, we initialize a weaker shock wave (Mach 1.2) in the domain. 
This is necessary, because testing entropy conservation can only achieved by disabling stabilizing dissipation terms. 
The post shock states ($r = \le 0.5$) are computed by the normal shock wave equations \cite{naca1951},

\begin{align*}
\rho &=
\begin{cases}
	\approx1.3416, 	& \mathrm{if} \ r \le 0.5 \\
	1,			& \mathrm{if} \ r > 0.5 
\end{cases},
& u_r &=
\begin{cases}
	\approx 0.3615, 	& \mathrm{if} \ r \le 0.5 \\
	0,			& \mathrm{if} \ r > 0.5 
\end{cases}, 
& p& =
\begin{cases}
	\approx1.5133 , 	& \mathrm{if} \ r \le 0.5 \\
	1,			& \mathrm{if} \ r > 0.5,
\end{cases},
\end{align*}
where $\spvec{r} = \spvec{x}-(1.5,1.5,1.5)^T$, $r = \norm{\spvec{r}}_2$ is the distance to the center of the blast, and $u_r = \norm{\spvec{u}}_2 = \spvec{u} \cdot \spvec{r}/r$ is the radial velocity. For this test we had to lower the CFL number to $0.3$.

Figure~\ref{fig:warped3Dmesh} illustrates the density and the blending coefficient at $t=0.4$ on a section cut of the heavily warped 3D mesh.

Table \ref{tab:3DcurvedConservation} shows the maximum deviation from the initial condition for all the conservative quantities when using the entropy conservative (EC) and the entropy stable (ES) surface numerical fluxes.
This deviation is computed for any quantity $u_i$ as
\begin{equation}
\mathrm{deviation}(u_i) = \max_{t \in T} \left\lbrace \left\lvert \int_{\Omega} u_i(t) \d V - \int_{\Omega} u_i(0) dV \right\rvert \right\rbrace,
\end{equation}
where $T$ is the set of time samples. 
The computation is conservative except for numerical noise.

\begin{table}[htb]
\centering
\begin{tabular}{c|ccccc}\toprule
 & $\rho$ & $\rho \ux$ & $\rho \uy$ & $\rho \uz$ & $\rho E$ \\

\midrule
$f^{*\EC}$ &
$6.64\cdot 10^{-13}$ & $7.65\cdot 10^{-15}$ & $7.22\cdot 10^{-15}$ & $1.03\cdot 10^{-14}$ & $6.11\cdot 10^{-13}$ \\
$f^{*\ES}$ & 
$6.61\cdot 10^{-13}$ & $8.32\cdot 10^{-15}$ & $8.49\cdot 10^{-15}$ & $8.98\cdot 10^{-15}$ & $6.25\cdot 10^{-13}$ \\
\bottomrule
\end{tabular}
\caption{Maximum deviation from the initial condition for all conservative variables}
\label{tab:3DcurvedConservation}
\end{table}

Finally, Table \ref{tab:3DcurvedECES} shows the maximum and minimum net entropy production in the domain for all time steps.
The RHS operator with EC fluxes only (left column) is entropy conservative at every Runge Kutta step of the computation ($T$ is the set of time samples for each Runge Kutta step). 
The entropy stable formulation (right column), using EC fluxes in the DG volume integral and and ES fluxes at surfaces and FV volume integral, is entropy stable on the whole domain.

\begin{table}[!htbp]
\centering
\begin{tabular}{c|cc}\toprule
 & $f^{*\EC}$ & $f^{*\ES}$ \\
\midrule
$\min_{t \in T} \dot{\eta}_{\Omega}$ & $-5.19 \cdot 10^{-16}$ & $-7.81 \cdot 10^{-1}$ \\
$\max_{t \in T} \dot{\eta}_{\Omega}$ & $1.88\cdot 10^{-16}$   & $-4.99 \cdot 10^{-3}$ \\
\bottomrule
\end{tabular}
\caption{Entropy conservation and stability test }
\label{tab:3DcurvedECES}
\end{table}

\section{Numerical Applications}
In this section we present four typical benchmark applications with strong shocks. We use the same setup of our troubled element indicator and the blending function $\alpha$ for all the test cases on a sequence of coarse, medium and fine grids. Besides visual inspection of the numerical simulation results, we also depict the distribution of the blending factor to assess if the shock capturing mechanism is only triggered where expected (and desired). 

\subsection{Shu Osher Shock}
\label{sec:shuOsherShock}

\begin{figure}[!htbp]
	\centering
	\subfigure[Density on $64$ elements ($320$ \ac{dof}) zoomed onto the post shock oscillations] {
		\includegraphics[trim={1cm 0.1cm 1cm 1cm},clip,width=0.48\linewidth]{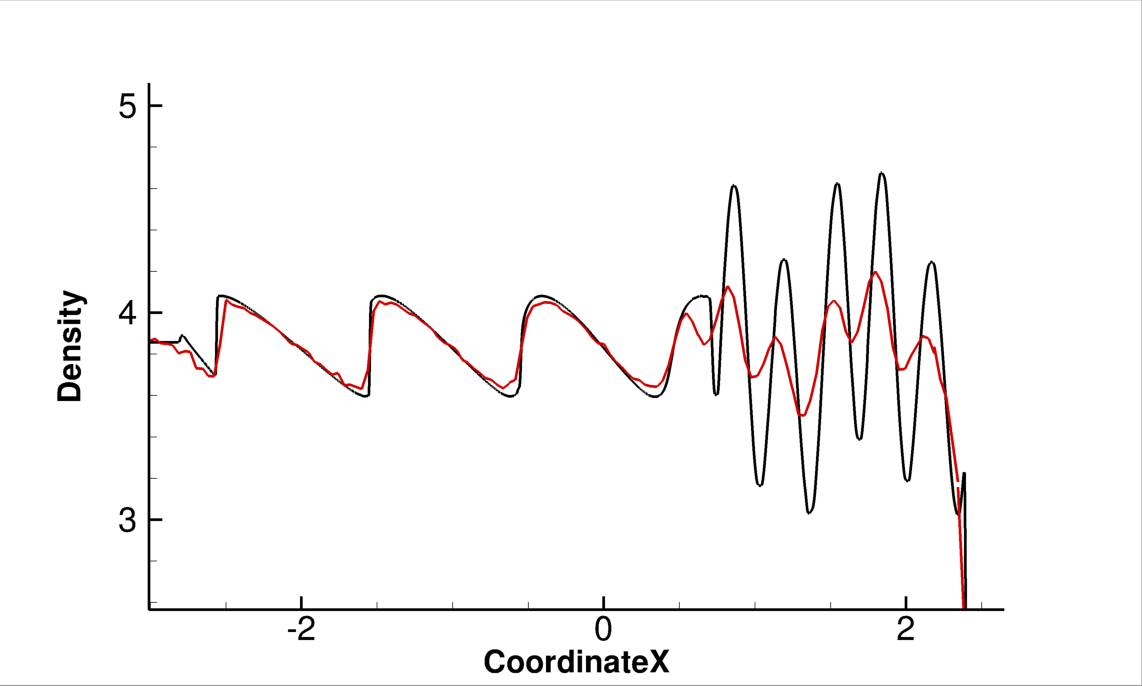}}
	\subfigure[Density and $\alpha$ on $64$ elements ($320$ \ac{dof})] {
		\includegraphics[trim={0cm 0.2cm 0.1cm 0cm},clip,width=0.48\linewidth]{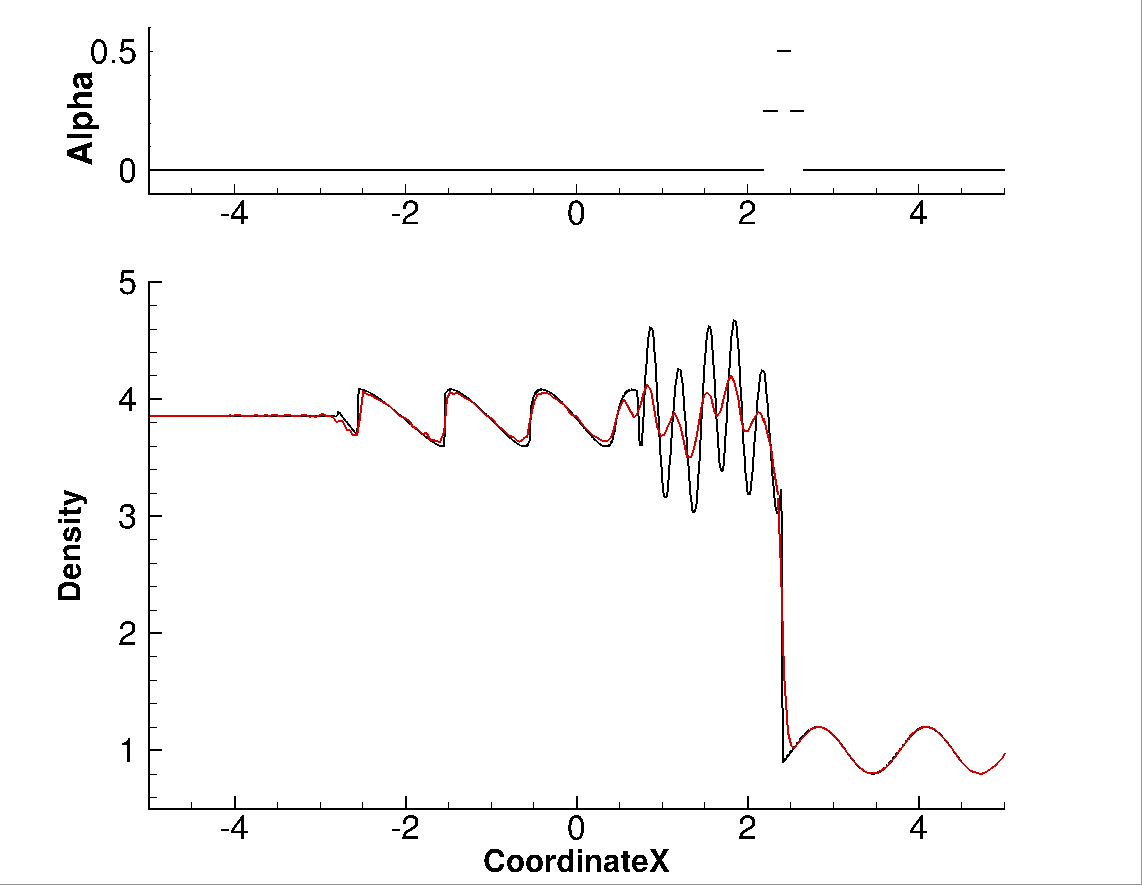}}
	\subfigure[Density on $128$ elements ($640$ \ac{dof}) zoomed onto the post shock oscillations] {
		\includegraphics[trim={1cm 0.1cm 1cm 1cm},clip,width=0.48\linewidth]{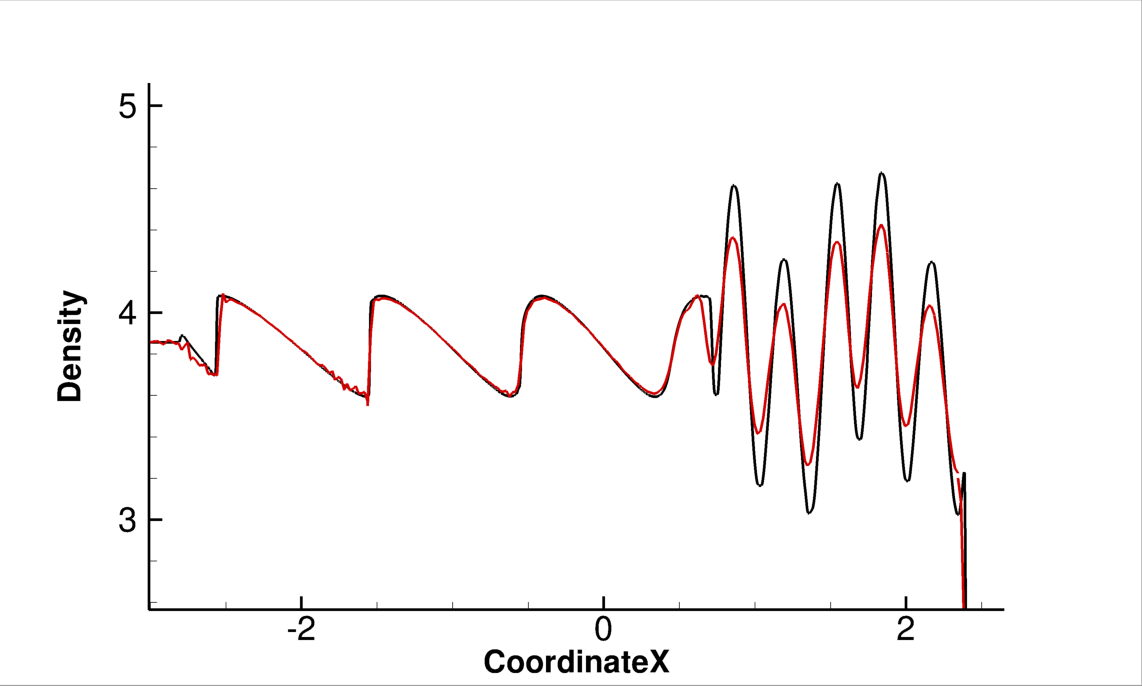}}
	\subfigure[Density and $\alpha$ on $128$ elements ($640$ \ac{dof})] {
		\includegraphics[trim={0cm 0.2cm 0.1cm 0cm},clip,width=0.48\linewidth]{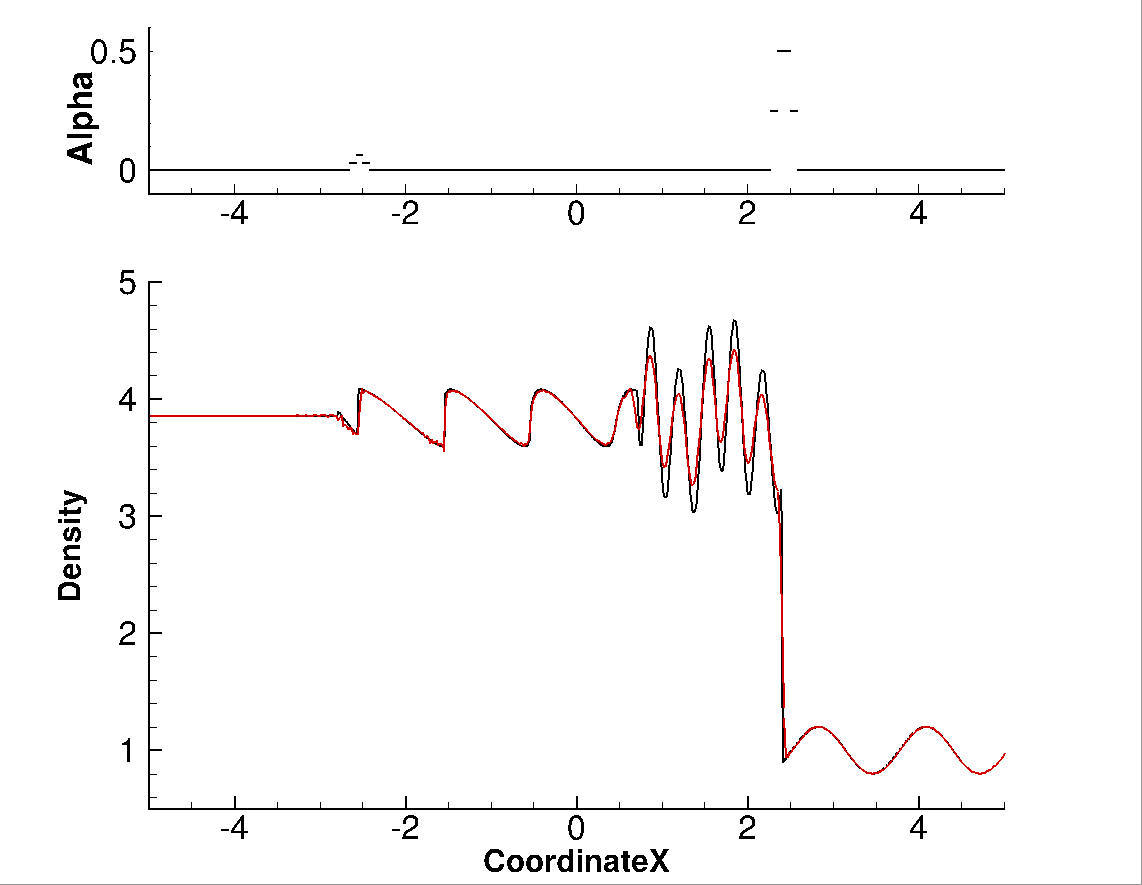}}
	\subfigure[Density on $256$ elements ($1280$ \ac{dof}) zoomed onto the post shock oscillations] {
		\includegraphics[trim={1cm 0.1cm 1cm 1cm},clip,width=0.48\linewidth]{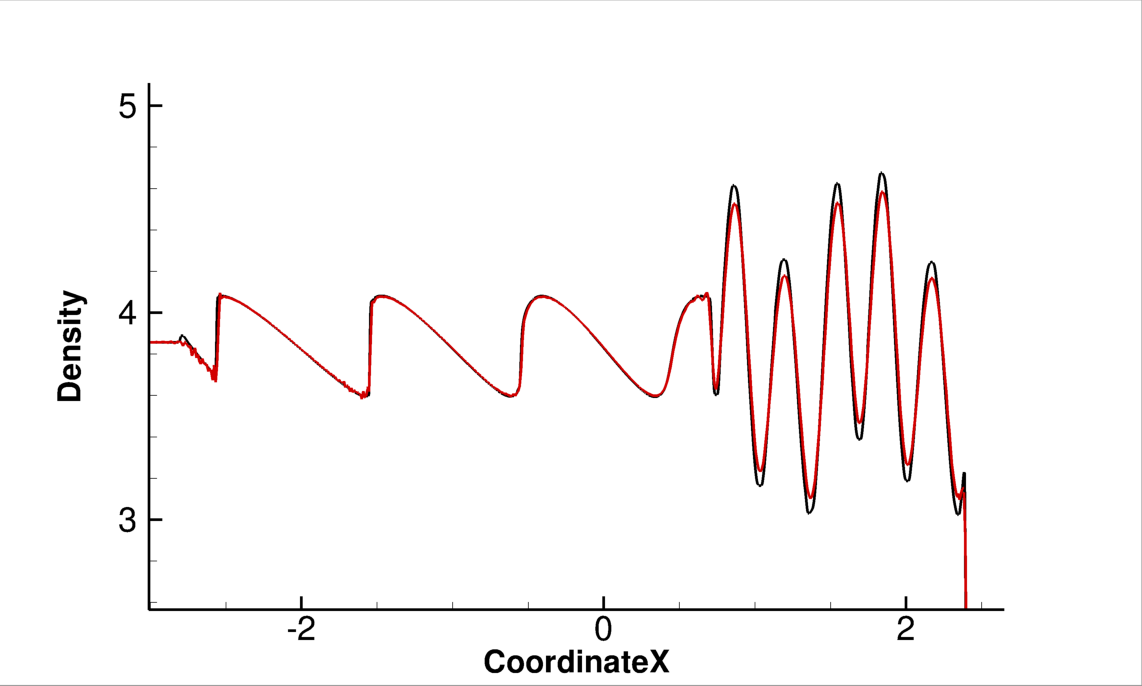}}
	\subfigure[Density and $\alpha$ on $256$ elements ($1280$ \ac{dof})] {
		\includegraphics[trim={0cm 0.2cm 0.1cm 0cm},clip,width=0.48\linewidth]{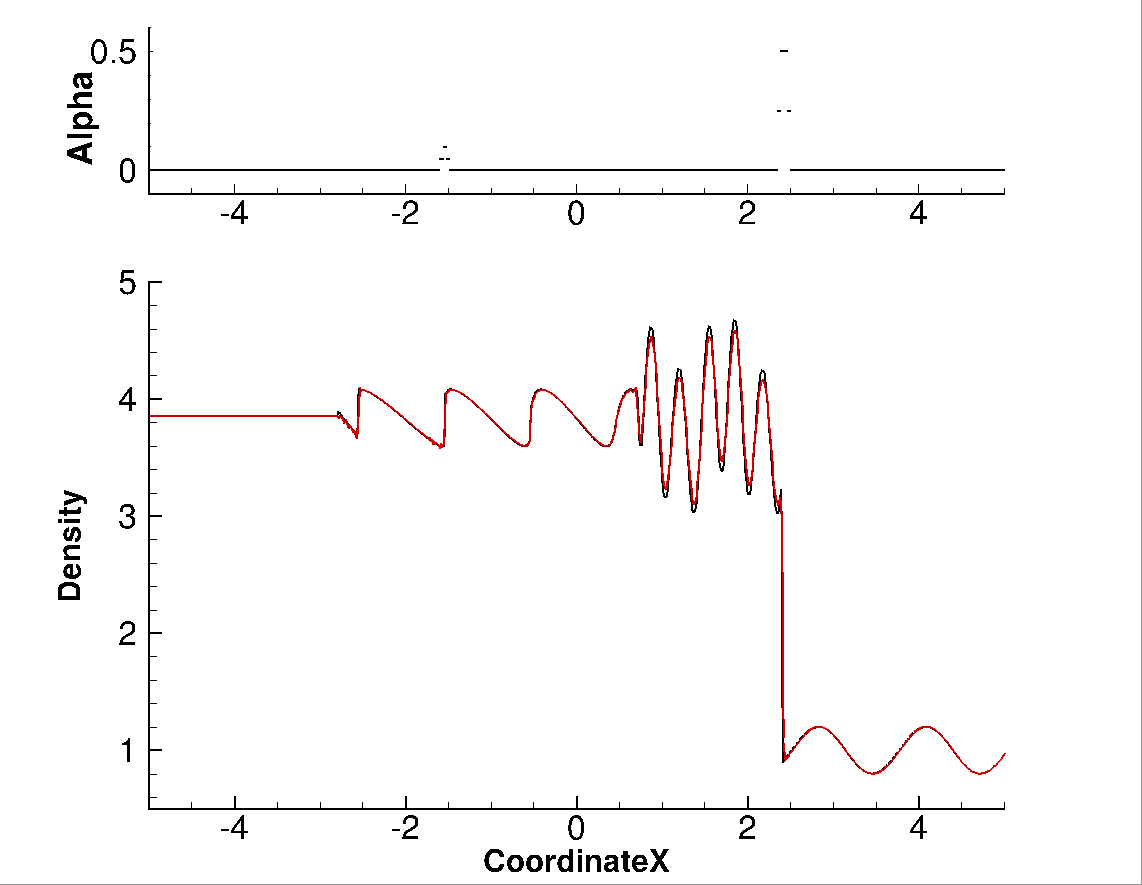}}
	\caption{The Shu Osher shock zoomed in the critical post shock region (Left column). Density and $\alpha$ are shown at $t=1.8$ for three different resolutions with the polynomial degree $N=4$. As a reference we plotted a simulation using $2048$ \ac{dof} in the slightly smaller domain $[-4.5, 4.5]$ with the \ac{ppm} \cite{colella1984piecewise} computed with the code FLASH \cite{fryxell2000flash}. The strong main shock is continuously detected in contrast to the minor post shocks.}
	\label{fig:shuOsherShock}
\end{figure}

A common benchmark for schemes regarding their ability to preserve physical oscillations across shocks is the 1D Shu-Osher shock test \cite[Example 8]{shuOsher1989}. A Mach 3 shock is initialized at $x=-4$ in the domain $[-5,5]$. Additionally the downstream state is distorted by a sine wave, which acts as physical oscillations. The simulation runs until $t=1.8$. The left boundary is set as supersonic inflow the right boundary is set as outflow.

\begin{align*}
\rho(x) &= \begin{cases}
\approx3.857 ,			& \text{if } x < -4 \\
1 + 0.2 \sin(5x),	& \text{if } x \ge 4
\end{cases},
& \ux(x) &= \begin{cases}
\approx2.629 ,	& \text{if } x < -4\\
0,	& \text{if } x \ge 4
\end{cases},  
& p(x) &= \begin{cases}
\approx10.333 ,	& \text{if } x < -4\\
1,	& \text{if } x \ge 4
\end{cases}.
\end{align*}
In Figure~\ref{fig:shuOsherShock} we can see the results of the computation for different resolutions ($320, 640$ and $1280$ \ac{dof}). The method converges towards the reference solution. The shock capturing mechanism mainly triggers at the main shock. Minor shocks in the post shock region are not continuously detected. Here we see that the parameters of the shock detector are calibrated to provide stability in the presence of strong shocks.

\subsection{Shock Diffraction on a Backward Facing Step}
\label{sec:bfs}
Originally this test case was used as a benchmark to compare shock capturing schemes \cite{takayama1991} and was further analysed in \cite{hillier1991} and \cite{bagabir2016}. We want to observe if the scheme can handle the corner, where the density can get very close to zero. To increase the stress on our proposed discretisation, we chose to increase the Mach number of the shock from the original $1.5$ to the higher value of $100$ and adapt the domain so that the shock wave propagation is in focus at the final time.

\begin{figure}[!htbp]
	\centering
	\subfigure{\includegraphics[width=0.49\linewidth]
		{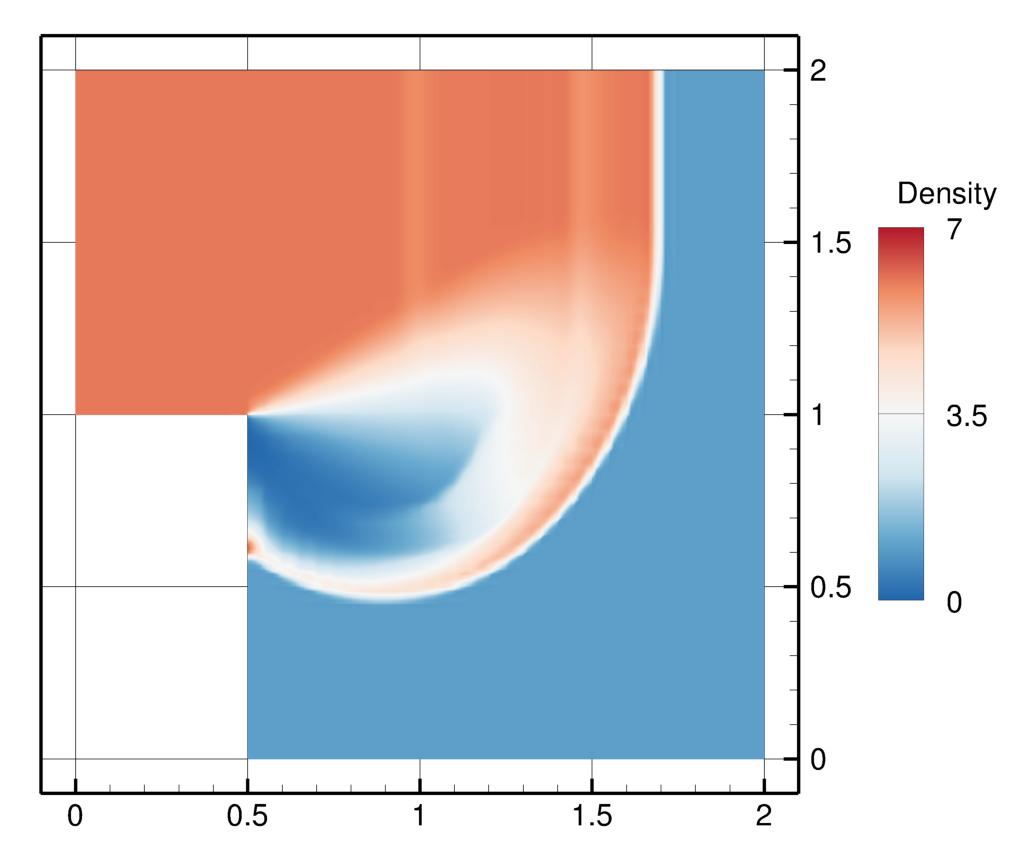}}
	\subfigure{\includegraphics[width=0.49\linewidth]
		{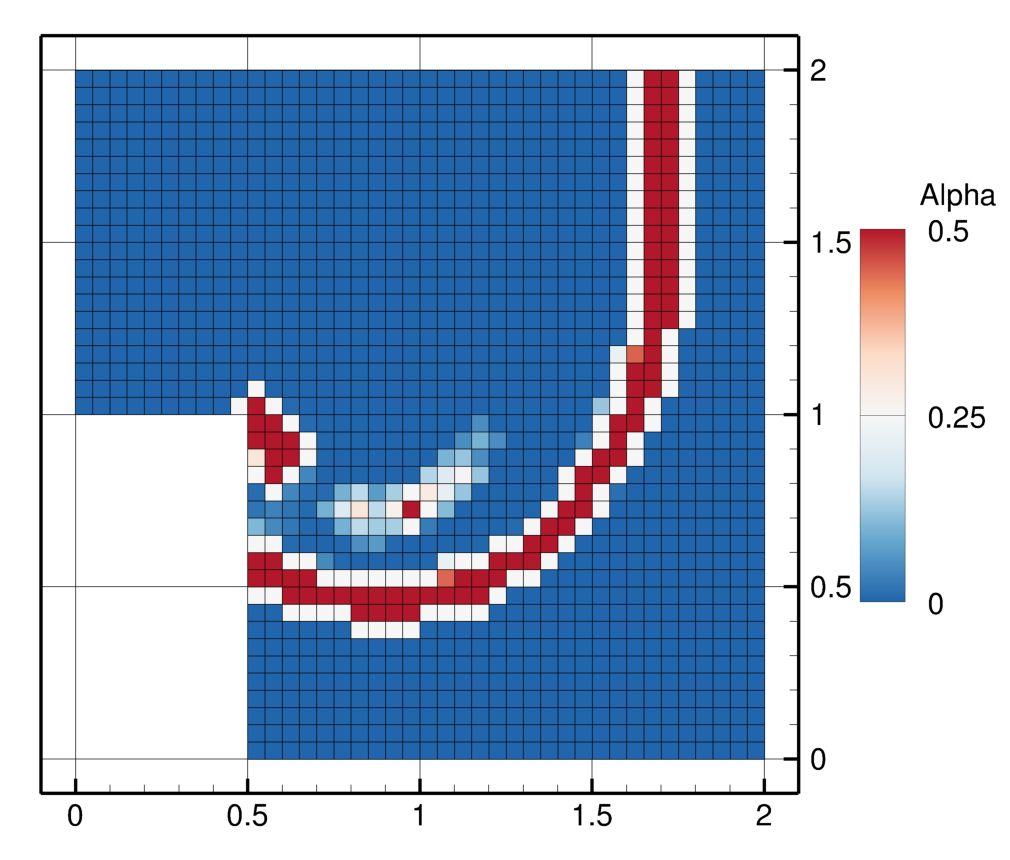}}
	\subfigure{\includegraphics[width=0.49\linewidth]
		{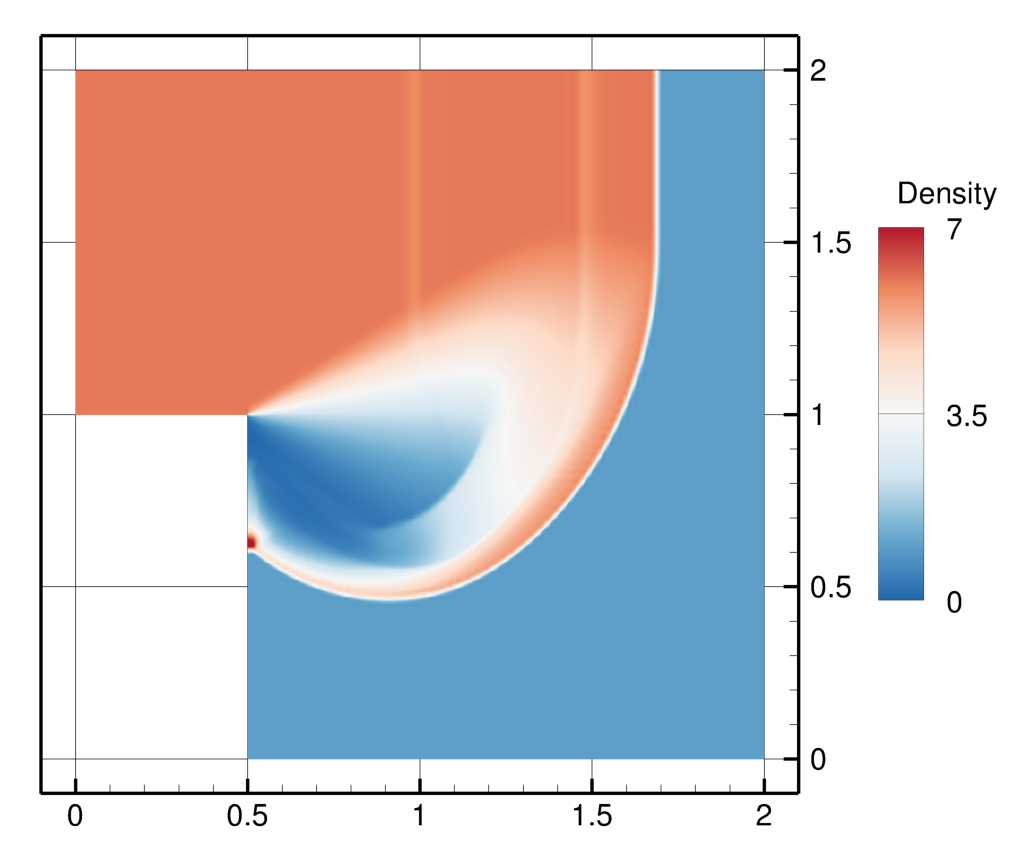}}
	\subfigure{\includegraphics[width=0.49\linewidth]
		{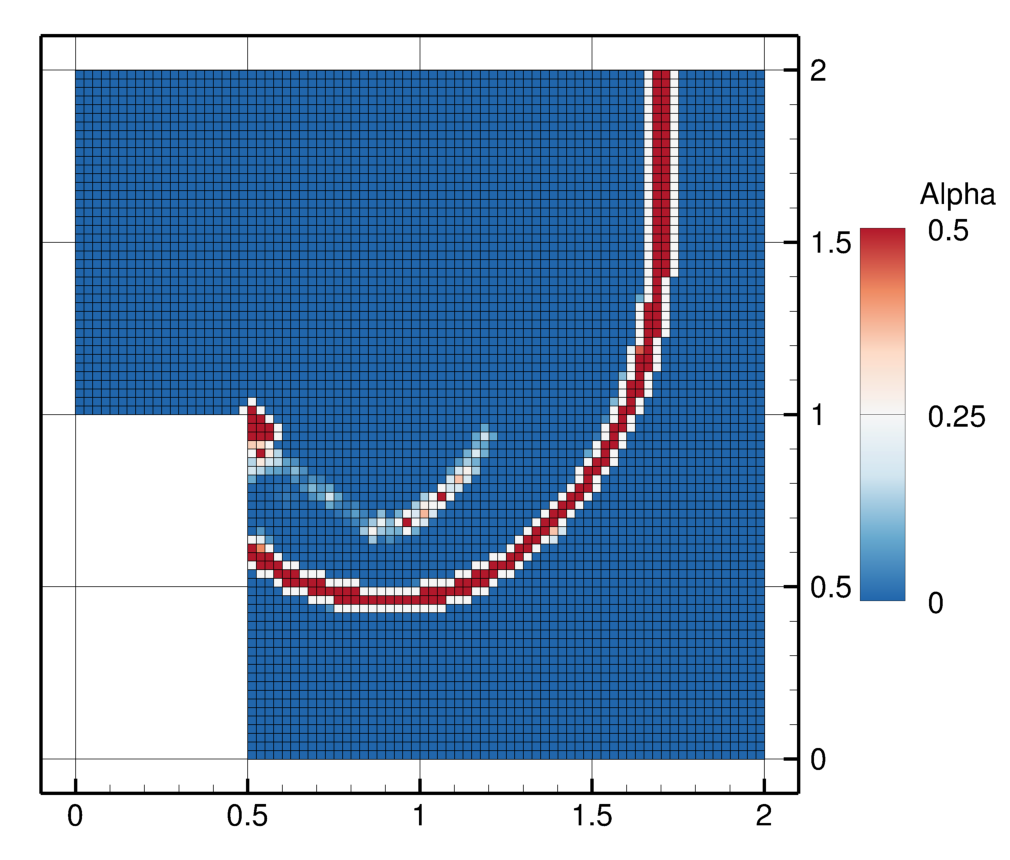}}
	\subfigure{\includegraphics[width=0.49\linewidth]
		{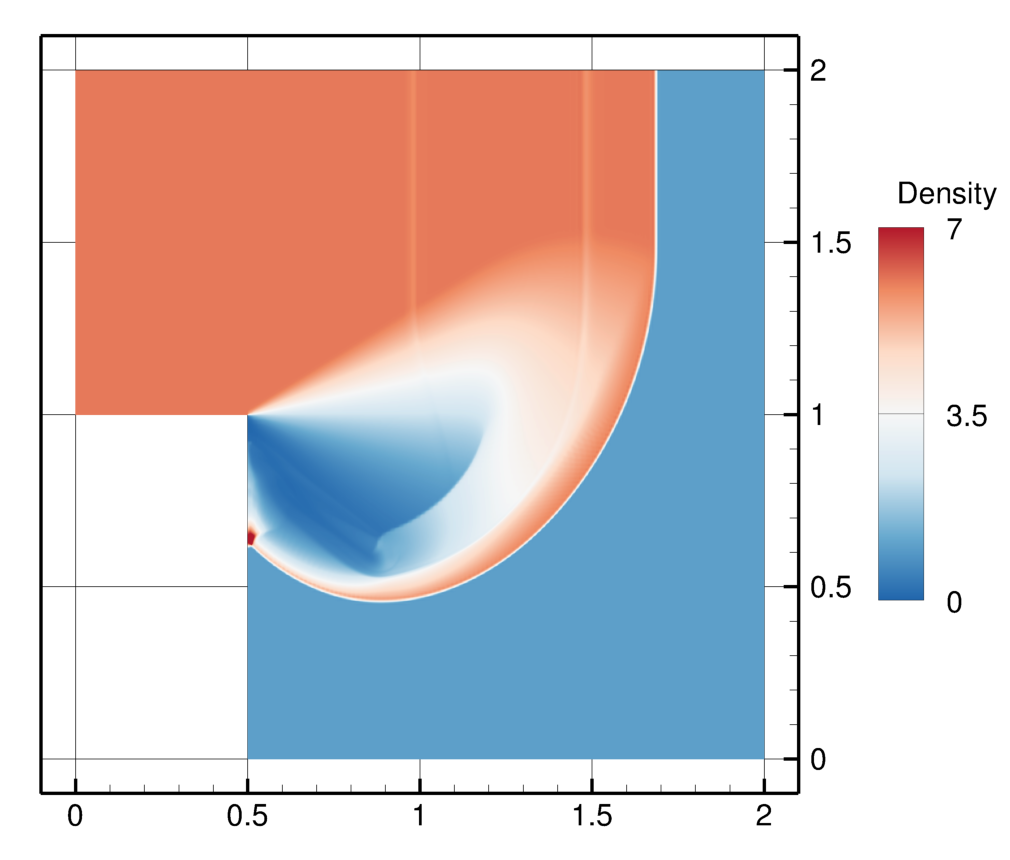}}
	\subfigure{\includegraphics[width=0.49\linewidth]
		{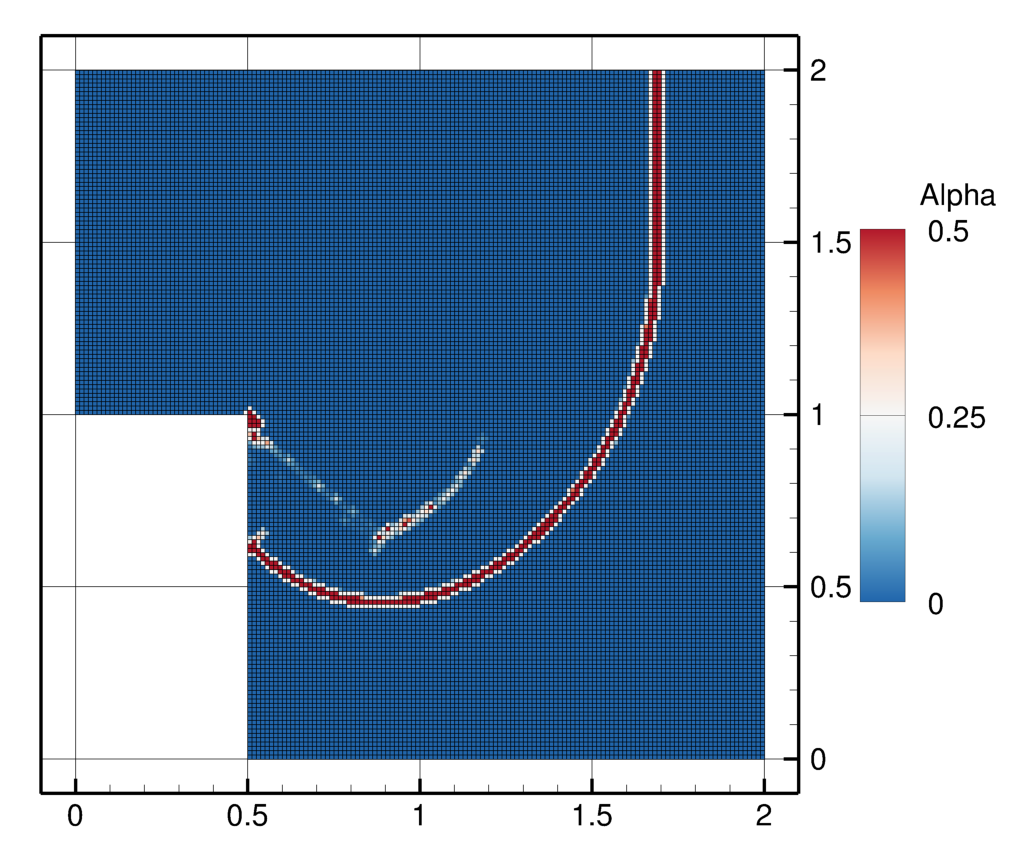}}
	\caption{Mach 100 shock wave diffraction over a backward facing step at simulation time $t=0.01$. Density $\rho$ in the left part and blending factor $\alpha$ with mesh lines in the right part. Corresponding total resolutions in \ac{dof} for polynomial degree $N=4$ is: 35k (top), 140k (mid), 560k (bottom).}
	\label{fig:bfs}
\end{figure}

The Mach $100$ shock travels through a channel of resting gas and hits at $t=0$ a backward facing step $\Omega = [0,2]^2\ \backslash\ ([0, 0.5]\times [0,1])$ at its corner $(0.5,1)$. The post shock states get computed by the normal shock wave equations \cite{naca1951}

\begin{align*}
\rho(x,y) &= \begin{cases}
\approx5.9970,		& \text{if } x \le 0.5 \\
1,		& \text{if } x > 0.5
\end{cases},
& \ux(x,y) &= \begin{cases}
\approx98.5914,	& \text{if } x \le 0.5\\
0,	& \text{if } x > 0.5
\end{cases}, \\
p(x,y) &= \begin{cases}
11666.5,	& \text{if } x \le 0.5\\
1,	& \text{if } x > 0.5
\end{cases},
& \uy(x,y) &= 0.
\end{align*}
The left boundary is set as supersonic inflow and the right boundary as supersonic outflow. All other boundaries are reflecting walls. The numerical results are depicted in Figure~\ref{fig:bfs}, where the density contours and the distribution of the blending factors $\alpha$ is plotted for three different simulations with coarse, medium and fine resolution. 

\subsection{Supersonic Flow over Forward Facing Step}

In this case, the classic forward facing step channel \cite{woodward1984} is considered. The tunnel with step gets initialised with a uniform supersonic flow. The flow dynamics generate several phenomena like a strong bow shock, shock reflections and a \changed{Kelvin-Helmholtz} instability. The idea is to use this test case to investigate the ability of a shock capturing scheme to distinguish between steep gradients, like the vortices of the Kelvin-Helmholtz instability, and shocks, which occur in various forms. The domain $\Omega = ([0,3]\times[0,1])\ \backslash\ ([0.6, 3]\times[0.2,1])$ is initialized with
\begin{equation}
\rho(x,y) = 1.4, \qquad
p(x,y) = 1, \qquad
\ux(x,y) = 3, \qquad
\uy(x,y) = 0.
\end{equation}
The left boundary is set as supersonic inflow and the right one as supersonic outflow. All other boundaries are reflecting walls. A known problem of this test case is that the step corner is a singularity for the velocity, since in the friction-less case the walls do not force velocity to be equal to 0, see Woodward and Colella \cite{woodward1984}. We account for this behaviour and refine the mesh around the corner to reduce strong numerical artefacts.

\begin{figure}[!htbp]
	\centering
	\subfigure{\includegraphics[width=0.49\linewidth]
		{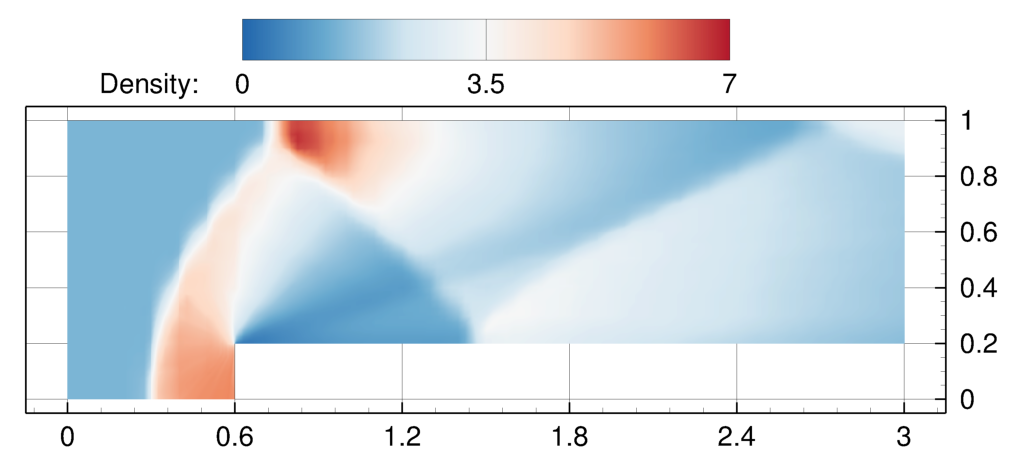}}
	\subfigure{\includegraphics[width=0.49\linewidth]
		{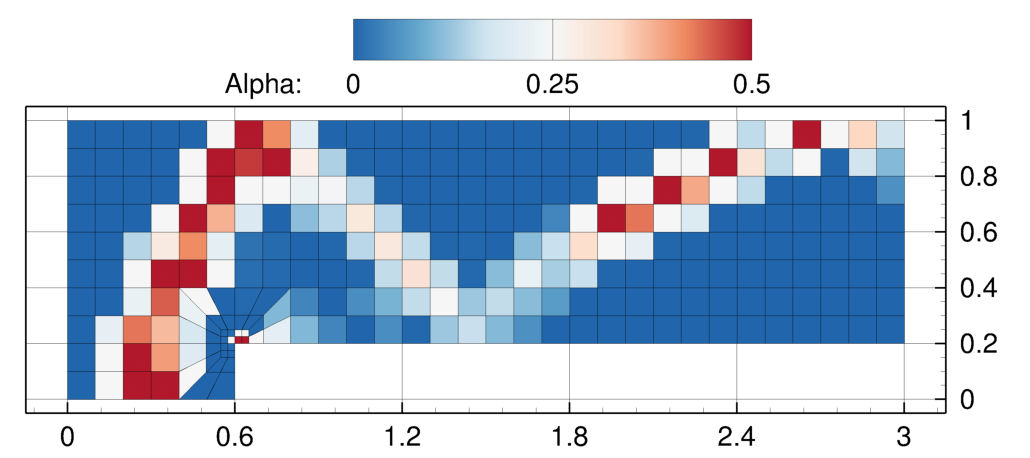}}
	\subfigure{\includegraphics[trim={0 0 0 3.5cm},clip,width=0.49\linewidth]
		{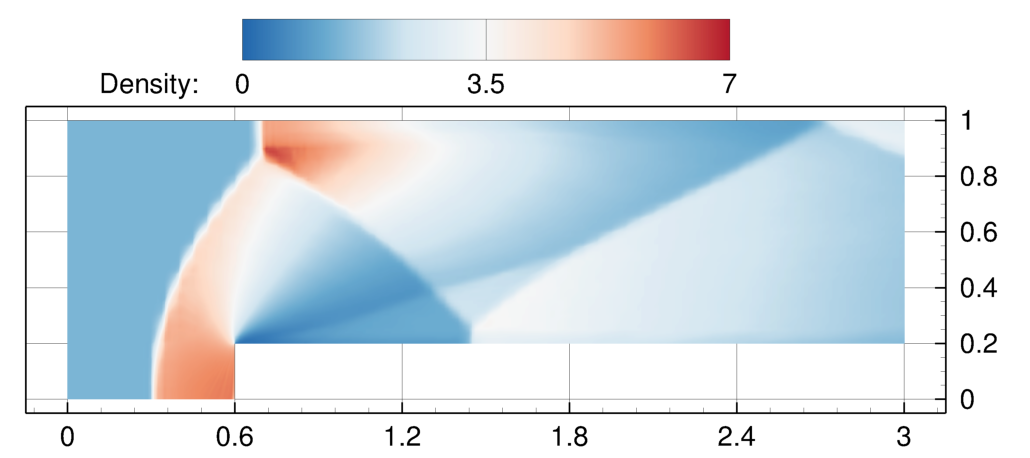}}
	\subfigure{\includegraphics[trim={0 0 0 3.5cm},clip,width=0.49\linewidth]
		{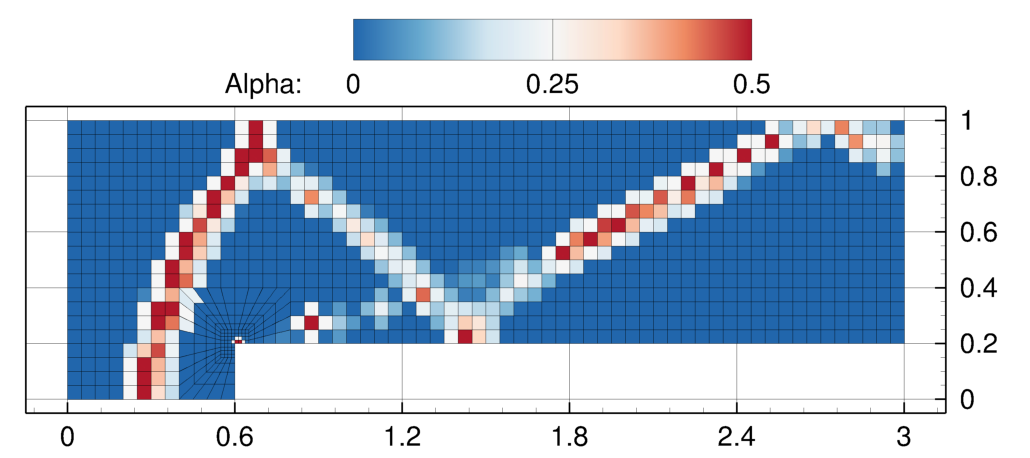}}
	\subfigure{\includegraphics[trim={0 0 0 3.5cm},clip,width=0.49\linewidth]
		{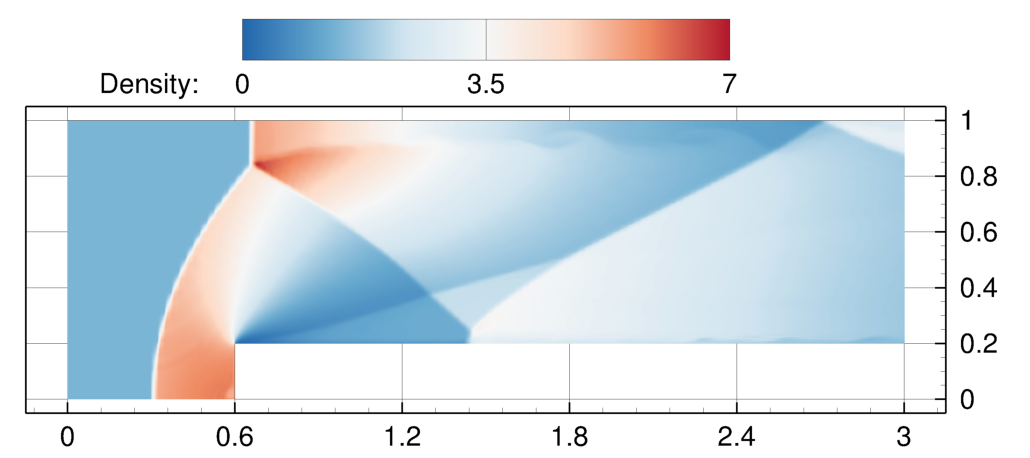}}
	\subfigure{\includegraphics[trim={0 0 0 3.5cm},clip,width=0.49\linewidth]
		{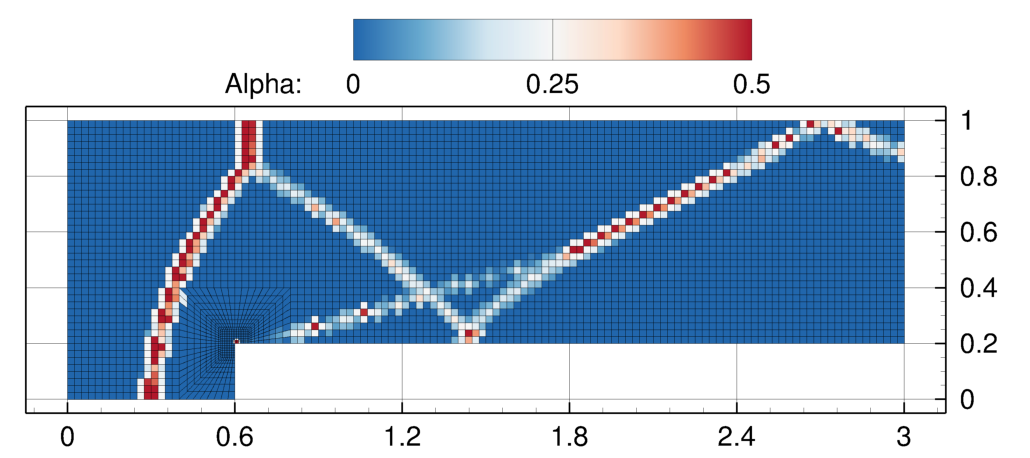}}
	\caption{Forward facing step at $t=3$. Density $\rho$ in the left and shock indicator $\alpha$ with mesh lines in the right column. Standard element size  $\Delta x_{\max}$ is $4$ times the element size $\Delta x_{\min}$ around the corner. Resolutions with polynomial degree $N=4$ in \ac{dof} is: 6.9k (top), 27.6k (mid), 110.4k (bottom).}
	\label{fig:ffs}
\end{figure}

We can observe in Figure~\ref{fig:ffs}, that the blending factor behaves very well and robustly detects the shocks, while keeping the contact discontinuities at minimum dissipation. At the finest grid resolution, the Kelvin-Helmholtz instability in the top of the channel gets triggered. Some numerical errors at the corner singularity still occur and propagate into the domain, causing a very thin artificial boundary layer. This layer hits the shock standing orthogonal on the lower wall at $x\approx 1.45$ and results in a small vortical driven boundary layer further downstream.

\subsection{Double Mach Reflection}
\label{sec:dmr}
\begin{figure}[!htbp]
	\centering
	\includegraphics[width=0.8\linewidth]{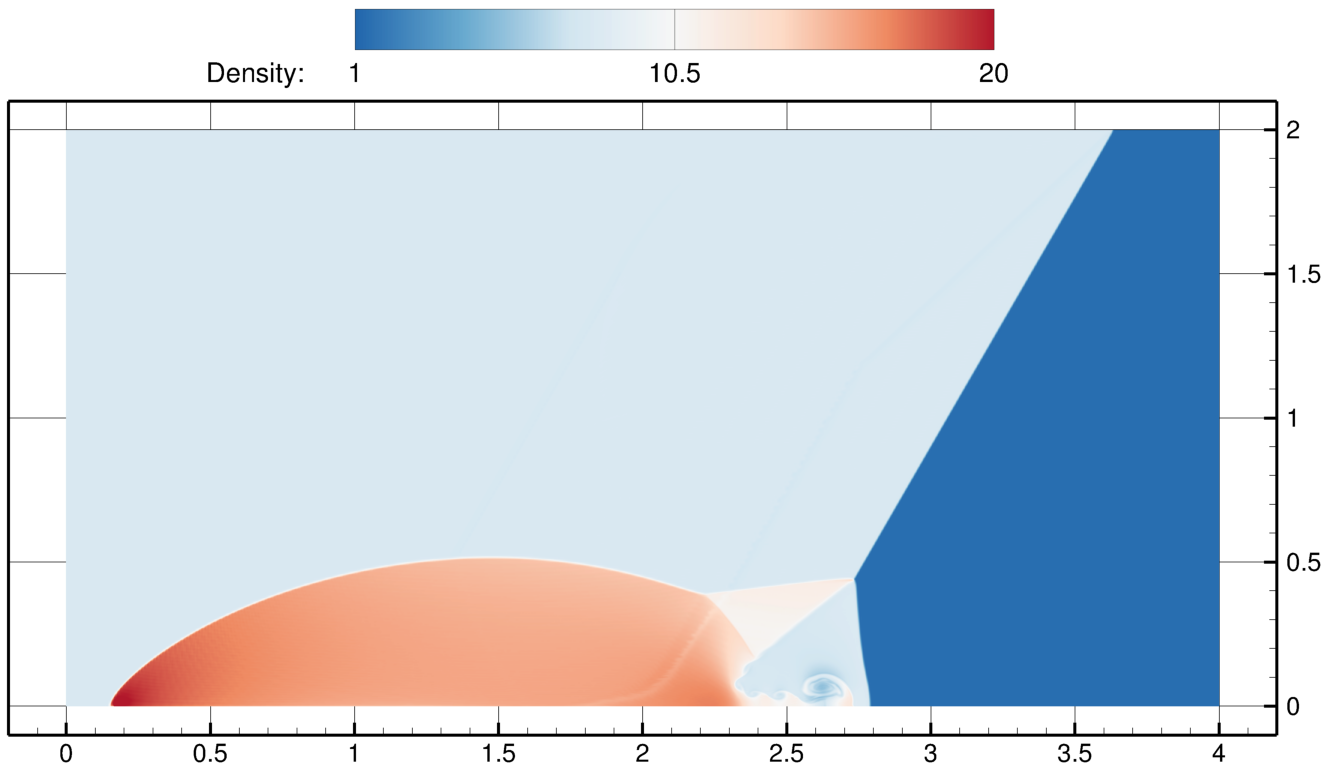}
	\caption{Double Mach Reflection problem: Density contours in the whole domain at simulation time $t=0.2$ with an equivalent Cartesian mesh resolution $\Delta x = 1/96$ (1.8432 M \ac{dof}).}
	\label{fig:dmr1}
\end{figure}

\changed{This is another common test case, which is discussed in detail in \cite{woodward1984} and \cite{kemm2016}.
The general setup and result are depicted in Figure~\ref{fig:dmr1}.}
A Mach $10$ shock hits a wedge of angle $\phi= \pi / 6$ at $t=0$. The domain is rotated to allow for a simple Cartesian mesh setup on a rectangular domain $\Omega = ([0,4]\times[0,2])$, see e.g. \cite{kemm2016} for a discussion. The initial conditions are given by

\begin{align*}
\rho(x,y)  &= \begin{cases}
8,		& \text{if } x \le g(y,0) \\
1.4,		& \text{if } x > g(y,0)
\end{cases}, &
\ux(x,y)  &= \begin{cases}
7.144709581221619,		& \text{if } x \le g(y,0) \\
0,		& \text{if } x > g(y,0)
\end{cases}, \\
p(x,y) &= \begin{cases}
116.5,		& \text{if } x \le g(y,0) \\
1,		& \text{if } x > g(y,0)
\end{cases}, &
\uy(x,y) &= \begin{cases}
-4.125,		& \text{if } x \le g(y,0) \\
0,		& \text{if } x > g(y,0)
\end{cases},
\end{align*}
where $g(y,t) = y \tan(\phi) + 1 / 6 + 10 / \cos(\phi) t$ is the analytical shock position. The wedge boundary $\{(x,0) | x \in [1/6,4]\}$ is a reflecting wall. All other boundaries are set as the analytical solution of the shock motion.

\begin{figure}[!htbp]
	\centering
	\subfigure{\includegraphics[width=0.32\linewidth]
		{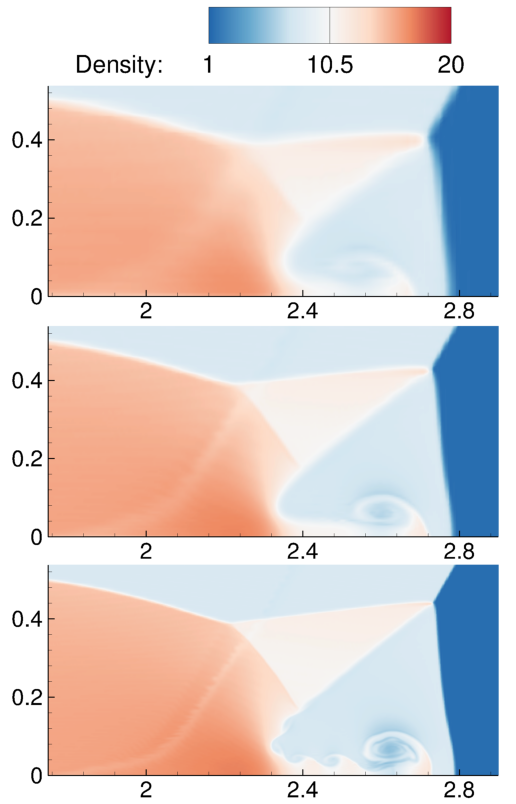}}
	\subfigure{\includegraphics[width=0.32\linewidth]
		{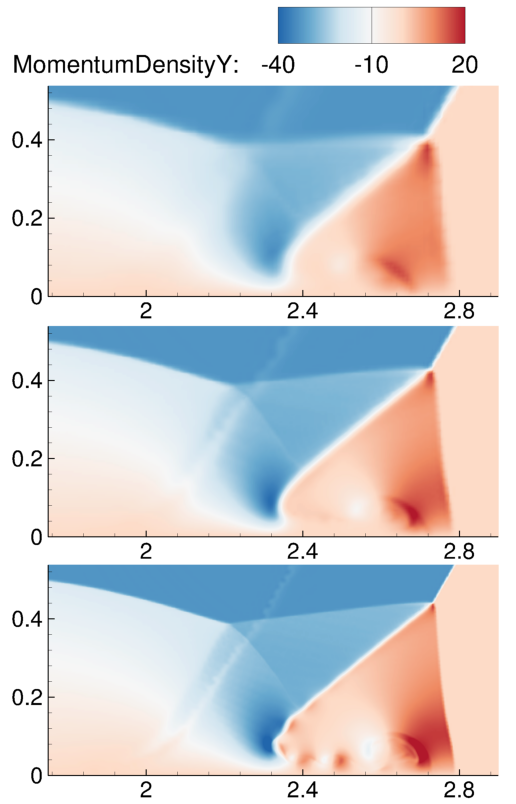}}
	\subfigure{\includegraphics[width=0.32\linewidth]
		{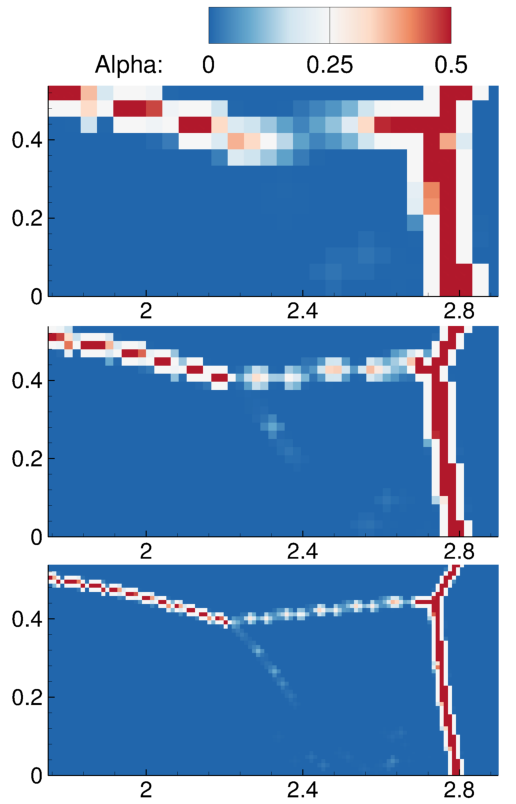}}
	\caption{Simulation result of the Double Mach Reflection problem at $t=0.2$ zoomed in the critical region of the rolled up vortex and the shock triple points. Density $\rho$ in the left, vertical momentum $\rho v$ in the middle and blending factor $\alpha$ in the right column. Resolutions with degree $N=4$ in \ac{dof} per unit square is: 14.4k (top), 57.6k (mid), 230.4k (bottom).}
	\label{fig:dmr}
\end{figure}

The numerical results plotted in Figure~\ref{fig:dmr} confirm that again the blending factor $\alpha$ tracks the shock dynamics quite well. Only some minor non zero blending values are assigned to elements in the vortex roll up region caused by the steep gradients. 

\subsection{Inviscid Bow Shock Upstream of a Blunt Body}
\begin{figure}[!htbp]
	\centering
	\subfigure{\includegraphics[width=0.15\linewidth]
		{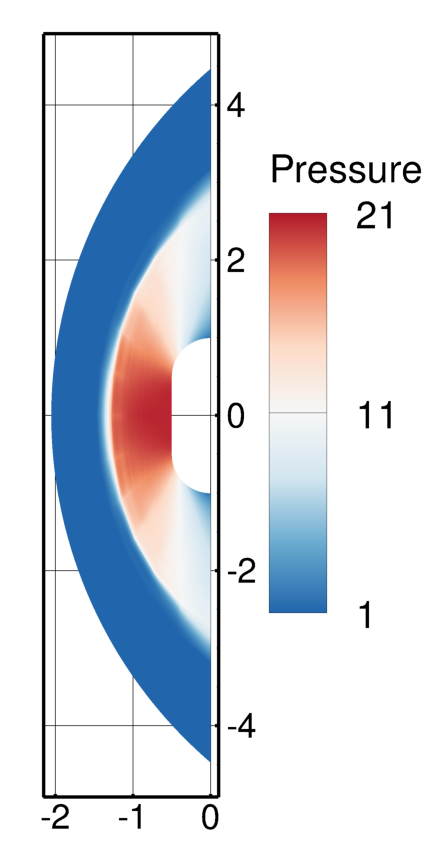}}
	\subfigure{\includegraphics[width=0.15\linewidth]
		{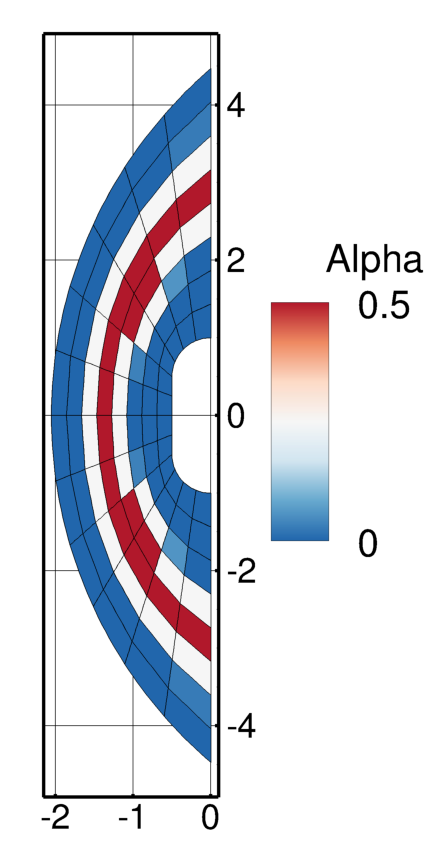}}
	\subfigure{\includegraphics[width=0.15\linewidth]
		{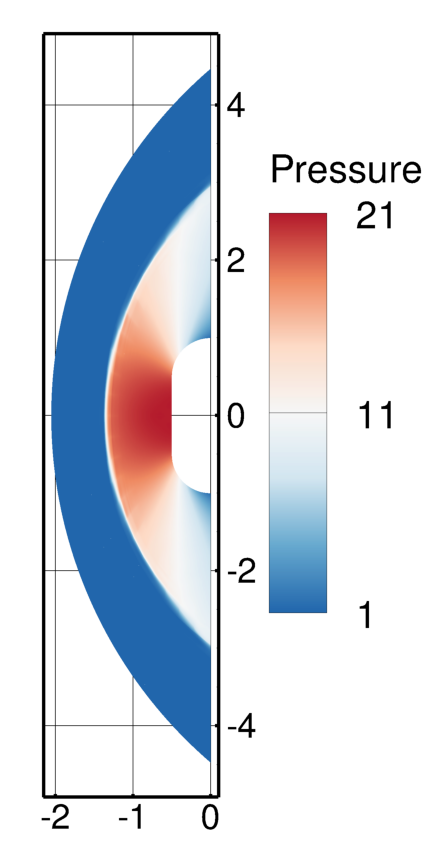}}
	\subfigure{\includegraphics[width=0.15\linewidth]
		{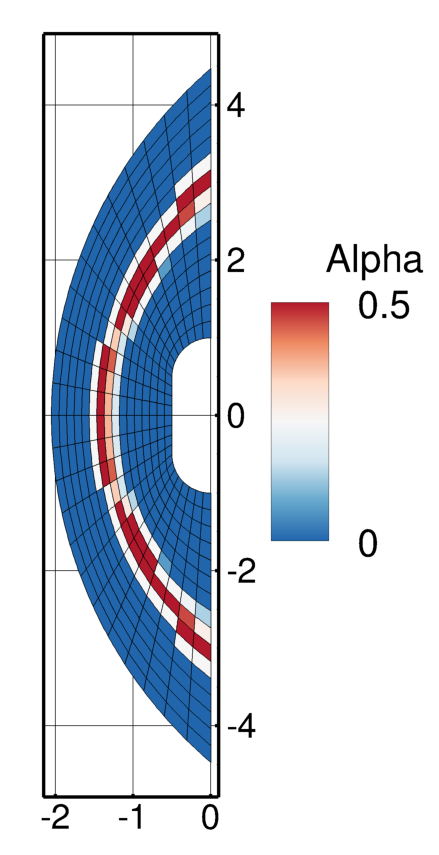}}
	\subfigure{\includegraphics[width=0.15\linewidth]
		{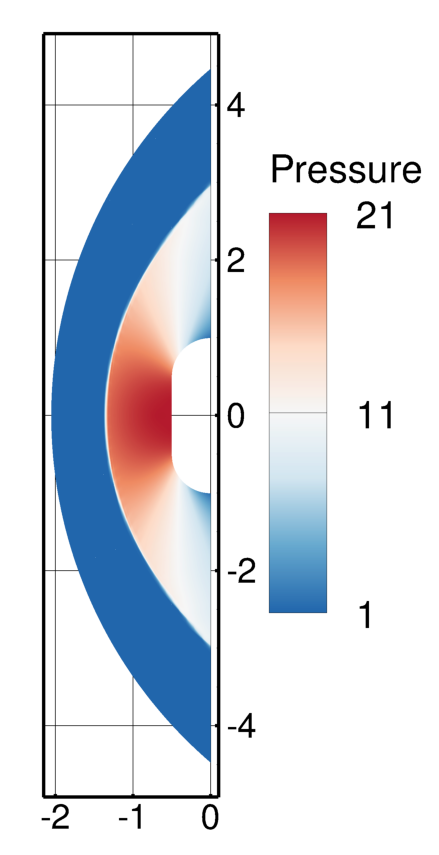}}
	\subfigure{\includegraphics[width=0.15\linewidth]
		{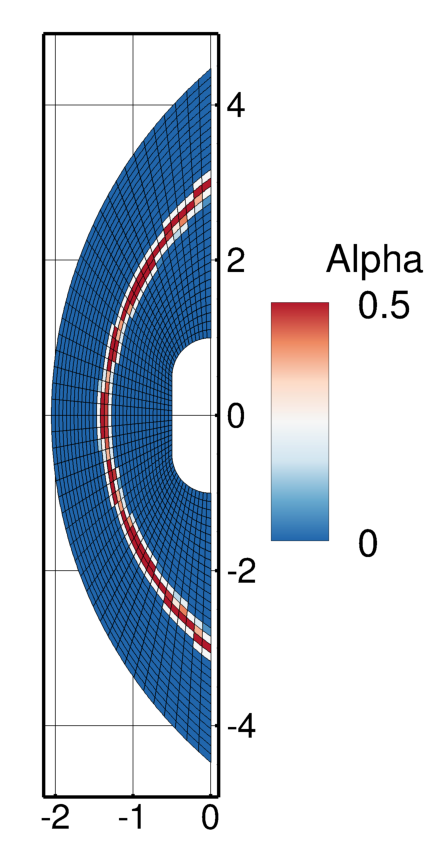}}
	\caption{Simulation results of a bow shock in front of a blunt body at simulation time $t=10$. Pressure $p$ is plotted as well as the blending factor $\alpha$ for three different mesh resolutions. Resolutions with degree $N=4$ in \ac{dof} is Left pair: 2.4k, Mid pair: 9.6k, Right pair: 38.4k.}
	\label{fig:bowshock}
\end{figure}

We follow the description of the test case proposed by the high order computational fluid dynamics workshop \cite{hiocfd5}. The aim of this subsection is to simulate a shock problem on a curvilinear grid. The left boundary is described as a circle with origin $(3.85,0)$ and radius $5.9$. The blunt body consists of a front of length $1$ and two quarter circles of radius $0.5$. The domain is initialized with
\begin{equation}
\rho(x,y) = 1.4, \qquad
p(x,y) = 1, \qquad
\ux(x,y) = 4, \qquad
\uy(x,y) = 0.
\end{equation}
The left boundary is set as supersonic inflow. The boundary on the body is a \changed{reflecting wall} and the right boundaries are supersonic outflows. The numerical results are plotted in Figure~\ref{fig:bowshock} and show again that the method works nicely without adjusting the shock capturing mechanism. 

\section{Conclusions}
In this work, we present the first subcell based shock capturing for \ac{dg} that is provably entropy stable. The main idea is to use a hybrid scheme: A careful blending of a low order and a high order discretisation based on the same \ac{lgl} node distribution allows to perform a common discrete entropy analysis. These theoretical results are available for hexahedral curvilinear meshes and for general diagonal norm SBP operators. In combination with a troubled element indicator, we use the hybrid framework to introduce a shock capturing mechanism with the idea to add the low order discretisation (dissipation) in case a shock is detected in the element. The choice of the troubled element indicator and the blending function $\alpha$ is robust in the sense that we were able to compute all the different test cases without adjusting the settings. Besides the desirable robustness, the numerical simulations with the novel scheme gave very accurate results and ongoing research is focused on the extension to shock turbulence applications.

\section*{Acknowledgements}
Special thanks to Michael Bergmann, who helped with the implementation aspects and is responsible for most of the existing \ac{dg} code in TRACE. Gregor Gassner and Andrés Rueda-Ramírez have been supported by the European Research Council (ERC) under the European Union's Eights Framework Program Horizon 2020 with the research project Extreme, ERC grant agreement no. 714487.

\appendix
\section{Proofs}
\subsection{Entropy Balance Law of First Order 1D Finite Volume}
\label{subsec: ProofFVentropyBalance}
Proof of \eqref{FVentropyBalance}. For a single interface $(j,j+1)$ holds
\begin{align}
	v_j f^*_{(j,j+1)} - \psi_j
	&= (v_j - \frac{1}{2} v_{j+1} + \frac{1}{2} v_{j+1}) f^*_{(j,j+1)} - (\psi_j - \frac{1}{2} \psi_{j+1} + \frac{1}{2} \psi_{j+1})
	\\&= \frac{1}{2} (v_{j+1} + v_j) f^*_{(j,j+1)} - \frac{1}{2} (\psi_{j+1} + \psi_j)
	\\&- \frac{1}{2} (v_{j+1} - v_j) f^*_{(j,j+1)} \changed{+} \frac{1}{2} (\psi_{j+1} - \psi_j)
	\\&= \avg{v}_{(j,j+1)} f^*_{(j,j+1)} - \avg{\psi}_{(j,j+1)}
	\\&- \frac{1}{2} \diff{v}_{(j,j+1)} f^*_{(j,j+1)} \changed{+} \frac{1}{2} \diff{\psi}_{(j,j+1)}
	\\&= q^*_{(j,j+1)} - \frac{1}{2} r_{(j,j+1)}
	\label{eq:entropyFluxRight}
\end{align}

and

\begin{align}
v_{j+1} f^*_{(j,j+1)} - \psi_{j+1}
&= (v_{j+1} - \frac{1}{2} v_j + \frac{1}{2} v_j) f^*_{(j,j+1)} - (\psi_{j+1} - \frac{1}{2} \psi_j + \frac{1}{2} \psi_j)
\\&= \frac{1}{2} (v_{j+1} + v_j) f^*_{(j,j+1)} - \frac{1}{2} (\psi_{j+1} + \psi_j)
\\&+ \frac{1}{2} (v_{j+1} - v_j) f^*_{(j,j+1)} - \frac{1}{2} (\psi_{j+1} - \psi_j)
\\&= \avg{v}_{(j,j+1)} f^*_{(j,j+1)} - \avg{\psi}_{(j,j+1)}
\\&+ \frac{1}{2} \diff{v}_{(j,j+1)} f^*_{(j,j+1)} - \frac{1}{2} \diff{\psi}_{(j,j+1)}
\\&= q^*_{(j,j+1)} + \frac{1}{2} r_{(j,j+1)}
\label{eq:entropyFluxLeft}
\end{align}

and therefore

\begin{align}
& v_j (f^*_{(j,j+1)} - f^*_{(j-1,j)}) 
\\=& v_j f^*_{(j,j+1)} - \psi_j - v_j f^*_{(j-1,j)} + \psi_j
\\=& (q^*_{(j,j+1)} - q^*_{(j-1,j)}) - \frac{1}{2} (r_{(j,j+1)} + r_{(j-1,j)}),
\end{align}
see also \cite[page 19]{Fjordholm2013}.

\subsection{Entropy Balance Law of the 1D DGSEM Split Form}
\label{subsec: ProofDGSEMentropyStab}
Proof of \eqref{eq:DG_entropyChange} (see also \cite[page 14]{chen2017} \changed{and \cite{fisher_phdthesis2012,carpenter2016}}). Performing the product $ v \cdot \eqref{eq:DGSEMconsLaw_nodal_symflux} $ and integrating over the element with the collocation \ac{lgl} quadrature leads to 

\begin{align}
\sum_{j=0}^N w_j J \dot{\eta}_j 
  &= \sum_{j=0}^N w_j v_j J \dot{u}_j
\\&= - \underbrace{\sum_{j=0}^{N} v_j w_j \sum_{k=0}^N 2 D_{jk} f^{*S}_{(j,k)}}_{\text{volume integral}}
- \underbrace{\sum_{j=0}^{N} v_j w_j \left[\frac{\delta_{jN}}{w_N}  (f^*_{(N,R)} - f_N) - \frac{\delta_{j0}}{w_0}  (f^*_{(L,0)} - f_0) \right] }_{\text{surface integral}}
\\&= -q^*_{(N,R)} + \frac{1}{2} r_{(N,R)} + q^*_{(L,0)} + \frac{1}{2} r_{(L,0)} + \sum_{j,k=0}^N Q_{jk} r_{(j,k)}.
\end{align}
The volume integral:
\begin{align}
	 \sum_{j=0}^{N} v_j w_j \sum_{k=0}^N 2 D_{jk} f^{*S}_{(j,k)}
	=& \sum_{j=0}^{N} v_j \sum_{k=0}^N 2 Q_{jk} f^{*S}_{(j,k)}
	\\ \text{(SBP property)} \qquad =& \sum_{j=0}^{N} v_j \sum_{k=0}^N (B_{jk} - Q_{kj} + Q_{jk}) f^{*S}_{(j,k)}
	\\ \text{(def. of $B$ \& consistent flux)} \qquad =& \sum_{j=0}^{N} \tau_j v_j f_j 
	+ \sum_{j,k=0}^N v_j (- Q_{kj} + Q_{jk}) f^{*S}_{(j,k)}
	\\ \text{(symm. flux \& re-index)} \qquad =& \sum_{j=0}^{N} \tau_j v_j f_j 
	+ \sum_{j,k=0}^N Q_{jk} (v_j - v_k) f^{*S}_{(j,k)} 
	\\ \text{(definition of $r$)} \qquad =& \sum_{j=0}^{N} \tau_j v_j f_j 
	+ \sum_{j,k=0}^N Q_{jk} (\psi_j - \psi_k - r_{(j,k)}) 
	\\ =& \sum_{j=0}^{N} \tau_j v_j f_j 
	+ \sum_{j=0}^N \psi_j \underbrace{\sum_{k=0}^N Q_{jk}}_{=0}
	- \sum_{k=0}^N \psi_k \underbrace{\sum_{j=0}^N Q_{jk}}_{=\tau_k}
	- \sum_{j,k=0}^N Q_{jk} r_{(j,k)}   
	\\ =& \sum_{j=0}^{N} \tau_j v_j f_j 
	 - \sum_{k=0}^N \tau_k \psi_k 
	 - \sum_{j,k=0}^N Q_{jk} r_{(j,k)} 
	\\ \text{(rename sum index)} \qquad =& \sum_{j=0}^{N} \tau_j v_j f_j 
	- \sum_{j=0}^N \tau_j \psi_j 
	- \sum_{j,k=0}^N Q_{jk} r_{(j,k)} 
	\\ =& \sum_{j=0}^{N} \tau_j (v_j f_j - \psi_j) - \sum_{j,k=0}^N Q_{jk} r_{(j,k)}
	\\ \text{(defintion of $\tau$)} \qquad =& v_N f_N - \psi_N - (v_0 f_0 -\psi_0) - \sum_{j,k=0}^N Q_{jk} r_{(j,k)} \label{eq:EntropyDGSEM_VolumeX}
\end{align}

The surface integral:

\begin{align}
&\sum_{j=0}^{N} v_j w_j \left[\frac{\delta_{jN}}{w_N} (f^*_{(N,R)} - f_N) - \frac{\delta_{j0}}{w_0}  (f^*_{(L,0)} - f_0) \right]
\\&= v_N w_N \frac{1}{w_N}  (f^*_{(N,R)} - f_N) - v_0 w_0 \frac{1}{w_0}  (f^*_{(L,0)} - f_0)
\\&= v_N (f^*_{(N,R)} - f_N) - v_0 (f^*_{(L,0)} - f_0)
\label{eq:EntropyDGSEM_SurfaceX}
\end{align}

Adding \eqref{eq:EntropyDGSEM_SurfaceX} and  \eqref{eq:EntropyDGSEM_VolumeX} leads to

\begin{align}
& v_N f_N - \psi_N - (v_0 f_0 -\psi_0)  - \sum_{j,k=0}^N Q_{jk} r_{(j,k)}
+ v_N (f^*_{(N,R)} - f_N) - v_0 (f^*_{(L,0)} - f_0)
\\=& (v_N f^*_{(N,R)} - \psi_N) - (v_0 f^*_{(L,0)} - \psi_0)  - \sum_{j,k=0}^N Q_{jk} r_{(j,k)}
\\=& q^*_{(N,R)} - \frac{1}{2} r_{(N,R)} - q^*_{(L,0)} - \frac{1}{2} r_{(L,0)} - \sum_{j,k=0}^N Q_{jk} r_{(j,k)} ,
\end{align}
using \eqref{eq:entropyFluxRight} and \eqref{eq:entropyFluxLeft}.

\section{Entropy Stability on 3D Curved Elements}
\label{sec:3d_es}
We start with the scalar conservation law in 3 space dimensions.
\begin{equation}
	\dot{u} + \nabla \cdot \spvec{f}(u) = 0,
\end{equation}
where 
\begin{equation}
	u : \mathbb{R}^3 \times \mathbb{R} \rightarrow \mathbb{R}, \quad u = u(x,t),
\end{equation}

\begin{equation}
	\dot{u} := \frac{\partial u}{\partial t},
\end{equation}

\begin{equation}
	\nabla = \left(\frac{\partial}{\partial x^1}, \frac{\partial}{\partial x^2},
	\frac{\partial}{\partial x^3}\right)^T,
\end{equation}

\begin{equation}
	\spvec{f}: \mathbb{R} \rightarrow \mathbb{R}^3, \quad u \mapsto (f^1(u), f^2(u), f^3(u))^T. \label{eq:3d_flux_physical}
\end{equation}

\subsection{Mapping the Equation}
Assume we have a mapping $\spvec{x}(\xi^1,\xi^2,\xi^3) : [0,1]^3 \rightarrow \mathbb{R}^3 $ that describes a single element $C$ of the mesh. The \changed{covariant basis vectors} are defined as
\begin{equation}
	\va_i = \frac{\partial \spvec{x}}{\partial \xi^i}, \quad i \in \{1,2,3\}
	\label{covariant_vectors}
\end{equation}
and the (volume weighted) contravariant vectors are
\begin{equation}
	J\va^i = \va_j \times \va_k, \quad (i,j,k) \text{ cyclic},
\end{equation}
where 
\begin{equation}
	J = \va_i \cdot (\va_j \times \va_k), \quad (i,j,k) \text{ cyclic},
\end{equation}
is the Jacobian of the transformation. Define the transformation matrix
\begin{equation}
	M := \left(\begin{array}{ccc}
	Ja^1_1 & Ja^2_1 & Ja^3_1 \\
	Ja^1_2 & Ja^2_2 & Ja^3_2 \\
	Ja^1_3 & Ja^2_3 & Ja^3_3 \\
	\end{array}\right).
\end{equation}
Then the transformation of a gradient is 
\begin{equation}
\nabla \cdot \spvec{f} = \frac{1}{J} \nabla_\xi \cdot (M^T \spvec{f}).
\end{equation}
We define 
\begin{equation}
	\spvec{\tilde{f}} := M^T \spvec{f}
\end{equation}
and get the conservation law in reference space
\begin{equation}
	J \dot{u} + \nabla_\xi \cdot \spvec{\tilde{f}}(u) = 0, \label{curvedEquation}
\end{equation}

\subsection{Entropy Balance Law of the 3D DGSEM Split Form}

Note: \textit{In the 3D tensor product context the spatial approximations are written with triple indices, e.g. $u(x,t) \approx \sum_{i,j,k=0}^{N} u_{ijk}(t) l_{ijk}(x)$. These will be written as subscript. Indices regarding the direction (in physical or reference space) will be written as superscript, like in foreshadowed in \eqref{eq:3d_flux_physical}. To readers not familiar with these notations or multi dimension tensor product methods, \cite[Appendix B]{gassner2016} is recommended.}

The \ac{dgsem} split form approximation in 3D reads

\begin{equation}
	J_{ijk} \dot{u}_{ijk} w_{ijk} + \left(\nabla_\xi \cdot \spvec{\tilde{f}}(u)\right)_{ijk} = 0, \label{eq:3d_dgsem}
\end{equation}
with
\begin{align}
\begin{split}
\frac{1}{w_{ijk}} \left(\nabla_\xi \cdot \spvec{\tilde{f}}(u)\right)_{ijk} 
\approx& 2\sum_{m=0}^{N} D_{im} \tilde{f}^{1*S}_{(i,m)jk} 
+\frac{1}{w_i} \left( \delta_{iN} \left[\tilde{f}^{1*}_{(N,R)jk} - \tilde{f}^1_{Njk}\right] - \delta_{i0} \left[\tilde{f}^{1*}_{(L,0)jk} - \tilde{f}^1_{0jk}\right]\right)
\\ +& 2\sum_{m=0}^{N} D_{jm} \tilde{f}^{2*S}_{i(j,m)k}  
+\frac{1}{w_j} \left( \delta_{jN} \left[\tilde{f}^{2*}_{i(N,R)k} - \tilde{f}^2_{iNk}\right] - \delta_{j0} \left[\tilde{f}^{2*}_{i(L,0)k} - \tilde{f}^2_{i0k}\right]\right) 
\\ +& 2\sum_{m=0}^{N} D_{km} \tilde{f}^{3*S}_{ij(k,m)} 
+\frac{1}{w_k} \left( \delta_{kN} \left[\tilde{f}^{3*}_{ij(N,R)} - \tilde{f}^3_{ijN}\right] - \delta_{k0} \left[\tilde{f}^{3*}_{ij(L,0)} - \tilde{f}^3_{ij0}\right]\right), \label{eq:3d_dgsem_split_form}
\end{split}
\end{align}
where the curvilinear symmetric two-point volume flux (in $\xi^1$-direction) is defined as
\begin{equation}
	\tilde{f}^{1*S}_{(i,m)jk} := \spvec{f}^{*S}(u_{ijk}, u_{mjk}) \cdot \avg{J\va^1}_{(i,m)jk}.
\end{equation}
Analogue definitions for the $\tilde{f}^{2*S}_{i(j,m)k}$ and $\tilde{f}^{3*S}_{ij(k.m)}$ hold.
The average and difference (in first direction) are defined as
\begin{align}
\avg{\cdot}_{(i,m)jk} &:= \frac{1}{2} \left[(\cdot)_{ijk} + (\cdot)_{mjk}\right], \\
\diff{\cdot}_{(i,m)jk} &:=(\cdot)_{mjk} - (\cdot)_{ijk}.
\end{align}

Note: \textit{Numerical fluxes with an index that exceeds the polynomial approximation ansatz ($\{0,\ldots,N\}$) like $L:=-1$ or $R:=N+1$ are interpreted as the opposing state of the Riemann problem, similar to \eqref{index_pair}}. \\

\begin{equation} \label{eq:EntNumFluxDG}
\tilde{q}^{1*}_{(i,m)jk} := \avg{v}_{(i,m)jk} \tilde{f}^{1*}_{(i,m)jk} - \avg{J\va^1}_{(i,m)jk} \cdot \avg{\vpsi}_{(i,m)jk}
\end{equation}
is the numerical entropy flux in $\avg{J\va^1}$ direction,
\begin{equation} \label{eq:EntProdDG}
\tilde{r}^1_{(i,m)jk} := \diff{v}_{(i,m)jk} \tilde{f}^{1*}_{(i,m)jk} - \avg{J\va^1}_{(i,m)jk} \cdot \diff{\vpsi}_{(i,m)jk}
\end{equation}
is the numerical entropy production term in $\avg{J\va^1}$ direction,
\begin{equation}
\vpsi  = v \spvec{f}(u(v)) - \spvec{q}(u(v))
\end{equation}
is the entropy flux potential and $v$ is the entropy variable. Analogue definitions for the $\avg{J\va^2}$ and $\avg{J\va^3}$ directions hold.

To determine the entropy balance law, the mesh has to be watertight, which means that the geometry is continuous across all element faces, and the metric identities \changed{\cite{fisher_phdthesis2012,carpenter2016,gassner2016,kopriva2006}} hold. We calculate the integral in $\xi^1$ direction of the fluxes in $\xi^1$ direction first:

\begin{align}
\sum_{i=0}^{N} v_{ijk} w_i \sum_{m=0}^N 2 D_{im} \tilde{f}^{1*S}_{(i,m)jk}
=& \sum_{i=0}^{N} v_{ijk} \sum_{m=0}^N 2 Q_{im} \tilde{f}^{1*S}_{(i,m)jk}
\\ \text{(SBP property)} \qquad =& \sum_{i=0}^{N} v_{ijk} \sum_{m=0}^N (B_{im} - Q_{mi} + Q_{im}) \tilde{f}^{1*S}_{(i,m)jk}
\\ \text{(def. of $B$ \& consistent flux)} \qquad =& \underbrace{\sum_{i=0}^{N} \tau_i v_{ijk} \tilde{f}^1_{ijk}}_{=: (a)} 
+ \sum_{i,m=0}^N v_{ijk} (- Q_{mi} + Q_{im}) \tilde{f}^{1*S}_{(i,m)jk} 
\\ \text{(symm. flux \& re-index)} \qquad =& (a) 
+ \sum_{i,m=0}^N Q_{im} (v_{ijk} - v_{mjk}) \tilde{f}^{1*S}_{(i,m)jk} 
\\ \text{(definition of $\tilde{r}^1$)} \qquad =& (a) 
+ \sum_{i,m=0}^N Q_{im} (\avg{J\va^1}_{(i,m)jk} \cdot (\vpsi_{ijk} - \vpsi_{mjk}) - \tilde{r}^1_{(i,m)jk}) 
\\ \text{(splitting the sum)} \qquad =& (a) + \sum_{i=0}^N \frac{1}{2} (J\va^1)_{ijk} \cdot \vpsi_{ijk} \underbrace{\sum_{m=0}^N Q_{im}}_{=0}
+ \sum_{i,m=0}^N \frac{1}{2} (J\va^1)_{mjk} \cdot \vpsi_{ijk} Q_{im}
\\ &- \sum_{m=0}^N \frac{1}{2} (J\va^1)_{mjk} \cdot \vpsi_{mjk}  \underbrace{\sum_{i=0}^N Q_{im}}_{=\tau_m}
- \sum_{i,m=0}^N \frac{1}{2} (J\va^1)_{ijk} \cdot \vpsi_{mjk} Q_{im}
 \underbrace{-\sum_{i,m=0}^N Q_{im} \tilde{r}^1_{(i,m)jk}}_{=: (b)}   
\\ =& (a) - \sum_{m=0}^N \tau_m \frac{1}{2} (J\va^1)_{mjk} \cdot \vpsi_{mjk} 
+ \sum_{i,m=0}^N \frac{1}{2} (J\va^1)_{mjk} \cdot \vpsi_{ijk} (Q_{im} - Q_{mi}) + (b) 
\\ (\text{SBP property}) \qquad =& (a) 
- \sum_{m=0}^N \tau_m \frac{1}{2} (J\va^1)_{mjk} \cdot  \vpsi_{mjk}
\\ &+ \underbrace{\sum_{i,m=0}^N (J\va^1)_{mjk} \cdot \vpsi_{ijk} Q_{im}}_{=: (c)}
- \sum_{i,m=0}^N \frac{1}{2} (J\va^1)_{mjk} \cdot \vpsi_{ijk} B_{im}
+ (b) 
\\ (\text{Definition of $B$}) \qquad =& (a)
- \sum_{m=0}^N \tau_m (J\va^1)_{mjk} \cdot \vpsi_{mjk}
+ (c)
+ (b) 
\\ \text{(rename sum index)} \qquad =& (a)
- \sum_{i=0}^N \tau_i (J\va^1)_{ijk} \cdot \vpsi_{ijk} 
+ (c)
+ (b) 
\\ =& \sum_{i=0}^{N} \tau_i (v_{ijk} \tilde{f}^1_{ijk} - (J\va^1)_{ijk} \cdot \vpsi_{ijk})
+ (c) 
+ (b).
\end{align}
The surface integral:
\begin{align}
&\sum_{i=0}^{N} v_{ijk} w_i 
\left[\frac{\delta_{iN}}{w_N}  
(\tilde{f}^{1*}_{(N,R)jk} - \tilde{f}^1_{Njk}) 
- \frac{\delta_{i0}}{w_0}  
(\tilde{f}^{1*}_{(L,0)jk} - \tilde{f}^1_{0jk})\right]
\\&= v_{Njk} (\tilde{f}^{1*}_{(N,R)jk} - \tilde{f}^1_{Njk}) 
- v_{0jk} (\tilde{f}^{1*}_{(L,0)jk}-\tilde{f}^1_{0jk}).
\end{align}
Volume and surface integrals added:
\begin{align}
& \sum_{i=0}^{N} \tau_i (v_{ijk} \tilde{f}^1_{ijk} - (J\va^1)_{ijk} \cdot \vpsi_{ijk}) + (c) + (b)
+ v_{Njk} (\tilde{f}^{1*}_{(N,R)jk} - \tilde{f}^1_{Njk}) 
- v_{0jk} (\tilde{f}^{1*}_{(L,0)jk} - \tilde{f}^1_{0jk})
\\=& v_{Njk} \tilde{f}^{1*}_{(N,R)jk} - (J\va^1)_{Njk} \cdot \vpsi_{Njk}  - (v_{0jk} \tilde{f}^{1*}_{(L,0)jk} - (J\va^1)_{0jk} \cdot \vpsi_{0jk}) + (c) + (b)
\\ \text{(watertight mesh)} \quad=& v_{Njk} \tilde{f}^{1*}_{(N,R)jk} - \avg{J\va^1}_{(N,R)jk} \cdot \vpsi_{Njk}  - (v_{0jk} \tilde{f}^{1*}_{(L,0)jk} - \avg{J\va^1}_{(L,0)jk} \cdot \vpsi_{0jk}) + (c) + (b)
\\=& \tilde{q}^{1*}_{(N,R)jk} - \frac{1}{2} \tilde{r}^1_{(N,R)jk} - \tilde{q}^{1*}_{(L,0)jk} - \frac{1}{2} \tilde{r}^1_{(L,0)jk} + (c) + (b),
\end{align}
using \eqref{eq:EntNumFluxDG} and \eqref{eq:EntProdDG}. 
\changed{In the last step, we added and subtracted half left and right neighbour values to the boundary values of $v$ and $\vpsi$.} Integrating over the $\xi^2$ and $\xi^3$ directions and adding the other two flux directions leads to
\begin{align}
\begin{split}
\sum_{i,j,k=0}^N w_{ijk} J_{ijk} \dot{\eta}_{ijk} 
=-& \sum_{j,k=0}^N w_{jk} \left[ \tilde{q}^{1*}_{(N,R)jk} - \tilde{q}^{1*}_{(L,0)jk} - \frac{1}{2} (\tilde{r}^1_{(N,R)jk} + \tilde{r}^1_{(L,0)jk}) \right]
\\ -& \sum_{i,k=0}^N w_{ik} \left[ \tilde{q}^{2*}_{i(N,R)k} - \tilde{q}^{2*}_{i(L,0)k} - \frac{1}{2} (\tilde{r}^2_{i(N,R)k} + \tilde{r}^2_{i(L,0)k}) \right]
\\ -& \sum_{i,j=0}^N w_{ij} \left[ \tilde{q}^{3*}_{ij(N,R)} - \tilde{q}^{3*}_{ij(L,0)} - \frac{1}{2} (\tilde{r}^3_{ij(N,R)} + \tilde{r}^3_{ij(L,0)}) \right]
\\ +& \sum_{i,j,k,m=0}^N w_{ijk} 
[D_{im} \tilde{r}^1_{(i,m)jk} + D_{jm} \tilde{r}^2_{i(j,m)k} + D_{km} \tilde{r}^3_{ij(k,m)}]
\\ -& \sum_{i,j,k,m=0}^N w_{ijk}  \vpsi_{ijk} \cdot
[D_{im} (J\va^1)_{mjk} + D_{jm} (J\va^2)_{imk} + D_{km} (J\va^3)_{ijm}].
\end{split} \label{eq:3d_dg_entropy_balance_integrated}
\end{align}
The last term can be rewritten component wise
\begin{align}
\sum_{i,j,k,m=0}^N w_{ijk} \vpsi_{ijk} \cdot[D_{im} (J\va^1)_{mjk} + D_{jm} (J\va^2)_{imk} + D_{km} (J\va^3)_{ijm}]
\\ = \sum_{i,j,k=0}^N w_{ijk} \sum_{d=1}^3 \psi^d_{ijk} \sum_{m=0}^N [D_{im} (Ja^1_d)_{mjk} + D_{jm} (Ja^2_d)_{imk} + D_{km} (Ja^3_d)_{ijm}],
\end{align}
which is equal zero as long the \changed{discrete} metric identities
\begin{equation}
\label{eq:metric_identities}
\sum_{m=0}^N [D_{im} (Ja^1_d)_{mjk} + D_{jm} (Ja^2_d)_{imk} + D_{km} (Ja^3_d)_{ijm}] = 
\left( \sum_{l=1}^{3} \frac{\partial}{\partial \xi^l} \mathbb{I}^N (Ja_d^l) \right)_{ijk},
\end{equation}
for $d\in \{1,2,3\}$ and $i,j,k \in \{0,\ldots,N\}$ hold. \changed{For example, they can be computed from a discrete curl, see \cite{kopriva2006}.}

\subsection{3D Curvilinear FV Subcell Formulation}

Next we construct the integral formulation of the conservation law for the subcell $C_{ijk}$, which is of size $w_{ijk} := w_i w_j w_k$, as
\begin{equation}
\int_{C_{ijk}} J \dot{u} \,dV + \int_{C_{ijk}} \nabla_\xi \cdot \spvec{\tilde{f}} \,dV = 0.
\end{equation}
We evaluate the time derivative integral with a rectangle quadrature rule at the \ac{lgl} point $\xi_{ijk}$.
\begin{equation}
J_{ijk} \dot{u}_{ijk} w_{ijk} + \int_{C_{ijk}} \nabla_\xi \cdot \spvec{\tilde{f}} \,dV = 0.
\end{equation}
With this rule we have exactly the same left hand side of the equation as in the DG formulation \eqref{eq:3d_dgsem}. Next we apply the gauss theorem to the spatial derivative integral to obtain

\begin{equation}
J_{ijk} \dot{u}_{ijk} w_{ijk} + \int_{\partial C_{ijk}} \spvec{\tilde{f}} \cdot \spvec{\tilde{n}} \,dA = 0,
\end{equation}
where $\spvec{\tilde{n}}$ is the reference normal vector on the subcell surface.
Now evaluate this surface integral and approximate the flux as first order reconstruction numerical fluxes, i.e.

\begin{align}
\begin{split}
\frac{1}{w_{ijk}} \int_{\partial C_{ijk}} \spvec{\tilde{f}} \cdot \spvec{\tilde{n}} \,dA
=&   \frac{1}{w_i} \left[ \tilde{f}^{1*}_{(i,i+1)jk} - \tilde{f}^{1*}_{(i-1,i)jk} \right]
\\+& \frac{1}{w_j} \left[ \tilde{f}^{2*}_{i(j,j+1)k} - \tilde{f}^{2*}_{i(j-1,j)k} \right]
\\+& \frac{1}{w_k} \left[ \tilde{f}^{3*}_{ij(k,k+1)} - \tilde{f}^{3*}_{ij(k-1,k)} \right] \label{eq:3d_fv}.
\end{split}
\end{align}
We describe a numerical flux in $\xi^1$ direction (others analogue) as 
\begin{equation}
\tilde{f}^{1*}_{(i,m)jk}
:= \norm{\vn} \spvec{f}^*\left(u_{ijk}, u_{mjk}; \frac{\vn}{\norm{\vn}} \right) = \sum_{d=1}^{3}  n^d \spvec{f}^*\left(u_{ijk}, u_{mjk}; \spvec{e}^d\right),
\end{equation}
where $\vn$ is the corresponding scaled normal vector on the subcell face, $\spvec{e}^d$ are the unit vectors in cartesian space and $f^*(u_L, u_R; \vn)$ is a general numerical flux approximation of a Riemann problem described with the two states $u_L, u_R$ and a normal direction $\vn$.  Note: \textit{In some literature this normal vector is not explicitly written. Depending on the context, the Riemann problem is then understood as 1D (``rotating the state") or assuming tensor product formulation and fluxes are described as $f, g$ and $h$ for the $\xi^1, \xi^2$ and $\xi^3$ direction respectively.} 

\subsection{Metric Terms of the 3D FV Subcells}
To compute the normal vector for the \ac{fv} method on the curvilinear subcells we take the \ac{dg} split form volume integral formulation as the basement. The contribution of the $\xi^1$ direction for the node $ijk$ reads

\begin{equation}
\frac{1}{w_i} (\tilde{\bar{f}}_{(i,i+1)jk} - \tilde{\bar{f}}_{(i-1,i)jk}) =
2 \sum_{m=0}^{N} D_{im} \tilde{f}^{1*S}_{(i,m)jk},
\end{equation}
in high order flux difference (left) and split form (right) formulations. Equivalently we get

\begin{equation}
\tilde{\bar{f}}_{(i,i+1)jk} = \tilde{\bar{f}}_{(i-1,i)jk} + 2 w_i \sum_{m=0}^{N} D_{im} \tilde{f}^{1*S}_{(i,m)jk}.
\end{equation}
Next substitute this recursive formulation by an explicit one

\begin{equation}
\tilde{\bar{f}}_{(i,i+1)jk} = \tilde{\bar{f}}_{(L,0)jk} + 2 \sum_{l=0}^{i} w_l \sum_{m=0}^{N} D_{lm} \tilde{f}^{1*S}_{(l,m)jk}.
\end{equation}
Since $\tilde{f}^{1*S}_{(l,m)jk} = \spvec{f}^{*S}_{(l,m)jk} \cdot \avg{J\va^1}_{(l,m)jk}$ and $\tilde{\bar{f}}_{(L,0)jk} = \spvec{f}^{*}_{(L,0)jk} \cdot (J\va^1)_{0jk}$ we can extract the metric terms to get

\begin{equation}
\tilde{\bar{f}}_{(i,i+1)jk} =: \spvec{\bar{f}}_{(i,i+1)jk} \cdot \vn_{(i,i+1)jk} = \spvec{f}^{*}_{(L,0)jk} \cdot (J\va^1)_{0jk} + 2 \sum_{l=0}^{i} w_l \sum_{m=0}^{N} D_{lm} \spvec{f}^{*S}_{(l,m)jk} \cdot \avg{J\va^1}_{(l,m)jk},
\end{equation}
where $\vn_{(i,i+1)jk}$ is the normal vector of the subcell interface $(i,i+1)jk$ we are looking for. We want our scheme to fulfill \ac{fsp}, so assuming a constant flux leads to an explicit formula for the normal vectors of the subcell interfaces
\begin{align}
\begin{split}
\label{eq:subcell_normal_vec}
\vn_{(i,i+1)jk} &= (J\va^1)_{0jk} + 2 \sum_{l=0}^{i} w_l \sum_{m=0}^{N} D_{lm} \avg{J\va^1}_{(l,m)jk}
\\&= (J\va^1)_{0jk} + 2 \sum_{l=0}^{i} \sum_{m=0}^{N} Q_{lm} \avg{J\va^1}_{(l,m)jk}
\\&= (J\va^1)_{0jk} + \sum_{l=0}^{i} \sum_{m=0}^{N} Q_{lm} (J\va^1)_{mjk}.
\end{split}
\end{align}
Note that the scaled normal vectors at the boundaries are consistent to the DG metric $\vn_{(L,0)jk}=(J\va^1)_{0jk}$ and $\vn_{(N,R)jk}=(J\va^1)_{Njk}$.

\subsection{Entropy Balance Law of the 3D FV Subcells}
\label{sec:3d_fv_entropy_balance}
Multiplying \eqref{eq:3d_fv} with the entropy variable $v_{ijk}$ leads to the entropy balance law 

\begin{align}
\begin{split}
\label{eq:3d_fv_entropy_balance}
J_{ijk} \dot{\eta}_{ijk} 
=-& \frac{1}{w_i} \left[ \tilde{q}^{1*}_{(i,i+1)jk} - \tilde{q}^{1*}_{(i-1,i)jk} - \frac{1}{2} (\tilde{r}^1_{(i,i+1)jk} + \tilde{r}^1_{(i-1,i)jk}) \right]
\\ -& \frac{1}{w_j} \left[ \tilde{q}^{2*}_{i(j,j+1)k} - \tilde{q}^{2*}_{i(j-1,j)k} - \frac{1}{2} (\tilde{r}^2_{i(j,j+1)k} + \tilde{r}^2_{i(j-1,j)k}) \right]
\\ -& \frac{1}{w_k} \left[ \tilde{q}^{3*}_{ij(k,k+1)} - \tilde{q}^{3*}_{ij(k-1,k)} - \frac{1}{2} (\tilde{r}^3_{ij(k,k+1)} + \tilde{r}^3_{ij(k-1,k)}) \right],
\end{split}
\end{align}
where 
\begin{equation}
\tilde{q}^{1*}_{(i,i+1)jk} := \avg{v}_{(i,i+1)jk} \tilde{f}^{1*}_{(i,i+1)jk} - \vn_{(i,i+1)jk} \cdot \avg{\vpsi}_{(i,i+1)jk}
\end{equation}
is the numerical entropy flux in $\vn_{(i,i+1)jk} $ direction,
\begin{equation}
\tilde{r}^1_{(i,i+1)jk} := \diff{v}_{(i,i+1)jk} \tilde{f}^{1*}_{(i,i+1)jk} - \vn_{(i,i+1)jk} \cdot \diff{\vpsi}_{(i,i+1)jk}
\end{equation}
is the numerical entropy production term in $\vn_{(i,i+1)jk} $ direction and
\begin{equation}
\vpsi = v \spvec{f}(u(v)) - \spvec{q}(u(v))
\end{equation}
is the entropy flux potential. Analogue definitions for the $\vn_{i(j,j+1)k} $ and $\vn_{ij(k,k+1)} $ directions hold. \\
Proof: For a right flux holds
\begin{align}
v_{ijk} \tilde{f}^{1*}_{(i,i+1)jk} - \vn_{(i,i+1)jk} \cdot \vpsi_{ijk}
&= (v_{ijk} - \frac{1}{2} v_{(i+1)jk} + \frac{1}{2} v_{(i+1)jk}) \tilde{f}^{1*}_{(i,i+1)jk} 
\\ &- \vn_{(i,i+1)jk} \cdot [\vpsi_{ijk} - \frac{1}{2} \vpsi_{(i+1)jk} + \frac{1}{2} \vpsi_{(i+1)jk}]
\\&= \frac{1}{2} (v_{(i+1)jk} + v_{ijk}) \tilde{f}^{1*}_{(i,i+1)jk} - \frac{1}{2} \vn_{(i,i+1)jk} \cdot [\vpsi_{(i+1)jk} + \vpsi_{ijk}]
\\&- \frac{1}{2} (v_{(i+1)jk} - v_{ijk}) \tilde{f}^{1*}_{(i,i+1)jk} \changed{+} \frac{1}{2} \vn_{(i,i+1)jk} \cdot [\vpsi_{(i+1)jk} - \vpsi_{ijk}]
\\&= \avg{v}_{(i,i+1)jk} \tilde{f}^{1*}_{(i,i+1)jk} - \vn_{(i,i+1)jk} \cdot \avg{\vpsi}_{(i,i+1)jk}
\\&- \frac{1}{2} \diff{v}_{(i,i+1)jk} \tilde{f}^{1*}_{(i,i+1)jk} \changed{+} \frac{1}{2} \vn_{(i,i+1)jk} \cdot \diff{\vpsi}_{(i,i+1)jk}
\\&= \tilde{q}^{1*}_{(i,i+1)jk} - \frac{1}{2} \tilde{r}^1_{(i,i+1)jk}
\label{eq:3d_entropyFluxRight}
\end{align}
and for a left flux
\begin{align}
v_{(i+1)jk} \tilde{f}^{1*}_{(i,i+1)jk} - \vn_{(i,i+1)jk} \cdot \vpsi_{(i+1)jk}
&= (v_{(i+1)jk} - \frac{1}{2} v_{ijk} + \frac{1}{2} v_{ijk}) \tilde{f}^{1*}_{(i,i+1)jk} 
\\&- \vn_{(i,i+1)jk} \cdot [\vpsi_{(i+1)jk} - \frac{1}{2} \vpsi_{ijk} + \frac{1}{2} \vpsi_{ijk}]
\\&= \frac{1}{2} (v_{(i+1)jk} + v_{ijk}) \tilde{f}^{1*}_{(i,i+1)jk} - \frac{1}{2} \vn_{(i,i+1)jk} \cdot [\vpsi_{(i+1)jk} + \vpsi_{ijk}]
\\&+ \frac{1}{2} (v_{(i+1)jk} - v_{ijk}) \tilde{f}^{1*}_{(i,i+1)jk} - \frac{1}{2} \vn_{(i,i+1)jk} \cdot [\vpsi_{(i+1)jk} - \vpsi_{ijk}]
\\&= \avg{v}_{(i,i+1)jk} \tilde{f}^{1*}_{(i,i+1)jk} - \vn_{(i,i+1)jk} \cdot \avg{\vpsi}_{(i,i+1)jk}
\\&+ \frac{1}{2} \diff{v}_{(i,i+1)jk} \tilde{f}^{1*}_{(i,i+1)jk} - \frac{1}{2} \vn_{(i,i+1)jk} \cdot \diff{\vpsi}_{(i,i+1)jk}
\\&= \tilde{q}^{1*}_{(i,i+1)jk} + \frac{1}{2} \tilde{r}^1_{(i,i+1)jk}.
\label{eq:3d_entropyFluxLeft}
\end{align}
This results in the 1D balance law
\begin{align}
\label{eq:1d_curvilinear_entropy_balance_fv}
\begin{split}
& \frac{1}{w_i} [ v_{ijk} (\tilde{f}^{1*}_{(i,i+1)jk} - \tilde{f}^{1*}_{(i-1,i)jk}) ] 
\\=\ & \frac{1}{w_i} [ v_{ijk} \tilde{f}^{1*}_{(i,i+1)jk} - \vn_{(i,i+1)jk} \cdot \vpsi_{ijk} - v_{ijk} \tilde{f}^{1*}_{(i-1,i)jk} + \vn_{(i-1,i)jk} \cdot \vpsi_{ijk} + \vn_{(i,i+1)jk} \cdot \vpsi_{ijk} - \vn_{(i-1,i)jk} \cdot \vpsi_{ijk} ]
\\=\ & \frac{1}{w_i} [ (\tilde{q}^{1*}_{(i,i+1)jk} - \tilde{q}^{1*}_{(i-1,i)jk}) - \frac{1}{2} (\tilde{r}^1_{(i,i+1)jk} + \tilde{r}^1_{(i-1,i)jk}) + \vpsi_{ijk} \cdot (\vn_{(i,i+1)jk} - \vn_{(i-1,i)jk})] .
\end{split}
\end{align}
We can insert the definition for the subcell normal vector \eqref{eq:subcell_normal_vec} into the surplus part of \eqref{eq:1d_curvilinear_entropy_balance_fv} to obtain
\begin{align}
\label{eq:subcell_normal_vec_difference}
	\frac{\vpsi_{ijk}}{w_i} \cdot [\vn_{(i,i+1)jk} - \vn_{(i-1,i)jk}] 
	= & \frac{\vpsi_{ijk}}{w_i} \cdot [(J\va^1)_{0jk} + \sum_{l=0}^{i} \sum_{m=0}^{N} Q_{lm} (J\va^1)_{mjk} - (J\va^1)_{0jk} - \sum_{l=0}^{i-1} \sum_{m=0}^{N} Q_{lm} (J\va^1)_{mjk}] 
	\\= & \frac{\vpsi_{ijk}}{w_i} \cdot \sum_{m=0}^{N} Q_{im} (J\va^1)_{mjk}
	\\= & \vpsi_{ijk} \cdot \sum_{m=0}^{N} D_{im} (J\va^1)_{mjk}
\end{align}
Following this procedure for all three directions and summing up the previously computed term leads again to equation \eqref{eq:metric_identities}
\begin{equation}
	\vpsi_{ijk} \cdot \sum_{m=0}^{N} D_{im} (J\va^1)_{mjk} + D_{jm} (J\va^2)_{imk}  +D_{km} (J\va^3)_{ijm} = 0,
\end{equation}
if the metric identities hold. Note: \textit{Replacing $\vpsi_{ijk}$ in \eqref{eq:subcell_normal_vec_difference} and following with a constant flux shows explicitly that these metric terms provide \ac{fsp}.} So in summary, if summing up all three directions of the 1D balance law \eqref{eq:1d_curvilinear_entropy_balance_fv} we get the 3D entropy balance law for the subcells \eqref{eq:3d_fv_entropy_balance}.

For the complete element integrated entropy we get

\begin{align}
\begin{split}
\sum_{i,j,k=0}^N w_{ijk} J_{ijk} \dot{\eta}_{ijk} 
=-& \sum_{j,k=0}^N w_{jk} \left[ \tilde{q}^{1*}_{(N,R)jk} - \tilde{q}^{1*}_{(L,0)jk} - \frac{1}{2} (\tilde{r}^1_{(N,R)jk} + \tilde{r}^1_{(L,0)jk}) - \sum_{i=0}^{N-1} \tilde{r}^1_{(i,i+1)jk} \right]
\\ -& \sum_{i,k=0}^N w_{ik} \left[ \tilde{q}^{2*}_{i(N,R)k} - \tilde{q}^{2*}_{i(L,0)k} - \frac{1}{2} (\tilde{r}^2_{i(N,R)k} + \tilde{r}^2_{i(L,0)k}) - \sum_{j=0}^{N-1} \tilde{r}^2_{i(j,j+1)k} \right]
\\ -& \sum_{i,j=0}^N w_{ij} \left[ \tilde{q}^{3*}_{ij(N,R)} - \tilde{q}^{3*}_{ij(L,0)} - \frac{1}{2} (\tilde{r}^3_{ij(N,R)} + \tilde{r}^3_{ij(L,0)}) - \sum_{k=0}^{N-1} \tilde{r}^3_{ij(k,k+1)} \right].
\end{split} \label{eq:3d_fv_entropy_balance_integrated}
\end{align}

We can separate the inner entropy production terms when reindexing sums to get form that splits surface and volume terms
\begin{align}
\begin{split}
\sum_{i,j,k=0}^N w_{ijk} J_{ijk} \dot{\eta}_{ijk} 
=-& \sum_{j,k=0}^N w_{jk} \left[ \tilde{q}^{1*}_{(N,R)jk} - \tilde{q}^{1*}_{(L,0)jk} - \frac{1}{2} (\tilde{r}^1_{(N,R)jk} + \tilde{r}^1_{(L,0)jk}) \right]
\\ -& \sum_{i,k=0}^N w_{ik} \left[ \tilde{q}^{2*}_{i(N,R)k} - \tilde{q}^{2*}_{i(L,0)k} - \frac{1}{2} (\tilde{r}^2_{i(N,R)k} + \tilde{r}^2_{i(L,0)k}) \right]
\\ -& \sum_{i,j=0}^N w_{ij} \left[ \tilde{q}^{3*}_{ij(N,R)} - \tilde{q}^{3*}_{ij(L,0)} - \frac{1}{2} (\tilde{r}^3_{ij(N,R)} + \tilde{r}^3_{ij(L,0)}) \right]
\\ +& \left( \sum_{j,k=0}^N w_{jk} \sum_{i=0}^{N-1} \tilde{r}^1_{(i,i+1)jk} + \tilde{r}^2_{j(i,i+1)k} + \tilde{r}^3_{jk(i,i+1)} \right).
\end{split} \label{eq:3d_fv_entropy_balance_integrated_reindex}
\end{align}

\subsection{Entropy Balance Law of the Hybrid Scheme}
The combined approximation ansatz for a single LGL-node/subcell reads
\begin{align}
\begin{split}
	- J_{ijk} \dot{u}_{ijk} =  \alpha \Bigl( &\frac{1}{w_i} \left[ \tilde{f}^{1*}_{(i,i+1)jk} - \tilde{f}^{1*}_{(i-1,i)jk} \right]
	\\+& \frac{1}{w_j} \left[ \tilde{f}^{2*}_{i(j,j+1)k} - \tilde{f}^{2*}_{i(j-1,j)k} \right]
	\\ +& \frac{1}{w_k} \left[ \tilde{f}^{3*}_{ij(k,k+1)} - \tilde{f}^{3*}_{ij(k-1,k)} \right] \Bigr)
	\\  + (1-\alpha)  \Bigl[& 2\sum_{m=0}^{N} D_{im} \tilde{f}^{1*\changed{S}}_{(i,m)jk} 
	+\frac{1}{w_i} \left( \delta_{iN} \left[ \tilde{f}^{1*}_{(N,R)jk} - \tilde{f}^1_{Njk}\right] - \delta_{i0} \left[\tilde{f}^{1*}_{(L,0)jk} - \tilde{f}^1_{0jk} \right]\right)
	\\ +& 2\sum_{m=0}^{N} D_{jm} \tilde{f}^{2*\changed{S}}_{i(j,m)k}  
	+\frac{1}{w_j} \left( \delta_{jN} \left[ \tilde{f}^{2*}_{i(N,R)k} - \tilde{f}^2_{ipk}\right] - \delta_{j0} \left[\tilde{f}^{2*}_{i(L,0)k} - \tilde{f}^2_{i0k} \right]\right) 
	\\ +& 2\sum_{m=0}^{N} D_{km} \tilde{f}^{3*\changed{S}}_{ij(k,m)} 
	+\frac{1}{w_k} \left( \delta_{kp} \left[ \tilde{f}^{3*}_{ij(N,R)} - \tilde{f}^3_{ijN}\right] - \delta_{k0} \left[\tilde{f}^{3*}_{ij(L,0)} - \tilde{f}^3_{ij0} \right]\right) \Bigr].
\end{split}
\label{eq:blendend_rhs_3d}
\end{align}

\begin{align}
\begin{split}
\sum_{i,j,k=0}^N w_{ijk} J_{ijk} \dot{\eta}_{ijk} 
= -& \sum_{j,k=0}^N w_{jk} \left[ \tilde{q}^{1*}_{(N,R)jk} - \tilde{q}^{1*}_{(L,0)jk} - \frac{1}{2} (\tilde{r}^1_{(N,R)jk} + \tilde{r}^1_{(L,0)jk}) \right]
\\ -& \sum_{i,k=0}^N w_{ik} \left[ \tilde{q}^{2*}_{i(N,R)k} - \tilde{q}^{2*}_{i(L,0)k} - \frac{1}{2} (\tilde{r}^2_{i(N,R)k} + \tilde{r}^2_{i(L,0)k}) \right]
\\ -& \sum_{i,j=0}^N w_{ij} \left[ \tilde{q}^{3*}_{ij(N,R)} - \tilde{q}^{3*}_{ij(L,0)} - \frac{1}{2} (\tilde{r}^3_{ij(N,R)} + \tilde{r}^3_{ij(L,0)}) \right]
\\ +& (1 - \alpha) \sum_{i,j,k,m=0}^N w_{ijk} 
[D_{im} \tilde{r}^1_{(i,m)jk} + D_{jm} \tilde{r}^2_{i(j,m)k} + D_{km} \tilde{r}^3_{ij(k,m)}]
\\ +& \alpha \left( \sum_{j,k=0}^N w_{jk} \sum_{i=0}^{N-1} \tilde{r}^1_{(i,i+1)jk} + \tilde{r}^2_{j(i,i+1)k} + \tilde{r}^3_{jk(i,i+1)} \right).
\end{split} \label{eq:3d_blended_entropy_balance_integrated}
\end{align}
Analogue to the 1D case, we can now choose \ac{ec} and \ac{es} fluxes to wipe out unwanted entropy production terms $\tilde{r}$.

\section{Some short Notes on other possible FV Representations}
\label{sec:otherFVrepresentations}

In contradiction to the natural subcell approach \eqref{FV_as_volume_surface_split}, the values could be chosen as integrals of the \ac{dg} polynomial for an arbitrary sub grid,
\begin{equation}
u^{\FV}_j = \int_{x_j^l}^{x_j^r} u^{\DG}(\xi) 	\mathrm{d\xi},
\end{equation}
$j\in \{0,\ldots,n\}$, $n\in\mathbb{N}$. Define this transformation as $T$, i.e.
\begin{equation}
\underline{u}^{\FV} = T \underline{u}^{\DG}.
\end{equation}
It would be desirable to define some operator similar to \eqref{eq:RHS_blended}, like
\begin{equation}
\underline{R}(\underline{u}^{\DG}) := \alpha T^{-1}\underline{R}^{\FV}(T\underline{u}^{\DG}) + (1-\alpha) \underline{R}^{\DG}(\underline{u}^{\DG}), \label{RHS_blended_general}
\end{equation}
but although the construction of the \ac{fv} values for the conservative variables happens in a conservative way, the blended \changed{residual} operator is not in a conservative form, without investing further effort on $T^{-1}$. Therefore we discard this general \ac{fv} subcell partitioning approach and focus the research on the natural subcells.\\

\addcontentsline{toc}{section}{References} % Literaturliste soll im Inhaltsverzeichnis auftauchen
\bibliographystyle{model1-num-names} %Vancouver style references.
\bibliography{literatur} % Literaturliste endgueltig anzeigen

\end{document}